\documentclass[preprint,12pt]{elsarticle}

\usepackage{amssymb}

\newcommand{\HiRep}{\texttt{HiRep}~}

\usepackage{comment}
\usepackage{physics}
\usepackage{graphicx}
\usepackage{color}
\usepackage{float}
\usepackage{booktabs}
\usepackage{tikz}
\usepackage{float}
\usepackage{listings}
\usepackage{fancyvrb}
\usepackage{subcaption}
\usepackage{bm}
\usepackage{dsfont}
\usepackage{slashed}
\usepackage{afterpage}
\usepackage{algorithm}
\usepackage{algpseudocode}
\usepackage{hyperref}
\usepackage[export]{adjustbox}
\usepackage{cancel}

\usetikzlibrary{decorations.markings}

\definecolor{myred}{RGB}{153,0,69}
\definecolor{myblue}{RGB}{22,83,126}
\definecolor{mylightblue}{RGB}{61,133,198}
\definecolor{mylightred}{RGB}{224,102,102}
\definecolor{mygray}{RGB}{240,240,240}

\definecolor{rgbred}{RGB}{242, 21, 10}
\definecolor{rgbgreen}{RGB}{94, 237, 33}
\definecolor{rgbblue}{RGB}{17, 29, 245}

\definecolor{palette1}{RGB}{242, 21, 10} 
\definecolor{palette2}{RGB}{17, 222, 245} 
\definecolor{palette3}{RGB}{252, 241, 23} 
\definecolor{palette4}{RGB}{21, 138, 17} 
\definecolor{palette5}{RGB}{17, 138, 118} 
\definecolor{palette6}{RGB}{131, 20, 179} 
\definecolor{palette7}{RGB}{107, 202, 242} 
\definecolor{palette8}{RGB}{94, 237, 33} 
\definecolor{palette9}{RGB}{252, 153, 23} 
\definecolor{palette10}{RGB}{255, 92, 171} 
\definecolor{palette11}{RGB}{16, 227, 132} 
\definecolor{palette12}{RGB}{199, 247, 7} 
\definecolor{palette13}{RGB}{14, 37, 171} 
\definecolor{palette14}{RGB}{171, 14, 106} 
\definecolor{palette15}{RGB}{17, 29, 245} 

\definecolor{pastelgreen}{RGB}{180, 222, 189}
\definecolor{beige}{RGB}{255, 247, 221}
\definecolor{softgray}{RGB}{240, 240, 240}
\definecolor{softblue}{RGB}{128, 161, 186}

\usepackage{soul}

\newcounter{bla}

\journal{Computer Physics Communications}

\begin{document}

\begin{frontmatter}



\title{High-Performance Simulations of Higher Representations of Wilson Fermions}


\author[a]{Vincent Drach}

\author[b]{Sofie Martins\corref{author}}

\author[b,c]{Claudio Pica}

\author[b]{Antonio Rago}

\cortext[author] {Corresponding author.\\\textit{E-mail address:}sofie.martins@uni-graz.at, \\\textit{permanent address:} Institute of Physics, NAWI Graz, University of Graz, Universitätsplatz 5, 8010 Graz, Austria}
\address[a]{Centre for Mathematical Sciences, University of Plymouth, England, UK}
\address[b]{QTC \& IMADA, University of Southern Denmark, Campusvej 55, 5230 Odense M, Denmark}
\address[c]{eScience Center, University of Southern Denmark, Campusvej 55, 5230 Odense M, Denmark}

\begin{abstract}
We present \HiRep v2, an open-source software suite for high-performance lattice field theory simulations with dynamical Wilson fermions in higher representations of $SU(N_g)$ gauge groups. This new version fully supports graphics processing unit (GPU) acceleration, optimizing both gauge configuration generation and measurements for NVIDIA and AMD GPUs. \HiRep v2 integrates improved gauge and fermionic lattice actions, advanced inverters, and Monte Carlo algorithms, including the (Rational) Hybrid Monte Carlo ((R)HMC) with Hasenbusch acceleration. It exhibits excellent scalability across multiple GPUs and nodes with minimal efficiency loss, making it a robust tool for large-scale simulations in physics beyond the Standard Model.
\\


\noindent \textbf{PROGRAM SUMMARY/NEW VERSION PROGRAM SUMMARY}

\begin{small}
\noindent
{\em Program Title:}  HiRep                                        \\
{\em CPC Library link to program files:} (to be added by Technical Editor) \\
{\em Developer's repository link:} https://github.com/claudiopica/hirep \\
{\em Licensing provisions(please choose one):} GPLv2  \\
{\em Programming language:} C, CUDA C, C++                                 \\
{\em Supplementary material:}                                 \\
{\em Journal reference of previous version:}*                  \\
{\em Does the new version supersede the previous version?:}*   \\
{\em Reasons for the new version:*}\\
{\em Summary of revisions:}*\\
{\em Nature of problem(approx. 50-250 words):}\\
Lattice Field Theory has proven indispensable for the quantitative understanding of strongly coupled quantum field theories, specifically in providing non-perturbative input to phenomenological models describing the dynamics of Quantum Chromodynamics (QCD) for precision tests of the Standard Model. Simulation software libraries for lattice calculations in QCD are readily available and optimized to run on heterogeneous CPU-GPU architectures with good scaling properties on modern supercomputers. 

In direct searches for physics beyond the Standard Model, software is needed that can simulate gauge groups other than $SU(3)$ and allow for fermions in higher representations, catering, among other things, to classes of composite Higgs and technicolor theories~\cite{Cacciapaglia:2020kgq}, and predictions in the large-$N_g$ limit~\cite{Schaich:2022xgy}.

There exists no other open-source library that implements the option for higher representations of Wilson fermions with general numbers of colors, that has as many capabilities in terms of actions and measurement code as \HiRep. \\

{\em Solution method(approx. 50-250 words):}\\
A central element of \HiRep is the implementation of a Dirac operator and optimized linear algebra routines that generalize to higher representations and general gauge groups. Since the application of the Dirac operator is one of the main bottlenecks of the numerical simulation, optimizations of the Dirac operator are a central part of any high-performance software implementations.

In this work, we present a series of significant developments and enhancements to the \HiRep suite. These include but are not limited to, the porting of the code to GPUs, see also \cite{Martins:2024dew, Martins:2024sdd} for previous progress reports, improvements in computational efficiency, and the introduction of new features to further support advanced lattice simulations. In particular, we show that independent of the theory chosen, our implementation of the Dirac operator reaches excellent performance on GPUs and that the software scales well on state-of-the-art supercomputers to a large number of compute nodes, and \HiRep is suitable for simulations of light fermionic masses on large lattices. Another recent performance improvement was achieved in \cite{openMpHiRep}, optimizing OpenMP support.\\
  
{\em Additional comments including restrictions and unusual features (approx. 50-250 words):}\\
None.\\


* Items marked with an asterisk are only required for new versions
of programs previously published in the CPC Program Library.\\
\end{small}
   \end{abstract}
\end{frontmatter}
\clearpage

\tableofcontents

\clearpage

\section{Introduction}\label{sec:introduction}

Lattice Field Theory has proven indispensable for the quantitative understanding of strongly coupled quantum field theories, specifically in providing non-perturbative input to phenomenological models describing the dynamics of Quantum Chromodynamics (QCD) for precision tests of the Standard Model. Simulation software libraries for lattice calculations in QCD are readily available and optimized to run on heterogeneous CPU-GPU architectures with good scaling properties on modern supercomputers, for example, in the GPU porting of \texttt{OpenQCD} \cite{openqcd, Luscher:2012av, openQCDgpu} and the development of the library \texttt{QUDA} \cite{Clark:2009wm, Babich:2011np}, which is used in the porting of suites such as \texttt{Chroma} \cite{Edwards:2004sx}, \texttt{tmLQCD} \cite{Jansen:2009xp, Abdel-Rehim:2013wba, tmLQCD, Kostrzewa:2022hsv}, and \texttt{OpenQ$^{\star}$D} \cite{Campos:2019kgw, Gruber:2022xnl, openQstarDgpu}. Newer libraries such as SIMULATeQCD \cite{HotQCD:2023ghu} are written for GPU accelerators specifically. 

In direct searches for physics beyond the Standard Model, software is needed that can simulate gauge groups other than $SU(3)$ and allow for fermions in higher representations, catering, among other things, to classes of composite Higgs and technicolor theories~\cite{Cacciapaglia:2020kgq}, and predictions in the large-$N_g$ limit~\cite{Schaich:2022xgy}.

This flexibility is offered by simulation suites that allow for GPU simulations of general gauge groups. \texttt{LatticeGPU} \cite{latticegpu, juliaLatticeGPU} written in \texttt{julia} allows pure gauge $SU(2)$ and $SU(3)$ simulations. The recent extension of \texttt{Grid} \cite{Boyle:2016lbp, Yamaguchi:2022feu, Bennett:2023gbe} supports single and mixed higher-representations of Wilson and Staggered fermions with $SU(N_g)$ gauge groups with $N_g=2,3,4,5$ with GPU acceleration. \texttt{HILA} \cite{Hila:2025} supports higher representations of Wilson fermions for general numbers of colors.
To this date, the most flexible open-source high-performance lattice software suite supporting simulations of higher representations of Wilson fermions with general gauge groups is \HiRep \cite{DelDebbio:2008zf, hirep_repo}.

A central element of \HiRep is the implementation of a Dirac operator and optimized linear algebra routines that generalize to higher representations and general gauge groups. Since the application of the Dirac operator is one of the main bottlenecks of the numerical simulation, optimizations of the Dirac operator are a central part of any high-performance software implementations.

In this work, we present a series of significant developments and enhancements to the \HiRep suite. These include but are not limited to, the porting of the code to GPUs, see also \cite{Martins:2024dew, Martins:2024sdd} for previous progress reports, improvements in computational efficiency, and the introduction of new features to further support advanced lattice simulations. In particular, we show that independent of the theory chosen, our implementation of the Dirac operator reaches excellent performance on GPUs and that the software scales well on state-of-the-art supercomputers to a large number of compute nodes, and \HiRep is suitable for simulations of light fermionic masses on large lattices. Another recent performance improvement was achieved in \cite{openMpHiRep}, optimizing OpenMP support.

The paper is organized as follows. In Sect.~\ref{sec:hirep}, we set the basic notation for the Wilson-Dirac operator used in this paper. In Sect.~\ref{sec:implementation}, we describe the detailed software implementation used in \HiRep v2. 
We present in Sect.~\ref{sec:algorithms} the algorithms supported by \HiRep v2 for the inverters, integrators, and the generation of gauge configurations and in Sect.~\ref{sec:actions} for the gauge, fermionic actions and boundary conditions.  
Finally, in Sect.~\ref{sec:tests}, numerical tests supporting the correctness of the \HiRep v2 code are shown, and in  Sect.~\ref{sec:benchmarks} we present representative benchmarks on modern supercomputer architectures.


\section{Basic Definitions}\label{sec:hirep}

The following basic definitions introduce the core elements of the software as concisely as possible. More depth in the origin of these formulas and possible extensions, such as Symanzik improvement or Hasenbusch acceleration, are discussed in the corresponding sections later in this paper. Note that the basic definitions are identical to the explanations and definitions in \cite{DelDebbio:2008zf}.

\subsection{Lattice Geometry}
\HiRep is fixed to simulate four-dimensional hypercubic lattices of finite spatial volume $V_3=XYZ$ and time extent $T$. Any sum over the lattice, often abbreviated as a sum over $x$ or a similar site dependence, has to be understood as a sum over the full four-dimensional lattice volume $V_4 =TXYZ$. We conventionally denote Lorentz indices to be running from $\mu = 0,\ldots,3$, with $\mu=0$ corresponding to the time direction. 

\subsection{Pauli matrices}

Using the Pauli matrices
\begin{equation}
\sigma_1 = \begin{pmatrix}0 & 1 \\ 1 & 0\end{pmatrix}\qquad\sigma_2=\begin{pmatrix}0 & -\mathrm{i} \\ \mathrm{i} & 0\end{pmatrix}\qquad\sigma_3 = \begin{pmatrix}1 & 0 \\ 0 & -1 \end{pmatrix}
\end{equation}
and using the definition $e_\mu$ as
\begin{equation}
e_{0} = -\mathds{1}_{2x2}\qquad e_{i} = -\mathrm{i}\sigma_i
\end{equation}
we define the Gamma matrices in the following basis
\begin{equation}
\gamma_{\mu} = \begin{pmatrix}0 & e_{\mu} \\ e^{\dagger}_{\mu} & 0\end{pmatrix}\,.
\end{equation}

\subsection{Wilson-Dirac operator}

Wilson fermions in the fundamental representation of the gauge group are defined via the discretized Wilson-Dirac operator:
\begin{equation}
D = \dfrac{1}{2}\sum_{\mu}\left[\gamma_\mu (\nabla_{\mu} + \nabla_{\mu}^{\star}) - a\nabla_{\mu}^{\star}\nabla_{\mu}\right]\,,
\end{equation}
with the backward and forward covariant derivatives $\nabla$ and $\nabla^{\star}$ defined as
\begin{equation}
\nabla_{\mu}f(x) = \dfrac{1}{a}\left[U(x,\mu)f(x + \mu)-f(x)\right]
\end{equation}
and
\begin{equation}
\nabla_{\mu}^{\star}f(x) = \dfrac{1}{a}\left[f(x) - U(x-\mu,\mu)^{\dagger}f(x-\mu)\right]\,,
\end{equation}
with the lattice gauge link variables $U(x,\mu)$ elements of the gauge group. The generalization of the Wilson-Dirac operator to general fermion representations makes use of the represented gauge field 

\begin{equation}
U^{R}(x, \mu) = \mathrm{exp}[\mathrm{i}\omega^{a}_{\mu}(x)T_R^a]\,,
\end{equation}
with $\omega^{a}_{\mu}(x)$ defined as the algebra vector of the gauge field transforming the fundamental representation. The represented gauge field $U^{R}(x, \mu)$

 acts on the spinor field $\psi^{R}(x)$ in the given representation $R$. The generalized operator is defined as 
\begin{align}
D_{m}^R \psi^{R}(x) &\equiv (D^R+m)\psi^{R}(x)
=(\tfrac{4}{a} + m)\psi^{R}(x) -\\\nonumber
&-\tfrac{1}{2a}\sum_{\mu}\bigl[(1-\gamma_\mu)\times U^{R}(x,\mu)\psi^{R}(x+\mu) \\
&\qquad+ (1+\gamma_{\mu})U^{R}(x-\mu, \mu)^{\dagger}\psi^{R}(x-\mu) \bigr]\,,
\end{align}
where a diagonal mass term for the fermion has also been introduced. In the following, the superscript $R$ is dropped for simplicity.

\begin{table}
\begin{tabular}{lccc}
\toprule
Rep. $R$ & Dim. $D_R$ & Normalization $T_R$ & Casimir $C_2(R)$\\
\midrule
FUND & $N_g$ & $1/2$ &  $(N^2-1)/(2N)$\\
ADJ & $N_g^2 - 1$ & $N$ &  $N$ \\
SYM & $N_g(N_g + 1)/2$ & $(N+2)/2$ & $C_2(\mathrm{FUND})2(N+2)/(N+1)$ \\
2AS & $N_g(N_g - 1)/2$ & $(N-1)/2$ &  $C_2(\mathrm{FUND})2(N-2)/(N-1)$\\
\bottomrule
\end{tabular}
\caption{Overview of supported representations $R$ denoted by their respective abbreviations, their dimensions $D_R$, their trace normalizations $T_R$, and Casimir operators $C_R$.}
\label{tbl:repsummary}
\end{table}

\texttt{HiRep} supports dynamical fermions in the fundamental representation, denoted in the following with $\mathrm{FUND}$, the adjoint representation, $\mathrm{ADJ}$, the two-index symmetric representation, $\mathrm{SYM}$, and the two-index antisymmetric representation and $\mathrm{2AS}$, see Tbl.~\ref{tbl:repsummary}. 

In this text, we further use the following variants of the Wilson-Dirac operator: The massive Wilson-Dirac operator
\begin{equation}
D_{m}\phi(x) = (D + m)\phi(x)\,,
\label{eq:Dm}
\end{equation}
and the massive Hermitian Wilson-Dirac operator
\begin{equation}
Q_m = \gamma_5 D_m\,.
\label{eq:Qm}
\end{equation}

\subsection{Actions}

The purpose of this library is to simulate theories with fermions in higher representations. The Lagrangian associated with these theories is
\begin{equation}
\mathcal{L} = -\dfrac{1}{4}G_{\mu\nu}^{a}G_{\mu\nu}^{a} + \overline{\Psi}_{i}^{R}(i\slashed{D}^{R}\delta_{ij} - M_{ij})\Psi_{j}^{R} + \text{h.c.}
\end{equation}
where $\Psi_i^{R}$ are Dirac fermions of $i=1,\ldots,N_f$ flavors, which transform under the representation $R$ and have masses defined by the mass matrix $M_{ij}$. There are many possibilities how to discretize this theory and tune the particle content to the desired application, including a pure gauge theory.

The simplest action that can be simulated in \HiRep is a Wilson pure gauge action. Wilson fermions can be added in addition. In the following, we will reference the plaquette, which is defined as
\begin{equation}
P_{\mu\nu}(U) = U_\mu(x)U_{\nu}(x+\mu)U_{\mu}^{\dagger}(x+\nu)U_{\nu}^{\dagger}(x)\,.
\label{eq:plaquette}
\end{equation}

\subsection{Hybrid Monte Carlo}

Simulations with fermions rely on the use of the Hybrid Monte Carlo (HMC). This performs a solution of Hamilton's equations describing a Molecular Dynamics evolution to produce an updated gauge configuration. To perform the Hamiltonian evolution, we introduce the Hamiltonian as the sum of the conjugate momenta, gauge, and fermionic action
\begin{equation}
H = H_{\pi} + H_{g} + H_f
\end{equation}
with
\begin{equation}
H_{\pi} = \dfrac{1}{2}T_{\mathrm{FUND}}\sum_{x,\mu}\pi^{a}_{\mu}(x)^2
\end{equation}
\begin{equation}
H_g = S_g \qquad H_f = S_f\,.
\end{equation}
Hamilton's equations will then be solved using a symplectic integrator, where, in the first step, the gauge fields are updated from the conjugate momenta
\begin{equation}
\dfrac{\mathrm{d}}{\mathrm{d}\tau}U^{\tau}_\mu = \pi^{\tau}_{\mu}(x)U^{\tau}_{\mu}(x)\,,
\end{equation}
where $\tau$ denotes the molecular dynamics evolution time, which we will only explicitly write here. The momenta are subsequently updated using the force associated with the Hamiltonian
\begin{equation}
\dfrac{\mathrm{d}}{\mathrm{d}\tau}\pi_{\mu}(x) = F_{\mu}(x)\,.
\end{equation}
For the gauge dynamics, the force is given by
\begin{equation}
F_{\mathrm{G},\mu}^{a} = \dfrac{\beta}{N_gT_{\mathrm{FUND}}}\mathrm{Re\,tr}[\mathrm{i}T_{R}^{a}U_{\mu}(x)V_{\mu}^{\dagger}(x)]\,,
\end{equation}
where $V$ denotes the staples attached to the link $U$. Using the spinor fields $\eta=(Q_m^2)^{-1}\psi$ and $\xi = Q_m \eta$ we can define the traced operator
\begin{equation}
T_{\mu}(x) = \mathrm{tr}_{\mathrm{Spin}}[\gamma_5(1-\gamma_\mu)\{\eta(x+\mu)\otimes\xi^{\dagger}(x) + \xi(x+\mu)\otimes \eta^{\dagger}(x)\}]\,,
\end{equation}
which allows the definition of the fermion forces via the projection
\begin{equation}
F_{\mathrm{F}}^{a}(x) = -\dfrac{1}{T_{\mathrm{FUND}}}\mathrm{Re\,tr}[U^{R}_{\mu}(x)T_{\mu}(x)]
\end{equation}

On this basis, algorithmic improvements are implemented in \texttt{HiRep}, which are discussed in further detail in Sect.~\ref{sec:algorithms} and \ref{sec:actions}.

\section{Code Organisation}

The code will be continuously developed to incorporate new research developments. However, we want to give an overview of which parts of this paper relate to which parts of this code in the current form. This may change in the future, so the code is described here in a central location to avoid interrupting the comprehension of permanent features by outdated links to the code.

At the top level, the code is organized as follows: The header files are located in \texttt{Include} and the library source code in \texttt{LibHR}. The abbreviation \textit{HR} for \HiRep is used throughout the code. Unit tests are located in \texttt{TestProgram} and benchmarking and performance testing are available from \texttt{Benchmarks}, which were used to obtain the results in Sect.~\ref{sec:benchmarks}. The directory \texttt{Doc} contains the \texttt{Doxygen} function reference setup and markdown files included in the \texttt{GitHub Pages}.

\HiRep implements different higher-level programs, the most important one is the code for simulations and gauge configuration generation, located in \texttt{HMC} and \texttt{PureGauge}. The HMC program also runs the RHMC, while the \texttt{PureGauge} function implements pure gauge algorithms such as heatbath and overrelaxation. When compiling, this program will be linked against the library. These higher-level programs, which only use the library, are later referred to as \textit{binaries} or \textit{programs}, as they implement a main function and produce an executable. 

There are other programs available in \texttt{HiRep}, which have the main purpose of extracting physical observables by analysing configurations, these are \texttt{Disconnected}, \texttt{GaugeFix}, \texttt{ModeNumber}, \texttt{RenormalizationFactors}, \texttt{Reweight}, \texttt{Scattering}, \texttt{Spectrum}, \texttt{StaticPotential}, and \texttt{WilsonFlow}. \texttt{Converter} allows for manipulating configurations to either be compatible with other libraries or produce other types of new useful configurations, for example, for thermalization.

In \texttt{LibHR}, the main implementations are located in \texttt{LibHR/Inverters}, \texttt{LibHR/Observables}, and \texttt{LibHR/Update}. In \texttt{LibHR/Inverters}, linear algebra and inverters of the Wilson-Dirac operator are implemented, which are described in Sect.~\ref{sec:linearalgebra} from the implementation point of view and Sect.~\ref{sec:inverters} from the algorithmic point of view. \texttt{LibHR/Observables} contains the core operator evaluation code for the material presented in Sect.~\ref{sec:observables}. The code that is focused on the gauge update during the simulation and therefore contains the implementation of the features in Sect.~\ref{sec:HMC} and Sect.~\ref{sec:puregaugesimulations} is located in \texttt{LibHR/Update}. This folder also includes implementations of monomials (\texttt{LibHR/Update/mon*}), which, depending on the action selected from Sect.~\ref{sec:actions}, can be flexibly used to build a target action term by term, and their associated forces (\texttt{LibHR/Update/force*}). Additionally, it contains the implementation of the Wilson-Dirac operator (\texttt{LibHR/Update/D*}), as described in Sect.~\ref{sec:diracimplementation}.

The core of lattice spectroscopy methods are implemented in different sources \texttt{LibHR/Observables/sources.c} with scattering tools available in \texttt{LibHR/Observables/measure\_scattering*}

Core implementations for optimization of memory accesses and arrangement, see Sect.~\ref{sec:geometry}, \ref{sec:optimizedaccess} and \ref{sec:memoryaccess}, of the field data in memory, both for single-threading and multi-threading setups, as well as buffers and halo regions that optimize MPI communications, see Sect.~\ref{sec:MPI}, are located in \texttt{LibHR/Geometry}.

\section{Implementation}\label{sec:implementation}
In this section, we present the details of the implementation used in \HiRep v2, starting from the basic data structures and linear algebra operations on fields. We then discuss how fields are laid out in memory and how communications across compute nodes occur.
\subsection{Data structures}
\subsubsection{Complex Numbers}
Field data is generally stored as double-precision numbers, but single-precision field implementations and corresponding operators are available. The central processing unit (CPU) implementation uses the C99 native complex numbers; however, CUDA (Compute Unified Device Architecture), which is a dialect of C++, requires a different implementation: in this case, complex numbers are implemented as custom types in our code, including all relevant operators, and do not require any external library.
\subsubsection{Elementary Data Types}
The links of the gauge field are defined by $SU(N_g)$ matrices for the fundamental gauge field and $SU(D_{R})$ matrices for the represented gauge field. Each of these matrices is stored as a lexicographically indexed vector of double-precision complex or real numbers, depending on the representation being complex or real.

The $SU(D_R)$ matrices act on $D_R$ dimensional complex vectors in color space. Dirac spinors have four components, so a single site of a spinor field consists of four vectors in color space.
\begin{table}[H]
\centering
\begin{tabular}{ccc}
\toprule
Type Name & Description & Vector length \\
\midrule
\texttt{suNg} & $SU(N_g)$ matrix &  $N_g^2$ \\ 
\texttt{suNf} & $SU(D_R)$ matrix &  $D_R^2$ \\
\texttt{suNg\_vector} & $SU(N_g)$ vector & $N_g$ \\
\texttt{suNf\_vector} & $SU(D_R)$ vector & $D_R$ \\
\texttt{suNg\_spinor} & Fundamental spinor & $4N_g$ \\
\texttt{suNf\_spinor} & Represented spinor & $4D_R$ \\
\bottomrule
\end{tabular}
\caption{Overview of elementary data types implemented for the Dirac operator for higher representations of Wilson fermions. Data length is given in consecutive double-precision complex numbers. If the representation is real, all represented types are saved as real vectors instead.}
\end{table}
\subsubsection{Field Types}
Field types are defined as data structures containing an array of elementary data types, one or more per lattice site. The gauge field type contains four gauge links per site corresponding to the positive space-time directions. The elementary types above contain the data for a single site or link. In practice, we map the gauge field links to the sites in the usual way, storing four links in the up-direction on a single site. 
The structure defining the field types also contains further information about the geometry, i.e., the mapping of the four-dimensional Euclidean sites to linear memory and additional fields necessary for the communication and memory management of the field.
\begin{table}[H]
\centering
\begin{tabular}{cccc}
\toprule
Type Name & Description & Site Type & Size \\
\midrule
\texttt{suNg\_field} & Fundamental gauge field & \texttt{suNg} & 4 \\
\texttt{suNf\_field} & Represented gauge field & \texttt{suNf} & 4 \\
\texttt{spinor\_field} & Spinor field & \texttt{suNf\_spinor} & 1 \\
\bottomrule
\end{tabular}
\caption{Overview of field types implemented for the Dirac operator for higher representations of Wilson Fermions}
\end{table}
The implementation of the higher representation is encapsulated in the functions that translate the fundamental gauge field to the represented one, as described in detail in the appendix of \cite{DelDebbio:2008zf}. 

At the implementation level, the represented field is allocated separately, except when the fermions are in the fundamental representation. While this implementation uses two fields, the representation of the gauge field could have been computed on the fly at each application, thereby trading more computations for less memory. This is similar to optimization suggestions such as computing neighboring indices on the fly instead of storing them and reading from memory. While the na\"ive expectation is that the memory bound kernel would be greatly sped up by shifting memory reads to the underutilized compute capabilities, no experimental sped up could be realized in our optimization attempts. This is mainly, because the kernel resources are greatly limited by register counts. The number of necessary registers is increased with $D_R$, see more discussion later in Sect.~\ref{sec:singlegputests} and especially Fig.~\ref{fig:bwlargeN}. Any more complicated implementation, such as the computing of neighbors and representation of matrices increases the number of registers, which limit kernel resources and, consequently, parallelism. The on-the-fly kernel we implemented was approximately 10\% slower than the kernel reading the indices from memory.

For the implementation of forces, scalars, and the different improvement terms, more data structures for elementary and field types are available.
\subsection{Linear Algebra}
\label{sec:linearalgebra}
The library implements many linear algebra operations acting on different data types. To keep the code generic, these operations are implemented using C macro templates. These C macro templates are classified into types of operations that apply to different kinds of fields. For example, basic operations such as a \textit{zero} and the identity or \textit{copy} operation are available for all fields, while gamma matrix operations acting on the field are only available for spinor fields.
\subsubsection{Implementation of Generic Operations}
Further, generic operations are available, meaning we can apply a certain operation to different field types, and the resolution to the correct implementation for the given type is done automatically. This can be, for example, the operation \texttt{add}: Whether we apply this operation to a spinor field or a field of $SU(N_g)$ matrices, the correct function is invoked to perform the addition.

The main complication in the implementation is that the C language has limited support for generics: if we want to define addition operators for different types, they must be distinguished by function name. A solution for this is implemented by the keyword \texttt{\_Generic} since C11. This allows to use a generic macro resolving to different functions depending on the type of the object passed to the macro at compile time. In this way it behaves differently than function overloading in C++. We implemented an interface that significantly simplifies linear algebra using a combination of aliasing, overloading, and \texttt{\_Generic}. 

When using GPU acceleration, another problem arises since CUDA is a C++ dialect and does not support native C99 complex numbers. The generic operations defined in the C code cannot be used in the CUDA C/C++ code. To complete the interface for the CUDA C++ modules, we duplicated the definitions for generic functions that make use of function overloading.

A detailed description of the implementation can be found in \ref{app:lina}.

\subsubsection{Operation Types}

\paragraph{Basic operations}
All field types in the library implement a \textit{zero} and \textit{copy} operation. To apply an operation to a field, we need to loop over the sites in the field data and apply the operation site-by-site. These loops can be parallelized using OpenMP threads or GPU threads in a kernel. In this case, the operation at each lattice site is completed by one thread.

The definition of the operations at site level is given by the underlying autogenerated macros for the elementary site types. However, the \textit{zero} and \textit{copy} operations are equivalent to the operations of a bitwise \texttt{memset} or \texttt{cudaMemset} to zero or \texttt{memcpy} or \texttt{cudaMemcpy} respectively. In practice, these operations are not implemented using these directives but in terms of a sitewise \textit{zero} and \textit{copy} operation. This is because, firstly, \texttt{memset} and \texttt{memcpy} are not multithreaded. Second, when compiling with GPU acceleration, \texttt{cudaMemset} and \texttt{cudaMemcpy} are memory operations, not kernels, which have a lower bandwidth than a kernel operation. As a result, the highest bandwidth can be achieved by assigning each GPU thread to a lattice site and then calling a manually unrolled elementary site macro that sets all components to zero.

The next more complicated operations implemented by almost all fields are addition and subtraction of two fields, multiplications by a global scalar, which can be real or complex, and combined mul-assign operations. A special case of this is also the application of a global minus sign to the field. All these operations are defined using the underlying macros for the elementary data types, which are defined as element-wise operations, e.g., multiplication by a real or complex scalar.

\paragraph{Linear Combinations}
For spinor fields, \HiRep defines specific functions for certain linear combinations to avoid the need to call more than one linear algebra operation. This type of operation is particularly useful in the inverters. This optimization corresponds to ``fusing" together one or more linear algebra kernels, and it generically decreases the total amount of memory moved. On GPUs, kernel fusion is also very useful for simulating smaller local lattices, where the runtime is significantly affected by the kernel call overhead. Given spinor fields $\xi, \eta$ and $\zeta$, $r_1,r_2\in\mathds{R}$ and $c_1,c_2\in\mathds{C}$, the supported linear combinations are
\begin{equation}
\texttt{lc:}\qquad \xi = r_1\eta + r_2\zeta \,,
\end{equation}
\begin{equation}
\texttt{clc:}\qquad \xi = c_1 \eta + c_2\zeta \,,
\end{equation}
and corresponding add-assign versions. For the eigenvalue solver, the library also supports
\begin{equation}
\texttt{lc1:}\qquad \xi \leftarrow \xi + c_1\eta\,,
\end{equation}
\begin{equation}
\texttt{lc2:}\qquad \xi \leftarrow c_1\xi + c_2\eta
\end{equation}
and
\begin{equation}
\texttt{lc3:}\qquad \xi \leftarrow c_1\eta + c_2\zeta - \xi
\end{equation}

\paragraph{Reductions}
Implemented reduction operations are products between two fields, the square and maximum norm of fields. Products are defined as the sum of the site-wise canonical product, that is, for two spinors $\xi$ and $\eta$
\begin{equation}
\langle \xi, \eta \rangle  = \sum_{a,\alpha}\xi^{\dagger,\alpha}_{a}\eta_{a}^{\alpha}
\end{equation}
for color index $a$ and spinor index $\alpha$, and correspondingly for two $SU(N)$ vectors $v$ and $t$, the product is defined as
\begin{equation}
\langle v, t \rangle  = \sum_{a}v_a^{\dagger}t_a\,,
\end{equation}
again with color index $a$. 

For two $SU(N)$ matrices, the product is defined to yield the Frobenius inner product of the two matrices, i.e., for the matrices $u$ and $v$
\begin{equation}
\langle u, v \rangle = \mathrm{tr}\,[u^{\dagger}v] = \sum_{a,b}u_{ab}^{\dagger}v_{ba}\,.
\end{equation}

Other fields are implemented in \HiRep than the standard fields needed for the Dirac operator. For scalar fields, two complex scalars $z_1,z_2$ have the usual inner product
\begin{equation}
\langle z_1, z_2\rangle = z_1^{\star}z_2
\end{equation}
and algebra vectors $\omega^{a}, \lambda^{a}$ are also simple scalar products of its real components
\begin{equation}
\langle \omega, \lambda \rangle = \sum_{a}\omega^a\lambda^a\,.
\end{equation}

All square norms are then defined over the inner product
\begin{equation}
||\xi||_{2} = \sqrt{\langle \xi, \xi\rangle}\,,
\end{equation}
with $\xi$ describing any of the objects mentioned above.

The maximum norm is defined as the maximum real component of the field. That is when casting the field to an array  of real values, the largest value in this array, or expressed differently

\begin{equation}
||\eta||_{\infty} = \max_{a,\alpha}[\max(|\mathrm{Re}(\eta_{a}^{\alpha})|, |\mathrm{Im}(\eta_{a}^{\alpha}))|]\,. 
\end{equation}

These maximum norms are useful for testing and in providing a norm for uniform convergence.

In principle, all reduction operations except the maximum norm are affected by the technical limitation that they accumulate the numerical error stemming from the use of finite double-precision. While non-reduction operations are usually precise up to the relative double precision of $10^{-15}$, the relative precision for reduction operations decreases with increasing lattice volume if na\"ive summation is used. This problem can be solved by performing the accumulation sum with higher precision than double precision. For this reason, in \texttt{HiRep} the accumulation sums are performed in nearly quadruple precision by using double-double arithmetic, which consists of a double precision accumulator and double precision compensation term, see details in Sect.~\ref{sec:quadreductions}.

\paragraph{Gamma Matrix Applications}
Applications of Dirac gamma matrices are only defined for spinor fields. Writing the spinor indices explicitly, these operations are defined by
\begin{equation}
\xi_{s_1}(x) = \gamma_{i,s_1s_2}\eta{s_2}(x)\,.
\end{equation}
The operations for $\gamma_5$ also feature versions that allow an add-assign or add-mul-assign to be performed together with the matrix application and and an optimization of the reduction $\langle\xi, \gamma_5\eta\rangle$.

\subsection{Elements of a Parallelized Dirac Operator Application}\label{sec:diracimplementation}

\begin{figure}
\begin{Verbatim}[frame=single]
void D(spinor_field *out, spinor_field *in) {
    sync_to_sendbuffer(in);
    start_communications(in);
    D_inner_points(out, in);
    complete_communications(in);
    D_boundary_points(out, in);
}
\end{Verbatim}
\caption{Pseudocode for the parallelized Hopping term operator. The parallelization of the remaining diagonal contributions to the Wilson-Dirac operator is trivial.}
\label{fig:pseudocode}
\end{figure}

On a single processing unit, the Dirac operator is a single function or kernel. However, when many processing units are used in parallel, the application consists of several steps: the computation of the inner lattice points, which only depend on data on the local processing unit, the execution of the data exchange between the processing units, and finally, the computation of the boundary points (all the lattice points not in the inner lattice). Independent of the specific implementation below, the steps follow the pseudocode in Fig.~\ref{fig:pseudocode}.

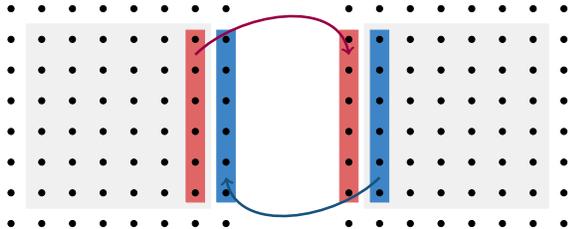
\begin{figure}[H]
\centering
\resizebox{3in}{!} {
\begin{tikzpicture}
    \draw[fill=mygray,color=mygray] (-1.5, -1.5) -- (1.5, -1.5) -- (1.5, 1.5) -- (-1.5, 1.5) -- (-1.5, -1.5);
     \draw[fill=mygray,color=mygray] (4, -1.5) -- (7, -1.5) -- (7, 1.5) -- (4, 1.5) -- (4, -1.5);

    \draw[color=mylightred,fill=mylightred] (1.1,-1.4) -- (1.4,-1.4) -- (1.4,1.4) -- (1.1,1.4) -- (1.1,-1.4);
    \draw[color=mylightblue,fill=mylightblue] (4.1,-1.4) -- (4.4,-1.4) -- (4.4,1.4) -- (4.1,1.4) -- (4.1,-1.4);
    \draw[color=mylightred,fill=mylightred] (3.6,-1.4) -- (3.9,-1.4) -- (3.9,1.4) -- (3.6,1.4) -- (3.6,-1.4);
    \draw[color=mylightblue,fill=mylightblue] (1.6,-1.4) -- (1.9,-1.4) -- (1.9,1.4) -- (1.6,1.4) -- (1.6,-1.4);
    
     \foreach \i in {0,...,7} {
        \foreach \j in {0,...,7} {
            \filldraw (-1.75+\i *0.5, -1.75 +\j *0.5) circle (0.05);
        }
    }
    
    \foreach \i in {0,...,7} {
        \foreach \j in {0,...,7} {
            \filldraw (3.75+\i *0.5, -1.75 +\j *0.5) circle (0.05);
        }
    }
    \draw[color=myred,->,very thick] (1.25, 1) to [out=45, in=100] (3.75, 1);
    \draw[color=myblue,->,very thick] (4.25, -1) to [out=-135, in=-80] (1.75, -1);
\end{tikzpicture}
}
\caption{Schematic exchange of boundary information between two processing units. This exchange is further split into even and odd pieces so that the total number of exchanged buffers is four. Figure as published in \cite{hirepdocs}.}
\label{fig:sendrecv}
\end{figure}

Fig.~\ref{fig:sendrecv} shows the exchange of data between neighboring processing units. In this case the red sites are saved in the send buffer of the local lattice to the left and sent to the receive buffer of the local lattice to the right. Now, operations that depend on a neighboring site can be performed on the inner points. These are the points in the shaded area.

If the execution is arranged in this order and the network is sufficiently fast, the only overhead is the synchronization to the send buffer, and the computation completely masks the computation. The communications are started before the computation of the inner points, and the processor waits for completion after the computation. In the ideal setting, the communications are faster than the computation and have already been completed by the time this barrier is reached. 

Additionally, \HiRep allows further parallelization of CPU-only setups through OpenMP, which parallelizes loops over contiguous lattice site elements for both geometries. This approach supports hybrid parallelization, leveraging additional intra-node parallelism to utilize shared memory hardware capabilities. Although the implementation is straightforward, further efficiency considerations are necessary to identify the optimal setup (see Sect.~\ref{sec:OpenMPaffinity})

\subsection{Geometry layout}\label{sec:geometry}
\begin{figure}[H]
\centering
\begin{minipage}[t]{0.3\linewidth}
\flushleft
\includegraphics[height=1.3in]{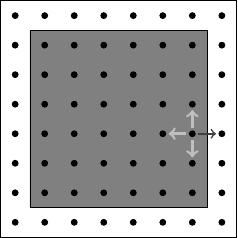}
\end{minipage}%
\begin{minipage}[t]{0.45\linewidth}
\flushright
\includegraphics[height=1.3in]{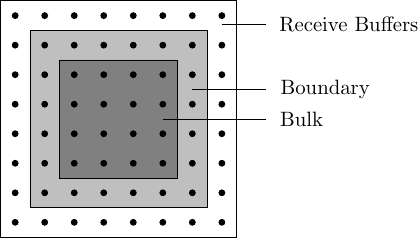}
\end{minipage}
\caption{A simple illustration in two dimensions for how the local lattice in each processing unit is partitioned depending on the geometry layout. The computation-optimized geometry (left) has only two categories (local lattice and receive buffers), while the memory-optimized geometry (right) has three regions (inner points, boundary, and receive buffers). Figure, as published in \cite{Martins:2024dew, hirepdocs}.}
\label{fig:comparisongeometries}
\end{figure}

As discussed previously, fields in \HiRep store their data as a single array of ``site-local" elementary types (one or more per site). The correspondence between the index in the array and the four-dimensional coordinates on the lattice is defined by the geometry layout described in this section.   

We implemented two different geometry layouts, illustrated in Fig.~\ref{fig:comparisongeometries}. The first one corresponds to the memory-optimized geometry also used in the previous version of \texttt{HiRep}, while the second one developed for the current version is referred to as the ``computation-optimized geometry" here. These two geometries differ in their treatment of communications among processing units. Since the message passing interface (MPI) needs a contiguous data buffer to send data across processing units and a simple lexicographic indexing of a four-dimensional Euclidean lattice does not allow for the boundary sites to be arranged contiguously, a synchronization (copy) of boundary data to a send buffer is necessary. The memory-optimized geometry is designed to minimize the total amount of memory that needs to be copied by arranging the synchronization buffers partially in place. This requires partitioning the lattice into inner sites, boundary sites, and receive buffer sites. The computation-optimized geometry is designed to maximize the amount of computation that can be done independently of communications and does not require partitioning the local lattice. However, this arrangement requires more memory to be synchronized to the send buffer.

\begin{figure}
\centering
\begin{tikzpicture}
    \draw[fill=lightgray, fill opacity=0.1, color=lightgray] (0.25,0.25) -- (0.25,5/2-0.25) -- (5/2-0.25,5/2-0.25) -- (5/2-0.25,0.25) -- (0.25,0.25);
    \foreach \i in {0,...,5} {
        \foreach \j in {0,...,5} {
            \draw[black] (\i/2,\j/2) circle (0.06);
        }
    }
    \begin{scope}[very thick,decoration={
        markings,
        mark=at position 0.8 with {\arrow{>}}}
        ]

        \draw[postaction={decorate}, red] (0,0+0.1) -- (0/2,1/2-0.1);
        \draw[postaction={decorate}, red] (0+0.1,0) -- (1/2-0.1,0);
        \draw[postaction={decorate}, black] (1/2,0+0.1) -- (1/2,1/2-0.1);
        \draw[postaction={decorate}, black] (0+0.1,1/2) -- (1/2-0.1,1/2);

        \draw[postaction={decorate}, black] (4/2,0+0.1) -- (4/2,1/2-0.1);
        \draw[postaction={decorate}, black] (4/2+0.1,0) -- (5/2-0.1,0);
        \draw[postaction={decorate}, red] (5/2,0+0.1) -- (5/2,1/2-0.1);
        \draw[postaction={decorate}, black] (4/2+0.1,1/2) -- (5/2-0.1,1/2);
        
        \draw[postaction={decorate}, black] (0,4/2+0.1) -- (0/2,5/2-0.1);
        \draw[postaction={decorate}, black] (0+0.1,4/2) -- (1/2-0.1,4/2);
        \draw[postaction={decorate}, black] (1/2,4/2+0.1) -- (1/2,5/2-0.1);
        \draw[postaction={decorate}, red] (0+0.1,5/2) -- (1/2-0.1,5/2);
        
        \draw[postaction={decorate}, black] (4/2,4/2+0.1) -- (4/2,5/2-0.1);
        \draw[postaction={decorate}, black] (4/2+0.1,4/2) -- (5/2-0.1,4/2);
        \draw[postaction={decorate}, black] (5/2,4/2+0.1) -- (5/2,5/2-0.1);
        \draw[postaction={decorate}, black] (4/2+0.1,5/2) -- (5/2-0.1,5/2);

    \end{scope}
    \draw[->, very thick] (-1,-1) -- (1,-1);
    \draw[->, very thick] (-1,-1) -- (-1,1);
    \draw node at (1+0.2,-1) {$t$};
    \draw node at (-1,1+0.2) {$x$};
\end{tikzpicture}
\caption{Since links are stored in the positive direction, some of the corners are necessary to be contained in the receive buffer, for example, for plaquette shapes that are located in the negative direction from the central site. They are necessary for clover improvement, see Sect.~\ref{sec:clover}. The necessary links are marked in red, while the black links are those belonging to the plaquette shape but are available without the corners.}
\end{figure}

A subtle issue to be pointed out in these geometries is the necessity to store the field values on the corners. In principle, the corners are not necessary to evaluate the Dirac operator or the plaquette, which we evaluate on each site in the positive direction and then average over the lattice. However, a clover term as introduced in Sect.~\ref{sec:clover}, as also present in the fermionic force, requires a plaquette evaluation in the negative direction. For this case, some, but not all, of the corners are necessary. Our geometry implementation includes only the necessary corners.

When not using GPUs for simulations, either geometry can be used. This means that any GPU benchmarks in Sect.~\ref{sec:benchmarks} were performed exclusively with the computation-optimized geometry. On CPU architectures, the question of which implementation is faster appeared to be very dependent on the system. The user should accordingly experiment with both to understand which geometry is ideal for their system.

\subsubsection{Memory-optimized geometry}

In the memory-optimized geometry, the sites of the local lattice are divided into two categories depending on whether the application of the Wilson-Dirac operator requires data from a neighboring processing unit: inner sites and boundary sites. 
When using the memory-optimized geometry, a computation requiring communications among processing units will start the communications, then update the inner points, complete the communications, and finally calculate the boundary points.
The memory-optimized geometry first indexes the inner sites, then proceeds to index the sites in the boundary region, and finally, the receive buffers. 

For both the boundary region and the receive buffers, a contiguous block of indices is used for sites that need to be communicated to the same destination. These blocks correspond to the faces and, when needed, the corners of the hypercubic extended lattice. The boundary region requires special care in the memory-optimized geometry because, in general, it is not possible to enumerate all the faces and corners of the local lattice and still maintain contiguous ranges of indices for all the faces and corners needed as send buffers. In the memory-optimized geometry, this is resolved by giving two or more indices to the same lattice site and declaring one of them a "master" copy while the remaining indices are to be kept in sync with the master site. Thanks to this mechanism the memory-optimized geometry does not require separate send buffers, which are instead embedded into the main local lattice structure, and it minimizes the amount of memory to copy before the data can be sent to other processing units via MPI.

\begin{figure}[H]
\centering
\begin{tikzpicture}
\draw[fill=lightgray] (0,0) -- (2,0) -- (2,0.5) -- (0,0.5) -- (0,0);
\draw[fill=lightgray] (2,0) -- (2.3,0) -- (2.3,0.5) -- (2, 0.5) -- (2,0); 
\draw[fill=lightgray] (2.3,0) -- (2.6,0) -- (2.6,0.5) -- (2.3, 0.5) -- (2.3,0); 
\draw[fill=lightgray] (2.6,0) -- (2.9,0) -- (2.9,0.5) -- (2.6, 0.5) -- (2.6,0); 
\draw[fill=lightgray] (2.9,0) -- (3.2,0) -- (3.2,0.5) -- (2.9, 0.5) -- (2.9,0); 
\draw[fill=gray] (3.2,0) -- (5.2,0) -- (5.2,0.5) -- (3.2,0.5) -- (3.2,0);
\draw[fill=gray] (5.2,0) -- (5.5,0) -- (5.5,0.5) -- (5.2, 0.5) -- (5.2,0); 
\draw[fill=gray] (5.5,0) -- (5.8,0) -- (5.8,0.5) -- (5.5, 0.5) -- (5.5,0); 
\draw[fill=gray] (5.8,0) -- (6.1,0) -- (6.1,0.5) -- (5.8, 0.5) -- (5.8,0); 
\draw[fill=gray] (6.1,0) -- (6.4,0) -- (6.4,0.5) -- (6.1, 0.5) -- (6.1,0); 
\draw[|-|] (0,0.7) -- (2, 0.7); 
\draw[|-|] (2,-0.2) -- (2.6,-0.2);
\draw[|-|] (2.6,0.7) -- (3.2,0.7);
\draw node at (1, 1) {\small Inner points};
\draw node[text width=1cm,align=center] at (2.3,-0.7) {\small Send \\buffers};
\draw node[text width=1cm,align=center] at (2.9,1.2) {\small Recv. \\buffers};
\end{tikzpicture}
\caption{Decomposition of the four-dimensional field in linear memory in a multi-process simulation with one direction parallelized, schematically. The light region is the even lattice, and the dark region is the odd lattice.}
\label{fig:decomplegacy}
\end{figure}
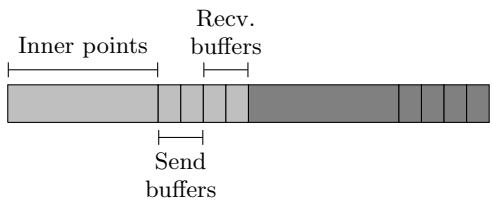

Since even-odd preconditioning is used, as customary in lattice simulations, the above construction is applied to the even and odd sites separately, with all the even sites indexed before the odd sites. This also allows to have fields that are only defined on the even or odd lattice, which is necessary with even-odd preconditioning. The decomposition in memory is illustrated in Fig.~\ref{fig:decomplegacy}, where after the even-odd decomposition, the sites are arranged so that the inner points are first, followed by the receive buffers or boundary points, and finally, the send buffers or points belonging to the halo region of the extended lattice.

Minimizing the amount of memory synchronized before sending the buffer is realized by numbering the sites from the innermost points to the surface of the extended local lattice. In this scheme, some points must appear in multiple positions to be duplicated in memory. The synchronization to the sendbuffer is now only a synchronization of these duplicated sites. 

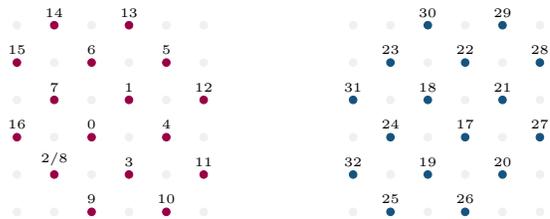
\begin{figure}
\centering
\resizebox{4in}{!}{
\begin{tikzpicture}
     \foreach \i in {0,...,5} {
        \foreach \j in {0,...,5} {
            \filldraw[color=mygray] (-1.25+\i *0.5, -1.25 +\j *0.5) circle (0.05);
        }
    }
    \foreach \i in {0,...,5} {
        \foreach \j in {0,...,5} {
            \filldraw[color=mygray] (3.25+\i *0.5, -1.25 +\j *0.5) circle (0.05);
        }
    }
    
    \filldraw[color=myred] (-0.25,-1.25) circle (0.05);
    \node[above] at (-0.25,-1.25) {\tiny 9};
    \filldraw[color=myred] (0.75,-1.25) circle (0.05);
    \node[above] at (0.75,-1.25) {\tiny 10};
    \filldraw[color=myred] (-0.75,-0.75) circle (0.05);
    \node[above] at (-0.75,-0.75) {\tiny 2/8};
    \filldraw[color=myred] (0.25,-0.75) circle (0.05);
    \node[above] at (0.25,-0.75) {\tiny 3};
    \filldraw[color=myred] (1.25,-0.75) circle (0.05);
    \node[above] at (1.25,-0.75) {\tiny 11};
    \filldraw[color=myred] (-0.25,-0.25) circle (0.05);
    \node[above] at (-0.25,-0.25) {\tiny 0};
    \filldraw[color=myred] (0.75,-0.25) circle (0.05);
    \node[above] at (0.75,-0.25) {\tiny 4};
    \filldraw[color=myred] (-1.25,-0.25) circle (0.05);
    \node[above] at (-1.25,-0.25) {\tiny 16};
    \filldraw[color=myred] (-0.75,0.25) circle (0.05);
    \node[above] at (-0.75,0.25) {\tiny 7};
    \filldraw[color=myred] (0.25,0.25) circle (0.05);
    \node[above] at (0.25,0.25) {\tiny 1};
    \filldraw[color=myred] (1.25,0.25) circle (0.05);
    \node[above] at (1.25,0.25) {\tiny 12};
    \filldraw[color=myred] (-1.25,0.75) circle (0.05);
    \node[above] at (-1.25,0.75) {\tiny 15};
    \filldraw[color=myred] (-0.25,0.75) circle (0.05);
    \node[above] at (-0.25,0.75) {\tiny 6};
    \filldraw[color=myred] (0.75,0.75) circle (0.05);
    \node[above] at (0.75,0.75) {\tiny 5};
    \filldraw[color=myred] (-0.75,1.25) circle (0.05);
    \node[above] at (-0.75,1.25) {\tiny 14};
    \filldraw[color=myred] (0.25,1.25) circle (0.05);
    \node[above] at (0.25,1.25) {\tiny 13};
    
    \filldraw[color=myblue] (3.75,-1.25) circle (0.05);
    \node[above] at (3.75,-1.25) {\tiny 25};
    \filldraw[color=myblue] (4.75,-1.25) circle (0.05);
    \node[above] at (4.75,-1.25) {\tiny 26};
    \filldraw[color=myblue] (3.25,-0.75) circle (0.05);
    \node[above] at (3.25,-0.75) {\tiny 32};
    \filldraw[color=myblue] (4.25,-0.75) circle (0.05);
    \node[above] at (4.25,-0.75) {\tiny 19};
    \filldraw[color=myblue] (5.25,-0.75) circle (0.05);
    \node[above] at (5.25,-0.75) {\tiny 20};
    \filldraw[color=myblue] (3.75,-0.25) circle (0.05);
    \node[above] at (3.75,-0.25) {\tiny 24};
    \filldraw[color=myblue] (4.75,-0.25) circle (0.05);
    \node[above] at (4.75,-0.25) {\tiny 17};
    \filldraw[color=myblue] (5.75,-0.25) circle (0.05);
    \node[above] at (5.75,-0.25) {\tiny 27};
    \filldraw[color=myblue] (3.25,0.25) circle (0.05);
    \node[above] at (3.25,0.25) {\tiny 31};
    \filldraw[color=myblue] (4.25,0.25) circle (0.05);
    \node[above] at (4.25,0.25) {\tiny 18};
    \filldraw[color=myblue] (5.25,0.25) circle (0.05);
    \node[above] at (5.25,0.25) {\tiny 21};
    \filldraw[color=myblue] (3.75,0.75) circle (0.05);
    \node[above] at (3.75,0.75) {\tiny 23};
    \filldraw[color=myblue] (4.75,0.75) circle (0.05);
    \node[above] at (4.75,0.75) {\tiny 22};
    \filldraw[color=myblue] (5.75,0.75) circle (0.05);
    \node[above] at (5.75,0.75) {\tiny 28};
    \filldraw[color=myblue] (4.25,1.25) circle (0.05);
    \node[above] at (4.25,1.25) {\tiny 30};
    \filldraw[color=myblue] (5.25,1.25) circle (0.05);
    \node[above] at (5.25,1.25) {\tiny 29};

\end{tikzpicture}
}
\caption{Two-dimensional example of a contiguous in-place memory arrangement to minimize synchronization. Figure as published in \cite{hirepdocs},}
\label{fig:sync}
\end{figure}

An example of a small 4$\times$4 two-dimensional local lattice with the indexing of the memory-optimized geometry is shown in Fig.~\ref{fig:sync}, with both directions parallelized. Notice that a single boundary site on the even lattice is duplicated (site index 2 synced to index 8).

\subsubsection{Computation-optimized geometry}

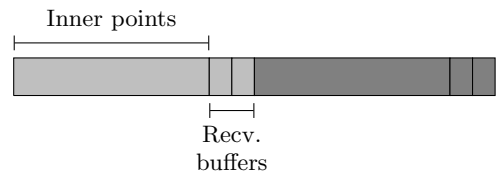
\begin{figure}[H]
\centering
\begin{tikzpicture}
\draw[fill=lightgray] (0,0) -- (2.6,0) -- (2.6,0.5) -- (0,0.5) -- (0,0);
\draw[fill=lightgray] (2.6,0) -- (2.9,0) -- (2.9,0.5) -- (2.6, 0.5) -- (2.6,0); 
\draw[fill=lightgray] (2.9,0) -- (3.2,0) -- (3.2,0.5) -- (2.9, 0.5) -- (2.9,0); 
\draw[fill=gray] (3.2,0) -- (5.8,0) -- (5.8,0.5) -- (3.2,0.5) -- (3.2,0);
\draw[fill=gray] (5.8,0) -- (6.1,0) -- (6.1,0.5) -- (5.8, 0.5) -- (5.8,0); 
\draw[fill=gray] (6.1,0) -- (6.4,0) -- (6.4,0.5) -- (6.1, 0.5) -- (6.1,0); 
\draw[|-|] (0,0.7) -- (2.6, 0.7); 
\draw[|-|] (2.6,-0.2) -- (3.2,-0.2);
\draw node at (1.3, 1) {\small Inner points};
\draw node[text width=1cm,align=center] at (2.9,-0.7) {\small Recv. \\buffers};
\end{tikzpicture}
\caption{Decomposition of the four-dimensional field in linear memory using the computation-optimized geometry in a multi-process simulation with one direction parallelize, schematically. The light region is the even lattice, and the dark region is the odd lattice. }
\label{fig:decompNew}
\end{figure}

The computation-optimized geometry is simpler: here, we keep the whole local lattice as a single index block (i.e., no boundary blocks), and the only decomposition needed is the even-odd one, see Fig~\ref{fig:decompNew}. The halo region of the extended local lattice corresponding to the receive buffers is the same as in the memory-optimized geometry. 

In the computation-optimized geometry, separate send buffers exist, and the synchronization operations then collect and copy all boundary sites from the local lattice and save them into the separate contiguous send buffer. 

The geometric location of the site on a specific boundary is not expressed in terms of the site index being in a particular block (range) of indices like in the memory-optimized geometry. Instead, we use an array of 8-bit (chars) as a look-up table to a bit-mask for the eight directions on the lattice (positive and negative). This additional global memory access is minimal, and the effect on the effective bandwidth is negligible. 

For operations requiring communications, such as the application of the Wilson-Dirac operator, the bit-mask can be used to avoid operations involving boundary terms while the boundary data is being communicated. 
The missing directions are executed only after the communications are completed in a second step. 

While this implementation iterates over the lattice sites twice instead of just once, the actual duration of the boundary kernel is rather short in comparison to the inner computation, so this is efficient in practice. On the other hand, more calculations on the local lattice can be performed compared to the memory-optimized geometry, which enables better masking of the communications.

\subsection{Memory and hardware affinity}\label{sec:optimizedaccess}

With Non-Uniform Memory Access (NUMA), we identify properties of systems equipped with more than one system bus. A platform with multiple processors on a single motherboard allows every processor to access all the memory on the board. The system defines a locality of memory through ACPI by identifying local and remote memory domains for each processor. When a processor accesses memory outside its own domain (remote memory), data must be transferred over the NUMA connection at a rate slower than accessing local memory. This implies that the speed of memory access will inherently depend on the location/proximity of the memory access to the accessing node.

Analogously to the memory access from the CPU, PCIe devices attached to the motherboard through a PCIe link are associated with one of the NUMA nodes, inheriting the same NUMA locality properties.

An example of this is given in Fig.~\ref{fig:nodetopology}. This part of a diagram generated using the hardware locality package (\texttt{hwloc}) shows the topology of an Intel Xeon Gold 6130 with a total memory of 377 GB. This memory is distributed to two different NUMA domains, with 188 GB and 189 GB respectively of memory and 16 CPU cores. The software needs to understand this topology and allocate the memory for data processed by a CPU in the same domain.

\begin{figure}
\centering
\begin{tikzpicture}
\draw (0,0) -- (13,0) -- (13,5) -- (0,5) -- (0,0);
\draw[fill=pastelgreen] (0.1,0.1) -- (6.45,0.1) -- (6.45,4.9) -- (0.1,4.9) -- (0.1,0.1);
\draw[fill=pastelgreen] (6.55,0.1) -- (12.9,0.1) -- (12.9,4.9) -- (6.55,4.9) -- (6.55,0.1);
\draw[fill=beige] (0.2,4.0) -- (6.35,4.0) -- (6.35,4.4) -- (0.2,4.4) -- (0.2,4.0);
\draw[fill=beige] (6.65,4.0) -- (12.8,4.0) -- (12.8,4.4) -- (6.65,4.4) -- (6.65,4.0);
\draw[fill=white] (0.2,4.0-0.5) -- (6.35,4.0-0.5) -- (6.35,4.4-0.5) -- (0.2,4.4-0.5) -- (0.2,4.0-0.5);
\draw[fill=white] (6.65,4.0-0.5) -- (12.8,4.0-0.5) -- (12.8,4.4-0.5) -- (6.65,4.4-0.5) -- (6.65,4.0-0.5);
\draw[fill=white] (0.2,3.0) -- (1.7,3.0) -- (1.7,3.4) -- (0.2,3.4) -- (0.2,3.0);
\draw[fill=white] (1.8,3.0) -- (3.3,3.0) -- (3.3,3.4) -- (1.8,3.4) -- (1.8,3.0);
\draw[fill=white] (4.85,3.0) -- (6.35,3.0) -- (6.35,3.4) -- (4.85,3.4) -- (4.85,3.0);
\draw[fill=white] (0.2,2.5) -- (1.7,2.5) -- (1.7,2.9) -- (0.2,2.9) -- (0.2,2.5);
\draw[fill=white] (1.8,2.5) -- (3.3,2.5) -- (3.3,2.9) -- (1.8,2.9) -- (1.8,2.5);
\draw[fill=white] (4.85,2.5) -- (6.35,2.5) -- (6.35,2.9) -- (4.85,2.9) -- (4.85,2.5);
\draw[fill=white] (0.2,2.0) -- (1.7,2.0) -- (1.7,2.4) -- (0.2,2.4) -- (0.2,2.0);
\draw[fill=white] (1.8,2.0) -- (3.3,2.0) -- (3.3,2.4) -- (1.8,2.4) -- (1.8,2.0);
\draw[fill=white] (4.85,2.0) -- (6.35,2.0) -- (6.35,2.4) -- (4.85,2.4) -- (4.85,2.0);

\draw[fill=white] (0.2+6.45,3.0) -- (1.7+6.45,3.0) -- (1.7+6.45,3.4) -- (0.2+6.45,3.4) -- (0.2+6.45,3.0);
\draw[fill=white] (1.8+6.45,3.0) -- (3.3+6.45,3.0) -- (3.3+6.45,3.4) -- (1.8+6.45,3.4) -- (1.8+6.45,3.0);
\draw[fill=white] (4.85+6.45,3.0) -- (6.35+6.45,3.0) -- (6.35+6.45,3.4) -- (4.85+6.45,3.4) -- (4.85+6.45,3.0);
\draw[fill=white] (0.2+6.45,2.5) -- (1.7+6.45,2.5) -- (1.7+6.45,2.9) -- (0.2+6.45,2.9) -- (0.2+6.45,2.5);
\draw[fill=white] (1.8+6.45,2.5) -- (3.3+6.45,2.5) -- (3.3+6.45,2.9) -- (1.8+6.45,2.9) -- (1.8+6.45,2.5);
\draw[fill=white] (4.85+6.45,2.5) -- (6.35+6.45,2.5) -- (6.35+6.45,2.9) -- (4.85+6.45,2.9) -- (4.85+6.45,2.5);
\draw[fill=white] (0.2+6.45,2.0) -- (1.7+6.45,2.0) -- (1.7+6.45,2.4) -- (0.2+6.45,2.4) -- (0.2+6.45,2.0);
\draw[fill=white] (1.8+6.45,2.0) -- (3.3+6.45,2.0) -- (3.3+6.45,2.4) -- (1.8+6.45,2.4) -- (1.8+6.45,2.0);
\draw[fill=white] (4.85+6.45,2.0) -- (6.35+6.45,2.0) -- (6.35+6.45,2.4) -- (4.85+6.45,2.4) -- (4.85+6.45,2.0);

\draw[fill=softgray] (0.2,1.9) -- (1.7,1.9) -- (1.7,0.2) -- (0.2,0.2) -- (0.2,1.9);
\draw[fill=softgray] (1.8,1.9) -- (3.3,1.9) -- (3.3,0.2) -- (1.8,0.2) -- (1.8,1.9);
\draw[fill=softgray] (4.85,1.9) -- (6.35,1.9) -- (6.35,0.2) -- (4.85,0.2) -- (4.85,1.9);

\draw[fill=softgray] (0.2+6.45,1.9) -- (1.7+6.45,1.9) -- (1.7+6.45,0.2) -- (0.2+6.45,0.2) -- (0.2+6.45,1.9);
\draw[fill=softgray] (1.8+6.45,1.9) -- (3.3+6.45,1.9) -- (3.3+6.45,0.2) -- (1.8+6.45,0.2) -- (1.8+6.45,1.9);
\draw[fill=softgray] (4.85+6.45,1.9) -- (6.35+6.45,1.9) -- (6.35+6.45,0.2) -- (4.85+6.45,0.2) -- (4.85+6.45,1.9);

\draw[fill=white] (3.55,2.5) -- (3.65,2.5) -- (3.65,2.4) -- (3.55,2.4) -- (3.55,2.5);
\draw[fill=white] (3.85,2.5) -- (3.95,2.5) -- (3.95,2.4) -- (3.85,2.4) -- (3.85,2.5);
\draw[fill=white] (4.15,2.5) -- (4.25,2.5) -- (4.25,2.4) -- (4.15,2.4) -- (4.15,2.5);
\draw[fill=white] (4.45,2.5) -- (4.55,2.5) -- (4.55,2.4) -- (4.45,2.4) -- (4.45,2.5);

\draw[fill=white] (3.55+6.45,2.5) -- (3.65+6.45,2.5) -- (3.65+6.45,2.4) -- (3.55+6.45,2.4) -- (3.55+6.45,2.5);
\draw[fill=white] (3.85+6.45,2.5) -- (3.95+6.45,2.5) -- (3.95+6.45,2.4) -- (3.85+6.45,2.4) -- (3.85+6.45,2.5);
\draw[fill=white] (4.15+6.45,2.5) -- (4.25+6.45,2.5) -- (4.25+6.45,2.4) -- (4.15+6.45,2.4) -- (4.15+6.45,2.5);
\draw[fill=white] (4.45+6.45,2.5) -- (4.55+6.45,2.5) -- (4.55+6.45,2.4) -- (4.45+6.45,2.4) -- (4.45+6.45,2.5);

\draw[fill=softblue] (0.3,1.5) -- (1.6,1.5) -- (1.6,0.3) -- (0.3,0.3) -- (0.3,1.5);
\draw[fill=softblue] (1.9,1.5) -- (3.2,1.5) -- (3.2,0.3) -- (1.9,0.3) -- (1.9,1.5);
\draw[fill=softblue] (4.95,1.5) -- (6.25,1.5) -- (6.25,0.3) -- (4.95,0.3) -- (4.95,1.5);

\draw[fill=softblue] (0.3+6.45,1.5) -- (1.6+6.45,1.5) -- (1.6+6.45,0.3) -- (0.3+6.45,0.3) -- (0.3+6.45,1.5);
\draw[fill=softblue] (1.9+6.45,1.5) -- (3.2+6.45,1.5) -- (3.2+6.45,0.3) -- (1.9+6.45,0.3) -- (1.9+6.45,1.5);
\draw[fill=softblue] (4.95+6.45,1.5) -- (6.25+6.45,1.5) -- (6.25+6.45,0.3) -- (4.95+6.45,0.3) -- (4.95+6.45,1.5);

\draw (0.3, 0.9) -- (1.6, 0.9);
\draw (1.9, 0.9) -- (3.2, 0.9);
\draw (4.95, 0.9) -- (6.25, 0.9);

\draw (0.3+6.45, 0.9) -- (1.6+6.45, 0.9);
\draw (1.9+6.45, 0.9) -- (3.2+6.45, 0.9);
\draw (4.95+6.45, 0.9) -- (6.25+6.45, 0.9);

\node[anchor=south west] at (0.2,4.37) {\texttt{Socket \#1}};
\node[anchor=south west] at (6.65,4.37) {\texttt{Socket \#2}};

\node[anchor=south west] at (0.2,3.87) {\footnotesize\texttt{Memory NUMA Domain \#1 188GB}};
\node[anchor=south west] at (0.2+6.45,3.87) {\footnotesize\texttt{Memory NUMA Domain \#2 189GB}};

\node[anchor=south west] at (0.2,3.45) {\footnotesize\texttt{L3 22MB}};
\node[anchor=south west] at (0.2+6.45,3.45) {\footnotesize\texttt{L3 22MB}};

\node[anchor=south west] at (0.2,2.93) {\footnotesize\texttt{L2}};
\node[anchor=south west] at (0.2+6.45,2.93) {\footnotesize\texttt{L2}};
\node[anchor=south west] at (1.80,2.93) {\footnotesize\texttt{L2}};
\node[anchor=south west] at (1.80+6.45,2.93) {\footnotesize\texttt{L2}};
\node[anchor=south west] at (4.85,2.93) {\footnotesize\texttt{L2}};
\node[anchor=south west] at (4.85+6.45,2.93) {\footnotesize\texttt{L2}};

\node[anchor=south west] at (0.2,2.43) {\footnotesize\texttt{L1 Data}};
\node[anchor=south west] at (0.2+6.45,2.43) {\footnotesize\texttt{L1 Data}};
\node[anchor=south west] at (1.80,2.43) {\footnotesize\texttt{L1 Data}};
\node[anchor=south west] at (1.80+6.45,2.43) {\footnotesize\texttt{L1 Data}};
\node[anchor=south west] at (4.85,2.43) {\footnotesize\texttt{L1 Data}};
\node[anchor=south west] at (4.85+6.45,2.43) {\footnotesize\texttt{L1 Data}};

\node[anchor=south west] at (0.2,1.97) {\footnotesize\texttt{L1 Inst}};
\node[anchor=south west] at (0.2+6.45,1.97) {\footnotesize\texttt{L1 Inst}};
\node[anchor=south west] at (1.80,1.97) {\footnotesize\texttt{L1 Inst}};
\node[anchor=south west] at (1.80+6.45,1.97) {\footnotesize\texttt{L1 Inst}};
\node[anchor=south west] at (4.85,1.97) {\footnotesize\texttt{L1 Inst}};
\node[anchor=south west] at (4.85+6.45,1.97) {\footnotesize\texttt{L1 Inst}};

\node[anchor=south west] at (0.2,1.43) {\footnotesize\texttt{Core\,0}};
\node[anchor=south west] at (1.80,1.43) {\footnotesize\texttt{Core\,1}};
\node[anchor=south west] at (4.85,1.43) {\footnotesize\texttt{Core\,15}};

\node[anchor=south west] at (0.2+6.45,1.43) {\footnotesize\texttt{Core\,16}};
\node[anchor=south west] at (1.80+6.45,1.43) {\footnotesize\texttt{Core\,17}};
\node[anchor=south west] at (4.85+6.45,1.43) {\footnotesize\texttt{Core\,31}};

\node[anchor=south west] at (0.3,0.9) {\footnotesize\texttt{PU\,0}};
\node[anchor=south west] at (0.3,0.325) {\footnotesize\texttt{PU\,1}};

\node[anchor=south west] at (1.9,0.9) {\footnotesize\texttt{PU\,2}};
\node[anchor=south west] at (1.9,0.325) {\footnotesize\texttt{PU\,3}};

\node[anchor=south west] at (4.95,0.9) {\footnotesize\texttt{PU\,30}};
\node[anchor=south west] at (4.95,0.325) {\footnotesize\texttt{PU\,31}};

\node[anchor=south west] at (0.3+6.45,0.9) {\footnotesize\texttt{PU\,32}};
\node[anchor=south west] at (0.3+6.45,0.325) {\footnotesize\texttt{PU\,33}};

\node[anchor=south west] at (1.9+6.45,0.9) {\footnotesize\texttt{PU\,34}};
\node[anchor=south west] at (1.9+6.45,0.325) {\footnotesize\texttt{PU\,35}};

\node[anchor=south west] at (4.95+6.45,0.9) {\footnotesize\texttt{PU\,62}};
\node[anchor=south west] at (4.95+6.45,0.325) {\footnotesize\texttt{PU\,63}};

\node at (4.05,2.65) {\tiny\texttt{$16\times$ total}};
\node at (4.05+6.45,2.65) {\tiny\texttt{$16\times$ total}};

\end{tikzpicture}
\caption{Part of the topology of an Intel Xeon Gold 6130 node with 32 cores after the diagram generated with \texttt{hwloc}. The memory and CPUs are distributed into two different NUMA domains. 22 MB L3 cache is shared among the cores in the same NUMA domain, while L2 (1024 KB) and L1 Data Cache and L1 Instruction Cache (each 32KB) are attached to each core, which in turn contain two processing units (PU).}
\label{fig:nodetopology}
\end{figure}

\subsubsection{OpenMP NUMA affinity and Data Placement}\label{sec:OpenMPaffinity}

To take advantage of NUMA's features while using OpenMP, our strategy
consists of two key components: 
\begin{enumerate}
    \item \textbf{Static Association}: Threads are statically
      associated with the memory layout using static scheduling. 
    \item \textbf{Correct Memory Placement}: Memory pages are
      correctly placed relative to the threads accessing them, a
      practice often referred to as optimal NUMA affinity. 
\end{enumerate}
OpenMP employs a \emph{First Touch} data placement policy. This policy
allocates a data page in the memory closest to the thread that
accesses the page for the first time, rather than the thread that
allocates it. This policy is the default on Linux and other operating
systems. 
To ensure correct affinity between threads and memory arrays,
\HiRep employs a simple strategy: parallelizing data
initialization through OpenMP. This approach minimizes NUMA
violations, thereby avoiding congestion at the memory controller and
ensuring efficient memory access. 

\subsubsection{GPU NUMA affinity}

The software can pin MPI processes to specific CPUs and GPUs to optimize execution. 
For a standard GPU simulation, each MPI process will manage a single GPU. The selected communicator, which can be a Cartesian communicator, defines the local rank of the process on the node, and the device with the matching local ID will be selected. It is possible to compile the code using the \texttt{hwloc} library, which allows to pin the MPI process to a free CPU located in the same NUMA domain as the selected GPU.

\subsubsection{Topology-Aware Communicators}
To further enhance code efficiency and prioritize intra-node
communication over inter-node communication, \HiRep provides the
ability to reshape the association of MPI processes with their
geometrical lattice sub-blocks.
While default Cartesian communicators use a row-major order of the MPI
processes, \HiRep offers the option to define communicators that are logically organized into hypercubic sub-blocks inspired by the implementation in \cite{openqcd}.
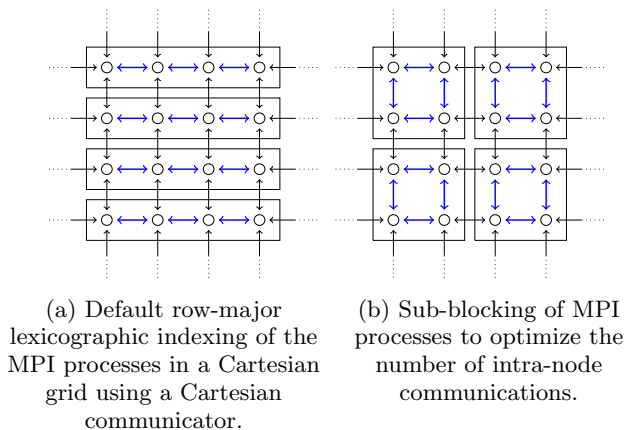
\begin{figure}
\centering
\begin{minipage}[t]{0.45\linewidth}
\centering
\resizebox{1.5in}{!}{
\begin{tikzpicture}
	\foreach \i in {1,...,4} {
		\foreach \j in {1,...,4} {
			\draw (\i,\j) circle (3pt);
		}
		\draw (0.6, 0.6+\i-1) -- (4.4,0.6+\i-1) -- (4.4,1.4+\i-1) -- (0.6,1.4+\i-1) -- (0.6,0.6+\i-1);
		
		\draw[<-] (\i,4.2) -- (\i,4.7);
		\draw[dotted] (\i,4.7) -- (\i,5.2);
		\draw[<-] (\i,0.8) -- (\i,0.3);
		\draw[dotted] (\i,0.3) -- (\i,-0.2);
		
		\draw[<-] (4.2, \i) -- (4.7,\i);
		\draw[dotted] (4.7,\i) -- (5.2,\i);
		\draw[<-] (0.8,\i) -- (0.3,\i);
		\draw[dotted] (0.3,\i) -- (-0.2,\i);
	}
	
	\foreach \i in {1,...,3} {
		\foreach \j in {1,...,4} {
			\draw[<->] (\j,\i+0.2) -- (\j,\i+1-0.2);
			\draw[<->,blue,thick] (\i+0.2,\j) -- (\i+1-0.2,\j);
		}
	}
\end{tikzpicture}
}
\subcaption{Default row-major lexicographic indexing of the MPI processes in a Cartesian grid using a Cartesian communicator. }
\end{minipage}%
\hfill
\begin{minipage}[t]{0.45\linewidth}
\centering
\resizebox{1.5in}{!}{
\begin{tikzpicture}
	\foreach \i in {1,...,4} {
		\foreach \j in {1,...,4} {
			\draw (\i,\j) circle (3pt);
		}
		\draw[<-] (\i,4.2) -- (\i,4.7);
		\draw[dotted] (\i,4.7) -- (\i,5.2);
		\draw[<-] (\i,0.8) -- (\i,0.3);
		\draw[dotted] (\i,0.3) -- (\i,-0.2);
		
		\draw[<-] (4.2, \i) -- (4.7,\i);
		\draw[dotted] (4.7,\i) -- (5.2,\i);
		\draw[<-] (0.8,\i) -- (0.3,\i);
		\draw[dotted] (0.3,\i) -- (-0.2,\i);
	}

	\foreach \i in {1,3} {
		\foreach \j in {1,...,4} {
			\draw[<->,blue,thick] (\j,\i+0.2) -- (\j,\i+1-0.2);
			\draw[<->,blue,thick] (\i+0.2,\j) -- (\i+1-0.2,\j);
		}
	}
	
	\foreach \i in {2} {
		\foreach \j in {1,...,4} {
			\draw[<->] (\j,\i+0.2) -- (\j,\i+1-0.2);
			\draw[<->] (\i+0.2,\j) -- (\i+1-0.2,\j);
		}
	}

	\draw (0.6,0.6) -- (0.6,2.4) -- (2.4,2.4) -- (2.4,0.6) -- (0.6,0.6);
	\draw (0.6+2,0.6) -- (0.6+2,2.4) -- (2.4+2,2.4) -- (2.4+2,0.6) -- (0.6+2,0.6);
	\draw (0.6,0.6+2) -- (0.6,2.4+2) -- (2.4,2.4+2) -- (2.4,0.6+2) -- (0.6,0.6+2);
	\draw (0.6+2,0.6+2) -- (0.6+2,2.4+2) -- (2.4+2,2.4+2) -- (2.4+2,0.6+2) -- (0.6+2,0.6+2);
\end{tikzpicture}
}
\subcaption{Sub-blocking of MPI processes to optimize the number of intra-node communications.}
\end{minipage}
\caption{Illustration of the MPI subblocking in two dimensions. The circles represent local lattices which are arranged on the MPI communicator. The squares represent the boundaries of the node. We can arrange the different lattices to be differently distributed to the nodes. Depending on this distribution, we affect the efficiency of the communications between them.}
\label{fig:mpisubblocking}
\end{figure}
This setup benefits both CPU and GPU implementations. On CPUs, it
increases intra-node communication and optimizes shared memory or bus
communication. On GPUs, where supported, it maximizes the use of direct GPU intra-node interconnections, further enhancing performance.

This is illustrated in Fig.~\ref{fig:mpisubblocking}. The figure shows \textit{cores} (CPUs or GPUs) as circles and \textit{nodes} as the blocks surrounding the circles.  The communications are displayed as arrows between the cores. If they are blue, the communications are between cores belonging to the same node, while the black ones require communications between different nodes. The communications within the node are generally substantially faster than those between nodes. At the same time, each core communicates only with a few of its neighbors, not with all other cores. As a result, the cores can be arranged on the different nodes such that the time spent on communications is minimal.

An example of a speed-up due to the topological arrangement that is not immediately obvious is displayed in Fig.~\ref{fig:mpisubblocking}. On the left-hand side, the default row-major indexing of the MPI processes results in three out of 13 communications of each node being inside the node. When blocking the MPI processes, there are four out of 12 communications that occur within the node. This can achieve a speed-up in the Dirac operator, provided that the communications are not completely masked by computations in both cases.

\subsection{Memory Access Patterns}\label{sec:memoryaccess}

\subsubsection{Reuse}

The arrangement of lattice sites in memory can be optimized for cache reuse in lattice local operations, which depend only on neighboring sites, such as when applying the Hopping term in the Wilson-Dirac operator. This improves performance on both CPUs and GPUs. We use a path-blocking arrangement, which means that the lattice sites are enumerated using a hierarchical scheme in which we first consider small sub-blocks (of size 2, 3, or 4, depending on the size of the local lattice) in lexicographic order and the sites inside each sub-block are themselves indexed in lexicographic order. 

\begin{figure}
\centering
\resizebox{1in}{!}{
\begin{tikzpicture}
\foreach \j in {0,2} {
  \filldraw (-1.75+0 *0.5, -1.75 +\j *0.5) circle (0.05);
}
\foreach \j in {1,3} {
  \filldraw (-1.75+1 *0.5, -1.75 +\j *0.5) circle (0.05);
}
\foreach \j in {0,2} {
  \filldraw (-1.75+2 *0.5, -1.75 +\j *0.5) circle (0.05);
}
\foreach \j in {1,3} {
  \filldraw (-1.75+3 *0.5, -1.75 +\j *0.5) circle (0.05);
}

\foreach \j in {1,3} {
  \filldraw[lightgray] (-1.75+0 *0.5, -1.75 +\j *0.5) circle (0.05);
}
\foreach \j in {0,2} {
  \filldraw[lightgray] (-1.75+1 *0.5, -1.75 +\j *0.5) circle (0.05);
}
\foreach \j in {1,3} {
  \filldraw[lightgray] (-1.75+2 *0.5, -1.75 +\j *0.5) circle (0.05);
}
\foreach \j in {0,2} {
  \filldraw[lightgray] (-1.75+3 *0.5, -1.75 +\j *0.5) circle (0.05);
}

\filldraw[red] (-1.75+1 *0.5, -1.75 + 1 *0.5) circle (0.05);
\filldraw[red] (-1.75+3 *0.5, -1.75 + 1 *0.5) circle (0.05);
\filldraw[red] (-1.75+2 *0.5, -1.75 + 2 *0.5) circle (0.05);
\filldraw[red] (-1.75+2 *0.5, -1.75 + 0 *0.5) circle (0.05);
\draw[thick, ->, red] (-1.75+1 *0.5, -1.75 + 1 *0.5) -- (-1.75+2 *0.5-0.1, -1.75 + 1 *0.5);
\draw[thick, ->, red] (-1.75+3 *0.5, -1.75 + 1 *0.5) -- (-1.75+2 *0.5+0.1, -1.75 + 1 *0.5);
\draw[thick, ->, red] (-1.75+2 *0.5, -1.75 + 2 *0.5) -- (-1.75+2 *0.5, -1.75 + 1 *0.5 + 0.1);
\draw[thick, ->, red] (-1.75+2 *0.5, -1.75 + 0 *0.5) -- (-1.75+2 *0.5, -1.75 + 1 *0.5 - 0.1);
\end{tikzpicture}
}\hspace{0.5in}\resizebox{1in}{!}{
\begin{tikzpicture}
\foreach \j in {0,2} {
  \filldraw (-1.75+0 *0.5, -1.75 +\j *0.5) circle (0.05);
}
\foreach \j in {1,3} {
  \filldraw (-1.75+1 *0.5, -1.75 +\j *0.5) circle (0.05);
}
\foreach \j in {0,2} {
  \filldraw (-1.75+2 *0.5, -1.75 +\j *0.5) circle (0.05);
}
\foreach \j in {1,3} {
  \filldraw (-1.75+3 *0.5, -1.75 +\j *0.5) circle (0.05);
}

\foreach \j in {1,3} {
  \filldraw[lightgray, fill opacity=0.1, opacity=0.2] (-1.75+0 *0.5, -1.75 +\j *0.5) circle (0.05);
}
\foreach \j in {0,2} {
  \filldraw[lightgray, fill opacity=0.1, opacity=0.2] (-1.75+1 *0.5, -1.75 +\j *0.5) circle (0.05);
}
\foreach \j in {1,3} {
  \filldraw[lightgray, fill opacity=0.1, opacity=0.2] (-1.75+2 *0.5, -1.75 +\j *0.5) circle (0.05);
}
\foreach \j in {0,2} {
  \filldraw[lightgray, fill opacity=0.1, opacity=0.2] (-1.75+3 *0.5, -1.75 +\j *0.5) circle (0.05);
}

\draw[thick,->, red,dashed] (-1.75+1 *0.5+0.1, -1.75 + 3 *0.5) -- (-1.75+3 *0.5-0.1, -1.75 +3 *0.5);
\draw[thick,->, red,dashed] (-1.75+3 *0.5, -1.75 +3 *0.5-0.05) --(-1.75+0 *0.5+0.05, -1.75 +2 *0.5+0.05);
\draw[thick,->, red,dashed] (-1.75+0 *0.5+0.1, -1.75 +2 *0.5) -- (-1.75+2 *0.5-0.1, -1.75 +2 *0.5);
\draw[thick,->, red,dashed] (-1.75+2 *0.5-0.05, -1.75 +2 *0.5-0.05) -- (-1.75+1 *0.5+0.05, -1.75 +1 *0.5+0.05);
\draw[thick,->, red,dashed] (-1.75+1 *0.5+0.1, -1.75 +1 *0.5) -- (-1.75+3 *0.5-0.1, -1.75 +1 *0.5);
\draw[thick,->, red,dashed] (-1.75+3 *0.5-0.05, -1.75 +1 *0.5-0.05)  -- (-1.75+0 *0.5+0.05, -1.75 +0 *0.5+0.05);
\draw[thick,->, red,dashed] (-1.75+0 *0.5+0.1, -1.75 +0 *0.5) -- (-1.75+2 *0.5-0.1, -1.75 +0 *0.5);

\end{tikzpicture}
}

\vspace{6mm}
\caption{Reuse of sites for efficient caching: The even sites are read from multiple threads when computing the odd sites of the Dirac Operator and vice versa and are therefore eligible to be found in the processor cache. This means that the effective bandwidth for some operations can exceed the theoretical peak bandwidth given by the manufacturer. A path-blocking memory arrangement, where the lattice is blocked into sub-blocks stored contiguously, can increase cache locality.}
\label{fig:reusepb}
\end{figure}
This is an advantage, as illustrated in Fig.~\ref{fig:reusepb}. The site on the left-hand side shows four points of the odd lattice sites (black), reusing a single even site (gray). The cache locality is improved when arranging the sites in a consecutive block in memory and shown on the right hand side. The size of the sub-blocks is set at compile time and can be tuned to the ideal value depending on the hardware.

\subsubsection{GPU Strided Accesses}
\begin{figure}
\centering
\includegraphics[width=0.8\linewidth]{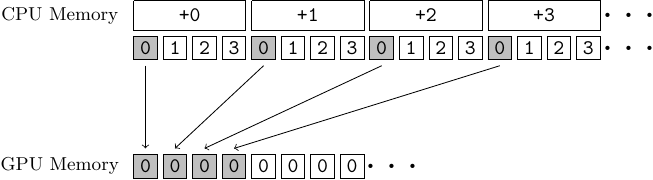}
\caption{Coalescing memory access for the GPU, figure, as published in \cite{Martins:2024dew}. In CPU memory, the data structures that contain the site data are stored site by site. This is inefficient for the GPU because the threads access the data structures component-wise. As a result, if we used the same memory layout for the GPU as for the CPU, the components would be accessed with a stride of the size of the struct between them. A more efficient layout arranges the components in blocks of the n-th component. This way, the GPU threads always access the memory contiguously and coalesced.}
\label{fig:gpumemcoalesce}
\end{figure}

\begin{figure}
\centering
\begin{minipage}[t]{0.7\linewidth}
\centering
\resizebox{2in}{!}{
\begin{tikzpicture}
\foreach \i in {1,...,8} {
	\draw[->, very thick] (\i, 1) -- (\i, 0);
	\draw[fill=lightgray] (\i-0.3,-0.2) -- (\i+0.3,-0.2) -- (\i+0.3,-0.7) -- (\i-0.3,-0.7) -- (\i-0.3,-0.2);
	\node at (\i,-0.45) {\Large\i};
	\node at (\i,1.4) {\Large\i};
	\node at (-0.2,-0.45) {byte};
	\node at (-0.2,1.4) {thread};
}
\end{tikzpicture}
}
\subcaption{Coalesced and contiguous}
\end{minipage}

\begin{minipage}[t]{0.7\linewidth}
\centering
\resizebox{2in}{!}{
\begin{tikzpicture}
\foreach \i in {1,...,8} {
	\draw[fill=lightgray] (\i-0.3,-0.2) -- (\i+0.3,-0.2) -- (\i+0.3,-0.7) -- (\i-0.3,-0.7) -- (\i-0.3,-0.2);
	\node at (\i,-0.45) {\Large\i};
	\node at (\i,1.4) {\Large\i};
	\node at (-0.2,-0.45) {byte};
	\node at (-0.2,1.4) {thread};
}

\draw[->, very thick] (1, 1) -- (3, 0);
\draw[->, very thick] (2, 1) -- (6, 0);
\draw[->, very thick] (3, 1) -- (2, 0);
\draw[->, very thick] (4, 1) -- (4, 0);
\draw[->, very thick] (5, 1) -- (5, 0);
\draw[->, very thick] (6, 1) -- (1, 0);
\draw[->, very thick] (7, 1) -- (8, 0);
\draw[->, very thick] (8, 1) -- (7, 0);

\end{tikzpicture}
}
\subcaption{Contiguous but not coalesced}
\end{minipage}

\begin{minipage}[t]{0.7\linewidth}
\centering
\resizebox{2in}{!}{
\begin{tikzpicture}
\foreach \i in {1,...,4} {
	\node at (-0.2,1.4) {thread};
	\node at (2*\i,1.4) {\Large\i};

}
\foreach \i in {1,...,8} {
	\draw[fill=lightgray] (\i-0.3,-0.2) -- (\i+0.3,-0.2) -- (\i+0.3,-0.7) -- (\i-0.3,-0.7) -- (\i-0.3,-0.2);
	\node at (\i,-0.45) {\Large\i};
	\node at (-0.2,-0.45) {byte};
}

\draw[->, very thick] (2, 1) -- (2, 0);
\draw[->, very thick] (4, 1) -- (4, 0);
\draw[->, very thick] (6, 1) -- (6, 0);
\draw[->, very thick] (8, 1) -- (8, 0);
\end{tikzpicture}
}
\subcaption{Strided access with one byte between the accesses}
\end{minipage}
\caption{Different types of memory accesses. The peak bandwidth of a GPU is only reached if all memory accesses are contiguous and coalesced.}
\label{fig:coalescence}
\end{figure}
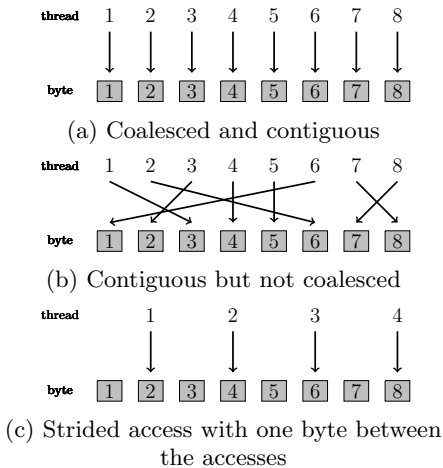

Lattice field theory simulations are memory bound, i.e., the time spent for a trajectory or inversion is limited by the time it takes to read the gauge field and fermion fields from memory more than the available computational capacity. In general, accessing a single contiguous block of memory linearly is more performant than scattered accesses, see Fig.~\ref{fig:coalescence}.

When simulating on CPUs without OpenMP, the iteration over the local lattice in the Dirac operator is implemented using a \texttt{for} loop. So, the coalesced way to access memory is component by component, one site after another.

A GPU executes kernels in groups of 32 concurrent threads, called warps, and we associate each thread with a lattice site. To achieve optimal performance via memory coalescence, instead of ordering the field data site-wise linearly in memory, we use blocks of 32 sites and stripe the site data in memory using real single-precision or double-precision floating point numbers, which corresponds to the components of the elements of the field, as illustrated in Fig.~\ref{fig:gpumemcoalesce}. 

During the application of linear algebra operations or the Wilson Dirac operator, the GPU will process warps of 32 GPU threads and simultaneously access the field data component by component for 32 consecutive sites. As a result of the memory layout above, these blocks of 32 numbers correspond to one contiguous block of coalesced data in GPU memory.

This GPU striping and blocking of memory is incompatible with the memory-optimized geometry implementation of the original \texttt{HiRep}, which contains many memory blocks, including the send buffers in place. Therefore, the GPU implementation only uses the computation-optimized geometry that contains only one block of sites for the local lattice and uses separate send buffers. As discussed above, the additional time for the copy operation to the send buffers is negligible. The computation-optimized geometry maximizes the number of operations that can be completed without any communications, maximizing our ability to mask communications with computations. 

\subsection{GPU Architecture}

The code was developed in CUDA C. Due to the similarities between CUDA C and HIP, the code is translated during compile time using macro redefinitions of the CUDA commands to HIP commands. While the Wilson-Dirac operator needed some tuning to account for the lower available registers on the tested AMD architecture, the code runtimes from this simple implementations are comparable between NVIDIA and AMD cards.

\subsection{MPI Communications}\label{sec:MPI}
\begin{figure}
\centering
\includegraphics[width=0.8\linewidth]{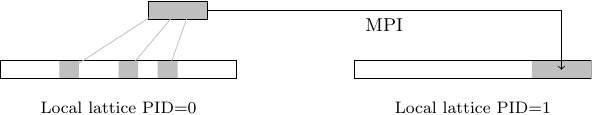}
\caption{Schematic illustration of MPI communications. For the computation-optimized geometry, all sites that need to be sent are synchronized to a send buffer. For the memory-optimized geometry, a synchronization minimizing scheme is used to synchronize in place and then send directly from the extended lattice. Figure, as published in \cite{Martins:2024dew}.}
\label{fig:mpiwsendbuffer}
\end{figure}

The communications are completed using CUDA-aware MPI and are implementation independent. Since the computation-optimized geometry is not optimized to minimize the amount of data synchronized to the send buffers, we need to synchronize all data to a separate send buffer, see Fig.~\ref{fig:mpiwsendbuffer}.

For the GPU implementation, depending on the compilation options, the communications are managed either from a separate Portable Operating System Interface (POSIX) thread using a blocking MPI \texttt{Sendrecv} operation or with non-blocking MPI asynchronous calls from the main thread. Using blocking calls has proven experimentally more stable on nodes that are not connected all-to-all between GPUs and have consequently bridges between GPUs that are utilized concurrently, but can only handle one data transfer at a time. When using non-blocking calls, memory transfers collide and pile up, causing long delays. Using blocking calls ensures that transfers are sequential, and, given the high bandwidth of intra-node memory, transfers can still be hidden by computation. Since blocking calls are managed in a separate thread, we increase the computational capacity of the transfer channel by allocating a dedicated thread exclusively to transfer execution. However, using non-blocking calls for all-to-all connected nodes is still slightly more performant.

For the evaluation of the hopping term of the Wilson-Dirac operator, only half-spinor data is required for each site. This means that memory transfer between neighboring nodes can be effectively reduced by a factor of two if only the necessary data is transmitted. Our implementation follows the approach described in \cite{openqcd}.

\section{Algorithms}\label{sec:algorithms}

\subsection{Inverters}\label{sec:inverters}

\subsubsection{Higher Precision Reductions}\label{sec:quadreductions}

To achieve higher precision for operations requiring the accumulation of sums over the lattice comparable to quadruple precision, we use double-double arithmetic, i.e., we represent the result of the summation using two IEEE 754 double precision numbers. In practice, we perform reduction operations by using compensated sums (a simple example is Kahan summation \cite{10.1145/512274.512287, Kahan1965PracniquesFR}), accumulating the sum in the first double-precision number and evaluating a compensation term in the second double-precision number. This is important in numerical inversion procedures since they rely on computing the norms and scalar products of vectors defined over the lattice. If performed na\"ively, the result of such operations will accumulate a numerical error in the worst case proportional to the lattice volume. 

Different algorithms for the accumulation are available in \texttt{HiRep}, achieving different trade-offs between precision and performance. Detailed descriptions for implementations of higher precision numbers can be found, e.g., in \cite{Dekker:1971,930115,richard1997adaptive} and references therein, from which we take the notation used in this section: We denote the regular exact addition and subtraction among real numbers as $\{+,-\}$, while the floating point addition and subtraction operators between IEEE 754 double precision numbers are denoted with $\{\oplus,\ominus\}$: i.e. $a\oplus b = fl(a+b)$ for one of the rounding rules in the standard. 

The algorithms below use the fact that the exact sum of two double precision numbers $a$ and $b$ can be represented exactly with an ordered pair of double precision numbers $(s,c)$, with $s = a \oplus b$ and $s + c = a + b$. The error or compensation term can be found using the so-called "fast two-sum" algorithm: assuming $\abs{a}\geq\abs{b}$ the error term is given by $c = b \ominus (s \ominus a)$.

\begin{algorithm}[H]
\caption{FastTwoSum(a,b)}\label{alg:cap}
\begin{algorithmic}
\State $s \gets a \oplus b$
\If{$\abs{a}\geq\abs{b}$}
    \State $c \gets b \ominus (s \ominus a)$
\Else
     \State $c \gets a \ominus (s \ominus b)$
\EndIf
\State return $(s, c)$
\end{algorithmic}
\end{algorithm}

A variant of the ``two-sum" algorithm also exists, which avoids the if branch at the expense of three extra floating point operations. Either variant can be used in the algorithms below.

The algorithms used in \HiRep for higher precision sums are described in the following.

The default implementation for higher precision accumulations in \HiRep is using a simple Neumaier summation \cite{Neumaier:1974}: the accumulation sum at each step is performed na\"ively, but the fast two-sum algorithm is used to find the compensation term at each step; the compensation terms so obtained are themselves summed together, and their final sum is added to the na\"ive result.

\begin{algorithm}[H]
\caption{Neumaier(double a[n])}
\begin{algorithmic}
\State $s \gets 0$, $c \gets 0$
\For{$i = 0\ldots n-1$}
\State $(s, e) \gets {\rm FastTwoSum}(s, a[i])$
\State $c \gets c \oplus e$
\EndFor
\State $s \gets s \oplus c$
\State return $s$
\end{algorithmic}
\end{algorithm}

Note that the accumulation of the compensation terms is done na\"ively in the above algorithm. One can instead use the same strategy by Neumaier to compute a compensated sum for the compensation terms, thus producing a "second-order" algorithm:
\begin{algorithm}[H]
\caption{SecondOrderNeumaier(double a[n])}
\begin{algorithmic}
\State $s \gets 0$, $c \gets 0$, $c2 \gets 0$
\For{$i = 0\ldots n-1$}
\State $(s, e) \gets {\rm FastTwoSum}(s, a[i])$
\State $(c, e) \gets {\rm FastTwoSum}(c, e)$
\State $c2 \gets c2 \oplus e$
\EndFor
\State $s \gets s \oplus (c \oplus c2)$
\State return $s$
\end{algorithmic}
\end{algorithm}

The advantage of the compensated sum algorithms above is that while they still have an error of order $O(\epsilon)$ where $\epsilon = 2^{-53} \simeq 1.1 \times 10^{-16}$ is the machine epsilon for IEEE double precision numbers, as in the na\"ive summation, the error does not grow with the number of terms in the sum but stays constant.

\HiRep v2 additionally supports computing the accumulation sum in double-double precision, corresponding to computing ``na\"ive" sums of double-double numbers with a higher precision of order $O(\epsilon^2)$, as also done in OpenQCD \cite{openqcd}:  
\begin{algorithm}[H]
\caption{DoubleDoubleSum(double a[n])}
\begin{algorithmic}
\State $s \gets 0$, $c \gets 0$
\For{$i = 0\ldots n-1$}
\State $(s, e) \gets {\rm FastTwoSum}(s, a[i])$
\State $(s, c) \gets {\rm FastTwoSum}(s, c \oplus e)$
\EndFor
\State return $s$
\end{algorithmic}
\end{algorithm}

In this context, it is important to note that an aggressive optimization by the compiler can remove the compensation. 

Finally, we note that higher precision summation in \HiRep is limited to the sums on the local lattice, while sums across computing units are performed as usual with MPI reductions.

\subsubsection{Supported Inverters}
For standard Monte-Carlo simulations, the Multi-Shift Conjugate Gradient (CG), the $\gamma_5$ Quasi Minimal Residual ($\gamma_5$QMR), and the stabilized Bi-Conjugate Gradient (BiCG)\cite{doi:10.1137/0914062, doi:10.1137/0913035} are supported for all architectures and efficient. The Conjugate Gradient is necessary for simulations with the RHMC, see Sect.~\ref{sec:rhmc}. However, the default inverter is the $\gamma_5$QMR \cite{Frommer:1995ik} for its good convergence properties. For ill-conditioned systems, the stabilized BiCG is often substantially faster but not guaranteed to converge.

\subsection{Integrators}

\HiRep implements multiple integrators of various orders of precision for the molecular dynamics evolution. Fig.~\ref{tbl:integrators} shows an overview of the available integrators. The integration scheme denotes the order of subsequent steps in updating the force (P) and the field (Q) if the integrator executes a single integration step. \HiRep supports hierarchical integrator structures \cite{Sexton:1992nu}, where each level can be assigned a different symplectic integrator, and monomials can be arranged on different levels. Discussion on ideal choices for integrators for lattice gauge theory simulations can be found in \cite{Kennedy:2009fe, Clark:2010qw, Kennedy:2007cb}.

\begin{table}
\centering
\begin{tabular}{lll}
\toprule
Integrator & Scheme & $\delta H\propto$ \\
\midrule
Leapfrog & PQP & $\delta \tau^2$\\
2nd-Order Omelyan & PQPQP & $\delta \tau^2$\\
4th-Order Omelyan & PQPQPQPQP & $\delta \tau^4$\\
\bottomrule
\end{tabular}
\caption{Overview over the available integrators in \texttt{HiRep}. We denote in the last column the scaling of the Hamiltonian violations $\delta H$ with the step size $\delta \tau$.}
\label{tbl:integrators}
\end{table}

\subsection{HMC}
\label{sec:HMC}

\subsubsection{Implementation of action monomials}
Dynamical lattice simulations aim to evaluate the general partition function, featuring a gauge field and fermions
\begin{equation}
\mathcal{Z} =\int \mathcal{D}U\prod_{f=1}^{N_f}\left[\mathcal{D}\overline{\psi}_f\mathcal{D}\psi_f\right]\,e^{-S_\mathrm{g}[U] - S_\mathrm{f}[U,\{\overline{\psi}_{f}\},\{\psi_f\}]}\,. 
\end{equation}
with a gauge action $S_g$ and a fermionic action $S_\mathrm{f}$. The Monte Carlo algorithms sample from the distribution $e^{-S_\mathrm{g}[U] - S_\mathrm{f}[U,\{\overline{\psi}_{f}\},\{\psi_f\}]}$ using variations of the HMC algorithm \cite{Duane:1987de}. 

The fermion action is given by
\begin{equation}
S_{\mathrm{f}}[U,\{\overline{\psi}_{f}\},\{\psi_f\}] = \sum_{f_1,f_2}\overline{\psi}_{f_1}M_{f_1,f_2}\psi_{f_2}
\end{equation}
where $M$ is a block diagonal matrix with $D^{R}_{m_f}$ on the diagonal. In \HiRep v2, the representation is the same for all fermions, and we often write $D^{R}_{m_f} = D_{m_f}$. The HMC can simulate an even number of fermions in the same representation with identical masses. 
In this case, we can simplify the action into a single determinant
\begin{align}
\begin{split}
\int \prod_{f=1}^{N_f}&\left[\mathcal{D}\overline{\psi}_f\mathcal{D}\psi_f\right]\,e^{-S_\mathrm{f}[U,\{\overline{\psi}_{f}\},\{\psi_f\}]} \\&= \prod_{f=1}^{N_f}\int\mathcal{D}\overline{\psi}_f\mathcal{D}\psi_f\,e^{-\overline{\psi}_fD_m\psi_f}= \mathrm{det}(D_m)^{N_f} = \mathrm{det}(Q_m^{N_f})\,.
\end{split}
\end{align} 
For even $N_f$, we can express this determinant as a functional integral over pseudofermion fields
\begin{align}
\mathrm{det}(Q_m^{N_f}) = \mathrm{det}((Q_m^{2})^{N_f/2})= \int\mathcal{D}\phi^{\dagger}\mathcal{D}\phi\,e^{\phi^{\dagger}(Q_m^{2})^{-N_f/2}\phi}\,,\qquad N_f/2 \in \mathds{N}\,,
\end{align}
this replaces the fermionic terms in the action with terms generated by pseudofermion (bosonic) fields, which is possible because the operator $Q_m^2 = Q^{\dagger}_mQ_m$ is Hermitian and positive definite and therefore, the integral converges. The inversion problem $Q_m^{-N_f/2}$ is for the standard HMC well defined for even numbers of fermions $N_f$. In addition, this inversion procedure is where the majority of time in the simulation is spent, and it is dominated by the execution time of the Wilson Dirac operator.

This software allows different definitions of these actions to yield different physical properties. The different actions are implemented in terms of monomials, corresponding to terms appearing in the action and functions that calculate the forces necessary for the molecular dynamics evolution (MDE) to canonically sample gauge configurations from the distribution.

\subsubsection{RHMC}
\label{sec:rhmc}
To simulate odd numbers of fermions in the theory, the RHMC algorithm \cite{Kennedy:1998cu, Clark:2003na} is available. The algorithm performs a rational approximation of the fermionic action, which is necessary since the regular HMC implementation only allows the power $\alpha\equiv N_f/2$ to be integer. The rational approximation
\begin{equation}
r(x) = \sum_{k=1}^{d}\gamma_k(x +\beta_k)^{-1} \simeq x^{-\alpha}
\end{equation}
is used to rewrite the fermion determinant as
\begin{equation}
\mathrm{det}\,Q_m^{\alpha}\simeq \int \mathcal{D}\phi\mathcal{D}\phi^{\dagger}e^{-\phi^{\dagger}r(Q_m^{2})\phi}\,,
\end{equation}
thus allowing any rational $\alpha$ to be used, as long the rational approximation is good enough on the spectral range of the operator $Q_m^2$, see \cite{Clark:2006wq}. The coefficients $\gamma_k$ and $\beta_k$ are evaluated to achieve the best precision at the given expansion order $d$ over the spectral range using the Remez algorithm. In \texttt{HiRep}, the coefficients of the rational approximation are precomputed and stored in a static table to be used at run time. 
The RHMC relies on a multi-shift inversion \cite{Frommer:1995ik} of the operator $(Q_m^2 +\beta_k)^{-1}$ to be efficient. 

\subsubsection{Even-Odd Preconditioning}
An even-odd preconditioning of the Dirac operator is often used for accelerating the HMC for lattice simulations, and its support in \HiRep can be switched on and off at compile time.

The preconditioner is defined given a decomposition of the lattice into even points, i.e. points where the coordinates fulfill $(t+x+y+z)\,\text{mod}\,2=0$, and the remaining odd points we can decompose any spinor field into an even and an odd component $\psi = \begin{pmatrix}\psi_{\mathrm{e}} &\psi_{\mathrm{o}}\end{pmatrix}$ we can rewrite the Dirac operator as
\begin{equation}
D_m = \begin{pmatrix}D_{\mathrm{ee},m} & D_{\mathrm{eo}} \\ D_{\mathrm{oe}} & D_{\mathrm{oo},m} \end{pmatrix}
\label{eq:dmeo}
\end{equation}

Given Eq. \ref{eq:dmeo}, the fermion determinant is given by
\begin{equation}
\mathrm{det}D_{m}  = \mathrm{det}(D_{\mathrm{oo},m})\mathrm{det}(\widehat{D}_m)\,,
\label{eq:deteo}
\end{equation}
with
\begin{equation}
\widehat{D}_{m} = D_{\mathrm{ee},m} - D_{\mathrm{eo}}D^{-1}_{\mathrm{oo},m}D_{\mathrm{oe}}\,.
\label{eq:deo}
\end{equation}

Accordingly, we can compute the determinant from its even-odd preconditioned version up to a prefactor that is constant in the case of unimproved Wilson fermions and with exponential clover improvement and computed separately for regular clover improvement. While most operators are implemented as in Eq.~\ref{eq:deo}, for simplicity in evaluating Eq.~\ref{eq:deteo} in \texttt{HiRep}, the even-odd preconditioned Wilson-Dirac operator without clover improvement is given by
\begin{equation}
\widehat{D}^{\mathrm{Wilson}}_{m} = (4+m)^2 - D_{\mathrm{eo}}D_{\mathrm{oe}}\,.
\end{equation}
absorbing explicitly another factor of $(4+m)$.

The Hermitian massive even-odd preconditioned Wilson-Dirac operator is in the following denoted $\widehat{Q}_m$ and defined as 
\begin{equation}
\widehat{Q}_{m} = \gamma_5(D_{\mathrm{ee},m} - D_{\mathrm{eo}}D^{-1}_{\mathrm{oo},m}D_{\mathrm{oe}})\,.
\end{equation}

\subsection{Pure Gauge Simulations}\label{sec:puregaugesimulations}
Pure gauge simulations using the HMC are slow compared to other methods, such as combining heat bath (HB) and overrelaxation (OR) steps. These pure gauge simulations are supported including multilevel algorithms and measurements of several gauge observables.

\subsubsection{Multi-level Algorithms}

\begin{figure}[H]
\centering
\resizebox{2.5in}{!}{
\begin{tikzpicture}

\draw[fill=orange] (0,2) -- (1,2) -- (1,3) -- (0,3) -- (0,2);
\draw[fill=orange] (4,2) -- (5,2) -- (5,3) -- (4,3) -- (4,2);

\draw[fill=orange] (0,0.5) -- (1,0.5) -- (1,1.5) -- (0,1.5) -- (0,0.5);
\draw[fill=orange] (2,0.5) -- (3,0.5) -- (3,1.5) -- (2,1.5) -- (2,0.5);
\draw[fill=orange] (4,0.5) -- (5,0.5) -- (5,1.5) -- (4,1.5) -- (4,0.5);
\draw[fill=orange] (6,0.5) -- (7,0.5) -- (7,1.5) -- (6,1.5) -- (6,0.5);

\node at (-1,4) {\Large Level 0};
\node at (-1,2.5) {\Large Level 1};
\node at (-1,1) {\Large Level 2};

\draw (0,3.5) -- (8,3.5) -- (8,4.5) -- (0,4.5) -- (0,3.5);
\draw (0,2) -- (8,2) -- (8,3) -- (0,3) -- (0,2);
\draw (0,0.5) -- (8,0.5) -- (8,1.5) -- (0,1.5) -- (0,0.5);

\foreach \i in {1,...,7} {
    \draw(\i,3.5) -- (\i,4.5);
    \draw(\i,2) -- (\i,3);
    \draw(\i,0.5) -- (\i,1.5);
}

\draw[->] (3.5,0) -- (4,0);
\node at (3.75,-0.2) {$t$};
\end{tikzpicture}
}
\caption{Illustration of the frozen time slices as in the implementation for multilevel boundary conditions. Here, a three-level algorithm is displayed, with a time extent of $8a$. The orange time slices are frozen, so the dynamical regions on the first level have size $3a$ and on the second level $1a$.}
\label{fig:ml}
\end{figure}
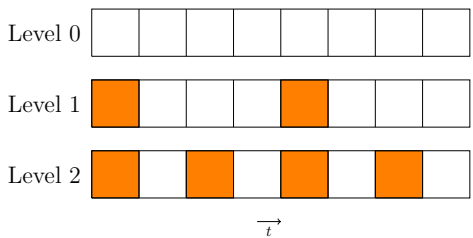

Statistical uncertainties of physical observables depend on the number of configurations generated, autocorrelation times, and the variance of the observable. While autocorrelation times depend on the algorithm, the variance is of the order of the operator. As a result, the signal-to-noise ratio degrades exponentially for large separations \cite{Parisi:1983ae, Lepage:1989hd}. 

The \HiRep simulation code has an implementation of the multilevel algorithm \cite{Luscher:2001up} as part of the \texttt{PureGauge} code as follows: The finite time extent is split in two, the first time slice of the width one is frozen, and updates are performed on the two decoupled sublattices. Higher levels are then performed by applying this logic to the sublattices separately. This also means that the length of the finite time extent limits the number of levels one can introduce.

An example of the application of fixed boundary conditions is shown in Fig.~\ref{fig:ml}. Here the orange colored slices are time slices are frozen when the integration on the respective level is performed. Note that the freezing applies to the links associated with the corresponding time slices, so, for example, as in the illustration, this cuts the links from time slice zero to time slice one and from time slice four to time slice five. Implementing a L{\"u}scher-Weisz gauge action would require freezing of two time slices and is not implemented \cite{Mykkanen:2012dv}. 

\section{Supported Actions}\label{sec:actions}

\subsection{Symanzik improvement}\label{sec:symanzik}

The library supports different improvement schemes to allow an $\mathcal{O}(a)$ or $\mathcal{O}(a^2)$ improvement in the fermion or gauge action.

\subsubsection{L{\"u}scher-Weisz Gauge Action}

The L{\"u}scher-Weisz Gauge action \cite{Luscher:1984xn} achieves an $\mathcal{O}(a^2)$ improvement by adding rectangular shapes of width $2\times 1$ $R_{\mu\nu}$ to the plaquette gauge action $P_{\mu\nu}$. 
\begin{equation}
S[U] = -\dfrac{\beta}{N}\sum_{x,\mu>\nu}c_0~\mathrm{Re\,tr}[P_{\mu\nu}(x)] + c_1\mathrm{Re\,tr}[R_{\mu\nu}(x) + R_{\nu\mu}(x)]
\end{equation}

\begin{figure}
\centering
\begin{tikzpicture}
    \draw[fill=lightgray, fill opacity=0.1, color=lightgray] (0.25,0.25) -- (0.25,5/2-0.25) -- (5/2-0.25,5/2-0.25) -- (5/2-0.25,0.25) -- (0.25,0.25);
    \foreach \i in {0,...,5} {
        \foreach \j in {0,...,5} {
            \draw[black] (\i/2,\j/2) circle (0.06);
        }
    }
    \begin{scope}[very thick,decoration={
        markings,
        mark=at position 0.8 with {\arrow{>}}}
        ]
        \draw[fill=black] (2/2,4/2) circle (0.06);
        \draw[fill=black] (2/2,5/2) circle (0.06);
        \draw[fill=red, color=red] (2/2,6/2) circle (0.08);
        \draw[fill=black] (3/2,4/2) circle (0.06);
        \draw[fill=black] (3/2,5/2) circle (0.06);

        \draw[postaction={decorate}, black] (2/2,4/2) -- (2/2,5/2-0.15);
        \draw[postaction={decorate}, black] (2/2,5/2) -- (2/2,6/2-0.15);
        \draw[postaction={decorate}, red] (2/2,6/2) -- (3/2-0.15,6/2);
        \draw[postaction={decorate}, black] (3/2,4/2) -- (3/2,5/2-0.15);
        \draw[postaction={decorate}, black] (3/2,5/2) -- (3/2,6/2-0.15);
        \draw[postaction={decorate}, black] (2/2,4/2) -- (3/2-0.15,4/2);
    \end{scope}
    \draw[->, very thick] (-1,-1) -- (1,-1);
    \draw[->, very thick] (-1,-1) -- (-1,1);
    \draw node at (1+0.2,-1) {$t$};
    \draw node at (-1,1+0.2) {$x$};

    \draw[fill=cyan, fill opacity=0.1, color=cyan] (0.25 + 4,0.25+1/2) -- (0.25 + 4,5/2-0.25+1/2) -- (5/2-0.25+4,5/2-0.25+1/2) -- (5/2-0.25+4,0.25+1/2) -- (0.25+4,0.25+1/2);
	\foreach \i in {0,...,5} {
        \foreach \j in {0,...,5} {
            \draw[black] (\i/2+4,\j/2) circle (0.06);
        }
    }
    
    \foreach \i in {0,...,5} {
        \foreach \j in {0,...,5} {
            \draw[cyan] (\i/2+4,\j/2+1/2) circle (0.1);
        }
    }
    \begin{scope}[very thick,decoration={
        markings,
        mark=at position 0.8 with {\arrow{>}}}
        ]
        \draw[fill=cyan, color=cyan] (2/2+4,4/2) circle (0.1);
        \draw[fill=cyan, color=cyan] (2/2+4,5/2) circle (0.1);
        \draw[fill=cyan, color=cyan] (2/2+4,6/2) circle (0.1);
        \draw[fill=cyan, color=cyan] (3/2+4,4/2) circle (0.1);
        \draw[fill=cyan, color=cyan] (3/2+4,5/2) circle (0.1);

        \draw[postaction={decorate}, cyan] (2/2+4,4/2) -- (2/2+4,5/2-0.15);
        \draw[postaction={decorate}, cyan] (2/2+4,5/2) -- (2/2+4,6/2-0.15);
        \draw[postaction={decorate}, cyan] (2/2+4,6/2) -- (3/2+4-0.15,6/2);
        \draw[postaction={decorate}, cyan] (3/2+4,4/2) -- (3/2+4,5/2-0.15);
        \draw[postaction={decorate}, cyan] (3/2+4,5/2) -- (3/2+4,6/2-0.15);
        \draw[postaction={decorate}, cyan] (2/2+4,4/2) -- (3/2+4-0.15,4/2);
    \end{scope}
\end{tikzpicture}
\caption{Implementation of L{\"u}scher-Weisz Gauge Action. Here, a naive implementation is causing issues in the MPI parallelization because, for some rectangles, a link is missing due to the single-width halo region; one example is the link marked in red. The solution to avoid changing the geometry implementation is to use a shifted field, in blue. The missing link is then located on the halo of the shifted field.}
\label{fig:lwimplementation}
\end{figure}

At the implementation level, this would require a halo region with a double width, as depicted in Fig.~\ref{fig:lwimplementation}. A single-width receive buffer region is insufficient to accommodate all rectangle shapes that are needed on the local lattice. The case depicted on the left-hand side of Fig.~\ref{fig:lwimplementation} shows the necessary links, all stored in the positive direction relative to the site. Here, one link is missing because it is outside the halo region. To solve this, we allocate an additional field shifted by one site as illustrated on the right-hand side in Fig.~\ref{fig:lwimplementation} for all eight directions. Now, rectangles located on the boundary can be computed from the shifted fields. This requires more memory but is still feasible for large lattices, even on modern GPUs with sufficient memory.

The fields are of type \texttt{staple\_field} and contain the three staples, which are $SU(N_g)$ matrices associated with each site and direction.

\subsubsection{Clover Improvement}\label{sec:clover}

Regular clover improvement \cite{Sheikholeslami:1985ij} adds a diagonal term to the Dirac Operator
\begin{equation}
D_{\mathrm{ee},m} + D_{\mathrm{oo},m} = (4+m) + c_{\mathrm{sw}}\dfrac{\mathrm{i}}{4}\sum_{\mu,\nu}\sigma_{\mu\nu}F_{\mu\nu}\,.
\end{equation}
The value $c_{\mathrm{sw}}$ can be non-perturbatively tuned to an ideal value that removes the $\mathcal{O}(a)$ effects produced by this action. We store the discretized version of the clover term in a separate field using 
\begin{align}
\begin{split}
C_{\mu\nu}(U,x) = P_{\mu\nu}(U, x) + P_{\nu\mu}(U,x-\hat{\mu}) + P_{\mu\nu}(U,x-\hat{\mu}-\hat{\nu}) + P_{\nu\mu}(U,x-\hat{\nu})
\label{eq:clover-term}
\end{split}
\end{align}
to define the field strength tensor
\begin{align}
\begin{split}
F_{\mu\nu} = \dfrac{1}{8}( C_{\mu\nu}(U,x) - C^{\dagger}_{\mu\nu}(U,x)) &= P_{\mu\nu}(U, x) + P_{\nu\mu}(U,x-\hat{\mu}) \\&+ P_{\mu\nu}(U,x-\hat{\mu}-\hat{\nu}) + P_{\nu\mu}(U,x-\hat{\nu})\,.
\label{eq:field-strength-tensor}
\end{split}
\end{align}

\subsubsection{Exponential Clover Improvement}

A stabilized version of the clover improvement can be achieved by using the exponential clover term \cite{Francis:2019muy}
\begin{equation}
D_{\mathrm{ee},m} + D_{\mathrm{oo},m} = (4+m_0)\,\mathrm{exp}\left(\dfrac{c_{\mathrm{sw}}}{4+m_0}\dfrac{\mathrm{i}}{4}\sum_{\mu,\nu}\sigma_{\mu\nu}F_{\mu\nu}\right)\,.
\end{equation}
While this improvement gives the same $\mathcal{O}(a)$ improvement as the regular clover improvement, the simulations show a reduced occurrence of exceptional configurations and, therefore, higher acceptance. Additionally, there is some empirical indication that the $\mathcal{O}(a^2)$ effects are generally smaller \cite{Francis:2019muy, Martins:2024ilw}.

\subsection{Twisted Mass Action}

\HiRep also supports the simulation of degenerate twisted mass
fermions\cite{Frezzotti:1999vv,Frezzotti:2000nk} by choosing the alternative twisted-mass monomial that corresponds to
\begin{align}
  \label{eq:2}
 D^{tm} = D_m + \tau^3 i \mu \gamma_5
\end{align}
where $D_m$ is the two flavour unimproved or exponential clover improved Wilson-Dirac operator.

\subsubsection{Hasenbusch Acceleration}

HMC simulations can be accelerated using Hasenbusch factorization~\cite{Hasenbusch:2001ne,Hasenbusch:2002ai} of the fermionic determinant. For each pair of flavors, the Hasenbusch factorization decomposes the fermion determinant as follows
\begin{align}
  \label{eq:1}
  \det{Q_m^2} = \det{W_{+} W_{- }} \det{Q_m \left(W_{+} W_{- }\right)^{-1} Q_m}\,,
\end{align}
where $W_\pm$ are chosen to speed up the simulations.

Two implementations are available. The first one is specific to unimproved Wilson fermions and is available with or without even-odd preconditioning. It corresponds to the choice:
\begin{align}
  \label{eq:23}
  W_+ = D_m + \delta_m,\quad\textrm{and}\quad W_- = W_+^\dagger\,.
\end{align}
where $\delta_m$ is an arbitrary real parameter. The fermionic determinant $\det{W_{+} W_{- }} $  corresponds to an HMC monomial with a shifted bare mass, while the other term is referred to as a Hasenbusch monomial which allows the user to set the parameter $\delta_m$.

Another implementation is available in the even-odd preconditioned
case. The fermionic determinant can be rewritten as
\begin{align}
  \label{eq:24}
  \det{Q_m^2}  = \det{\gamma_5 D_{ee}}^2\det{\widehat{Q}_{m}^2}\,.
\end{align}

This implementation is compatible with clover or exponential clover versions of the Dirac operator. The determinant $\det{\gamma_5D_{ee}}$ is irrelevant to the following discussion. One can then write:
\begin{align}
  \label{eq:25}
  \det{\widehat{Q}_{m}^2}  =  \det{W_{+} W_{- }} \det{\widehat{Q}_{m} \left(W_{+} W_{- }\right)^{-1} \widehat{Q}_{m}}\,.
\end{align}
Choosing
\begin{align}
  \label{eq:26}
  W_\pm(\mu) = \widehat{Q}_{m} \pm i \mu
\end{align}
the determinant $\det{W_{+} W_{- }}$ corresponds to an alternative twisted-mass monomial in the action, adding the alternative Hasenbusch monomial allows to include a term in the action defined by
\begin{align}
  \label{eq:27}
  \det  {  W_-(\mu)
                ( W_+
                 (\mu +\delta_\mu) W_-  (\mu + \delta_\mu))^{-1}  W_+  (\mu) }
\end{align}
where $\mu$ and $\delta_\mu$ are arbitrary real parameters set in the input file. The code therefore allows for multiple levels of factorization.

While a Hasenbusch-like trick is not directly available in the implementation for the RHMC, it is worth noting that for relatively light quark masses, a reduction of the fermion forces can be achieved by a similar manual splitting of the monomials into smaller fractions. A speed-up can already be achieved when simulating a two-fermion mass-degenerate theory with two RHMC monomials instead of a single HMC monomial. One can also split up a single fermion into monomials corresponding to four-fourths of a fermion, see \cite{Hansen:2017zly}, Figure 2.

\subsection{Boundary Conditions}

\subsubsection{Periodic and Antiperiodic Boundary Conditions}

The software supports periodic and antiperiodic boundary conditions in any direction. Boundary conditions are set at compile time. Periodic boundary conditions for gauge and fermion fields in some direction $i=0,\ldots,3$ are defined as
\begin{align}
\Psi_f(x + L_i\hat{e}_i) &= \Psi_f(x), \\
\overline{\Psi}_f(x + L_i\hat{e}_i) &= \overline{\Psi}_f(x) \\
U_{\mu}(x + L_i\hat{e}_i) &= U_{\mu}(x) 
\end{align}
$\forall\mu=0,\ldots,3$, and $f=1,\ldots,N_f$ and $L_i$ the (global) extent of the lattice in direction $i$. Antiperiodic boundary conditions introduce a negative sign for the fermion fields:
\begin{align}
\Psi_f(x + L_i\hat{e}_i) = -\Psi_f(x) \\
\overline{\Psi}_f(x + L_i\hat{e}_i) = -\overline{\Psi}_f(x) \\
U_{\mu}(x + L_i\hat{e}_i) = U_{\mu}(x)\,.
\end{align}

Boundary conditions in the fermions need to be correctly applied during the application of the Dirac operator. \HiRep implements antiperiodic boundary conditions by applying a negative sign to all links of one time slice of the represented gauge field, as illustrated in Fig.~\ref{fig:antiperiodicbc}. This effectively imposes antiperiodic boundary conditions, but only in the fermions, and the application is only necessary after every gauge update. 
\begin{figure}
    \centering
    \begin{tikzpicture}
    \foreach \i in {0,...,3} {
        \foreach \j in {0,...,3} {
            \draw[fill=black] (\i,\j) circle (0.06);
            \draw[->, very thick, lightgray] (\i,\j+0.2) -- (\i,\j+1-0.2);
            \draw[->, very thick, lightgray] (\j+0.2,\i) -- (\j+1-0.2,\i);
        }

        \draw[->, very thick, red] (0+0.2,\i) -- (1-0.2,\i);
        \draw[red] node at (0.41, 0.22+\i) {\tiny$\cdot(-1)$};
    }
    
    \draw[->, very thick] (-1,-1) -- (1,-1);
    \draw[->, very thick] (-1,-1) -- (-1,1);
    \draw node at (1+0.2,-1) {$t$};
    \draw node at (-1,1+0.2) {$x$};
    \end{tikzpicture}
    \caption{2D illustration of antiperiodic boundary conditions being applied on the represented gauge field. The $\cdot (-1)$ denotes a scalar multiplication of the correspondingly marked link with $-1$.}
    \label{fig:antiperiodicbc}
\end{figure}

\subsubsection{Open Boundary Conditions}

\paragraph{Fermions and fundamental gauge field}
Open boundary conditions in the temporal direction are supported in \HiRep. They are defined by setting all fields to zero on the boundary, i.e.
\begin{align}
 \Psi_f(x)\big|_{x_0=T-1} = 0\\
 \overline{\Psi}_f(x)\big|_{x_0=T-1} = 0\\
 U_{0}(x)\big|_{x_0=T-1} = 0\,,
\end{align}
for all $f=1,\ldots,N_f$. As before, the boundary conditions are implemented by applying them to the represented gauge field for fermions. However, we now want them to be applied to the fundamental gauge field. As a result, we first apply them to the fundamental gauge field and then represent this field. An illustration can be seen in Fig.~\ref{fig:openbc}.
\begin{figure}
\centering
    \begin{tikzpicture}

    \foreach \i in {0,...,3} {
        \foreach \j in {0,...,3} {
            \draw[fill=black] (\i,\j) circle (0.06);
            \draw[->, very thick, lightgray] (\i,\j+0.2) -- (\i,\j+1-0.2);
            \draw[->, very thick, lightgray] (\j+0.2,\i) -- (\j+1-0.2,\i);
        }
    }

    \foreach \i in {0,...,3} {
        \draw[->, very thick, red] (0+0.2+3,\i) -- (1-0.2+3,\i);
        \draw[red] node at (0.41+3, 0.22+\i) {\tiny$\cdot 0$};

        \draw[fill=red, color=red] (0+3,\i) circle (0.06);
        \draw[red] node at (0.2+3,\i-0.2) {\tiny$\cdot 0$};

    }
    
    \draw[->, very thick] (-1,-1) -- (1,-1);
    \draw[->, very thick] (-1,-1) -- (-1,1);
    \draw node at (1+0.2,-1) {$t$};
    \draw node at (-1,1+0.2) {$x$};
    \end{tikzpicture}
    \caption{2D illustration of open boundary conditions applied on the fundamental gauge and spinor fields. The red links in this illustration are set to a zero SU($N_g$) matrix, and the red sites are set to a zero spinor. Effectively usable for measurements are only $T-1$ time slices. The $\cdot 0$ denotes a scalar multiplication of the correspondingly marked links and sites with $0$ and the $\mathds{1}$ denotes setting the corresponding links to a unit matrix. }
    \label{fig:openbc}
\end{figure}

\paragraph{Plaquette Weights}
The behavior at the boundary can be improved by setting the plaquette weights and, when using the L{\"u}scher-Weisz gauge action instead of the pure plaquette action, rectangular weights. The coefficients to set the weights in the temporal and spatial direction, respectively $c_t$ and $c_s$, are read from the input file. 

\begin{figure}
\centering
\begin{minipage}[c]{0.8\linewidth}
\centering
\begin{tikzpicture}
    \foreach \i in {0,...,5} {
        \foreach \j in {0,...,5} {
            \draw[black] (\i/2,\j/2) circle (0.06);
        }
    }
    \begin{scope}[very thick,decoration={
        markings,
        mark=at position 0.65 with {\arrow{>}}}
        ] 
        \draw[fill=black] (0/2,1/2) circle (0.06);
        \draw[postaction={decorate}, red] (0/2,1/2)--(0/2,2/2);
        \draw[postaction={decorate}, red] (0/2,2/2)--(1/2,2/2);
        \draw[postaction={decorate}, red] (1/2,2/2)--(1/2,1/2);
        \draw[postaction={decorate}, red] (1/2,1/2)--(0/2,1/2);
        \draw[red] node at (0/2+0.25,1/2+0.25) {$c_t$}; 

        \draw[fill=black] (1/2,3/2) circle (0.06);
        \draw[postaction={decorate}, blue] (1/2,3/2) -- (1/2,5/2);
        \draw[postaction={decorate}, blue] (1/2,5/2) -- (2/2,5/2);
        \draw[postaction={decorate}, blue] (2/2,5/2) -- (2/2,3/2);
        \draw[postaction={decorate}, blue] (2/2,3/2) -- (1/2,3/2);
        \draw[blue] node at (1/2+0.25,3/2+0.5) {$1$}; 

        \draw[fill=black] (3/2,3/2) circle (0.06);
        \draw[postaction={decorate}, green] (3/2,3/2) -- (3/2,4/2);
        \draw[postaction={decorate}, green] (3/2,4/2) -- (4/2,4/2);
        \draw[postaction={decorate}, green] (4/2,4/2) -- (4/2,3/2);
        \draw[postaction={decorate}, green] (4/2,3/2) -- (3/2,3/2);
        \draw[green] node at (3/2+0.25,3/2+0.25) {$c_t$};

        \draw[fill=black] (2/2,1/2) circle (0.06);
        \draw[postaction={decorate}, yellow] (2/2,1/2) -- (2/2, 2/2);
        \draw[postaction={decorate}, yellow] (2/2,2/2) -- (3/2, 2/2);
        \draw[postaction={decorate}, yellow] (3/2,2/2) -- (3/2, 1/2);
        \draw[postaction={decorate}, yellow] (3/2,1/2) -- (2/2, 1/2);
        \draw[yellow] node at (2/2+0.25,1/2+0.25) {$1$}; 

        \draw[fill=black] (4/2,1/2) circle (0.06);
        \draw[postaction={decorate}, orange] (4/2,1/2) -- (4/2,3/2);
        \draw[postaction={decorate}, orange] (4/2,3/2) -- (5/2,3/2);
        \draw[postaction={decorate}, orange] (5/2,3/2) -- (5/2,1/2);
        \draw[postaction={decorate}, orange] (5/2,1/2) -- (4/2,1/2);
        \draw[orange] node at (4/2+0.25,1/2+0.5) {$0$};
    \end{scope}
    \draw[->, very thick] (-1,-1) -- (1,-1);
    \draw[->, very thick] (-1,-1) -- (-1,1);
    \draw node at (1+0.2,-1) {$t$};
    \draw node at (-1,1+0.2) {$x$};
\end{tikzpicture}
\end{minipage}
\caption{Examples of plaquette and rectangle weights used for open boundary conditions in evaluating observables, illustrated on a $6^2$ global lattice.}
\label{fig:plaqweightsbc}
\end{figure}

As illustrated in Fig.~\ref{fig:plaqweightsbc}, plaquettes or rectangles evaluated in the respective action have weights from 0 to 1, which corrects the finite volume effects through the boundary. The weights for the plaquettes are defined as in the following:

\begin{enumerate}
    \item If one of the links of the plaquette is located on the open boundary, that is, the corresponding site is located on timeslice $T-1$ with $\mu$ or $\nu=0$, then the weight is zero.
    \item If the plaquette is adjacent and perpendicular to the open boundary, that is, located on timeslice $0$ or $T-2$ with $\mu$ or $\nu=0$, the weight is $c_t$
    \item If the plaquette is adjacent and parallel to the open boundary, the weight is $c_s/2$.
    \item For all other plaquettes, the weight is $1$.
\end{enumerate}

For the rectangles, the weights are slightly different
\begin{enumerate}
    \item If one of the links of the rectangle is located on the open boundary, that is, the corresponding site is located on timeslice $T-1$ with $\mu$ or $\nu =0$ or $T-2$, with $\mu=0$ (i.e. the longer edge), then the weight is zero.
    \item If the rectangle is adjacent and parallel to the open boundary, the weight is $c_s/2$.
    \item All other rectangles have the weight $1$.
\end{enumerate}

\paragraph{Clover term}

The field containing the clover term defined by the field strength tensor definition in Eq.~\ref{eq:field-strength-tensor} also needs to obey open boundary conditions, but the term itself has non-zero contributions from the gauge field on the boundary. To fully impose the boundary conditions, the clover term needs to be set to zero on the boundary, which is achieved by imposing
\begin{equation}
C_{\mu\nu}(U,x)\big|_{x_0=0} = C_{\mu\nu}(U,x)\big|_{x_0=T-1} = 0\,, \text{for~}\mu,\nu=0,\ldots,3\,.
\end{equation}

\subsubsection{Schr{\"o}dinger Functional}

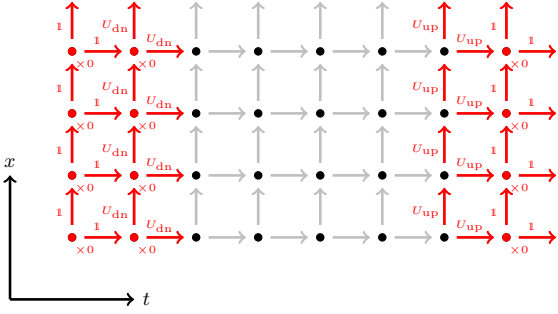
\begin{figure}[H]
\centering
\resizebox{3in}{!}{
    \begin{tikzpicture}

    \foreach \i in {0,...,5} {
        \foreach \j in {0,...,3} {
            \draw[fill=black] (\i,\j) circle (0.06);
        }
    }

    \foreach \i in {0,...,3} {
        \foreach \j in {0,...,3} {
            \draw[->, very thick, lightgray] (\i,\j+0.2) -- (\i,\j+1-0.2);
            \draw[->, very thick, lightgray] (\j+0.2,\i) -- (\j+1-0.2,\i);
        }
    }
    
    \foreach \i in {0,...,1} {
        \foreach \j in {0,...,3} {
            \draw[->, very thick, lightgray] (\i+4,\j+0.2) -- (\i+4,\j+1-0.2);
            \draw[->, very thick, lightgray] (\i+0.2+4,\j) -- (\i+1-0.2+4,\j);
        }
    }

    \foreach \i in {0,...,3} {
        \draw[->, very thick, red] (0+0.2,\i) -- (1-0.2,\i);
        \draw[red] node at (0.41, 0.18+\i) {\tiny$\mathds{1}$};
        \draw[->, very thick, red] (0,\i+0.2) -- (0,\i+1-0.2);
        \draw[red] node at (-0.2, 0.4+\i) {\tiny$\mathds{1}$};

        \draw[->, very thick, red] (0+1,\i+0.2) -- (0+1,\i+1-0.2);
        \draw[red] node at (-0.3+1, 0.4+\i) {\tiny$U_{\mathrm{dn}}$};

        \draw[->, very thick, red] (0+0.2+5,\i) -- (1-0.2+5,\i);
        \draw[red] node at (0.41+5, 0.18+\i) {\tiny$\mathds{1}$};
        \draw[->, very thick, red] (0+5,\i+0.2) -- (0+5,\i+1-0.2);
        \draw[red] node at (-0.3+5, 0.4+\i) {\tiny$U_{\mathrm{up}}$};

        \draw[fill=red, color=red] (0,\i) circle (0.06);
        \draw[red] node at (0.2,\i-0.2) {\tiny$\cdot 0$};

        \draw[fill=red, color=red] (1,\i) circle (0.06);
        \draw[red] node at (1.2,\i-0.2) {\tiny$\cdot 0$};

        \draw[fill=red, color=red] (5,\i) circle (0.06);
        \draw[red] node at (5.2,\i-0.2) {\tiny$\cdot 0$};
    }
    
    \draw[->, very thick] (-1,-1) -- (1,-1);
    \draw[->, very thick] (-1,-1) -- (-1,1);
    \draw node at (1+0.2,-1) {$t$};
    \draw node at (-1,1+0.2) {$x$};
    \end{tikzpicture}
}
    \caption{2D illustration of Schr{\"o}dinger functional boundary conditions are applied on the fundamental gauge and spinor fields. The links marked red in this illustration are set to two different SU($N_g$) matrices as defined below, and the sites marked red are set to a zero spinor. Note that for Schr{\"o}dinger functional boundary conditions, measurements can only be performed on $T-2$ time slices. }
    \label{fig:sfbc}
\end{figure}

\paragraph{Fermions}
It is also possible to simulate with Schr{\"o}dinger functional boundary \cite{Sint:1993un, Luscher:1996ug} conditions, which are defined on the spinor fields by
\begin{align}
P_{+}\Psi_f(x)\big|_{x_0=0} = \rho(x) \\ 
\overline{\Psi}_f(x)P_{-}\big|_{x_0=0} = \overline{\rho}(x) \\ 
P_{+}\Psi_f(x)\big|_{x_0=T-1} = \eta(x)\\
\overline{\Psi}_f(x)P_{-}\big|_{x_0=T-1} = \overline{\eta}(x)
\end{align}
for all $f=1,\ldots,N_f$, with the projector $P_{\pm} = \tfrac{1}{2}(1\pm\gamma_0)$, $\rho(x), \eta(x)$ two fixed charge distributions and $\Psi$ the fermionic fields in the theory. In \texttt{HiRep}, these charge distributions are set to zero globally. This avoids the appearance of the zero mode in the range of possible momenta, introducing an unphysical particle of mass $1/L$, with $L$ being the shortest spatial lattice extent.

\paragraph{Fundamental gauge field}
For the gauge fields, the spatial links are set to matrices of the type
\begin{align}
U_j(n,0) = \mathrm{diag}\left(\left\{e^{\mathrm{i}\tfrac{\theta_i}{T - 2}}\right\}_{i\in 1,\ldots,N_g}\right) = U_{\mathrm{dn}}
\end{align}

\begin{align}
U_j(n,T-1) = \mathrm{diag}\left(\left\{e^{\mathrm{i}\tfrac{\phi_i}{T - 2}}\right\}_{i\in 1,\ldots,N_g}\right) := U_{\mathrm{up}}
\end{align}
with $\sum_i \theta_i = 0$ and $\sum_i \phi_i = 0$. 

These boundary conditions are a type of fixed boundary conditions, that induce a static background field to the theory. The purpose of such boundary conditions is to tune $c_{\mathrm{sw}}$ to achieve $\mathcal{O}(a^2)$-scaling of discretization effects, without any contamination by terms at $\mathcal{O}(a)$.

The implementation only supports the specific ideal choices for the angles given the number of colors, which minimizes discretization effects by choosing the field so that it is weak at the cut-off scale \cite{Luscher:1996ug}. It is also possible to use Schr{\"o}dinger functional boundary conditions without a background field. In this case, all angles $\bm{\theta}$ and $\bm{\phi}$ are zero, and the matrices $U_{\mathrm{up}}$ and $U_{\mathrm{dn}}$ are unit matrices. 

When simulating with a background field, $N_g=2, 3$ and $4$ are supported. $SU(2)$ is set to the angles
\begin{equation}
\bm{\theta} = \left[0, 0\right] \qquad\bm{\phi} = \left[-\dfrac{\pi}{2},\dfrac{\pi}{2}\right]\,.
\end{equation}
$SU(3)$ supplies the angles
\begin{equation}
\bm{\phi} = \left[-\dfrac{5\pi}{3},\dfrac{2\pi}{3}, \pi\right]\,,
\end{equation}
\begin{equation}
\bm{\theta}  = \left[-\dfrac{\pi}{3}, 0, \dfrac{\pi}{3}\right]\,.
\end{equation}
For $SU(4)$ these angles are set to
\begin{equation}
\bm{\phi} = \left[-\dfrac{\sqrt{2}\pi}{4} - \dfrac{\pi}{2}, -\pi + \dfrac{\sqrt{2}\pi}{4}, \pi - \dfrac{\sqrt{2}\pi}{4}, \dfrac{\pi}{2} + \dfrac{\sqrt{2}\pi}{4}\right]\,,
\end{equation}
\begin{equation}
\bm{\theta} = \left[-\dfrac{\sqrt{2}\pi}{4},\dfrac{\sqrt{2}\pi}{4} - \dfrac{\pi}{2},\dfrac{\pi}{2} - \dfrac{\sqrt{2}\pi}{4}, \dfrac{\sqrt{2}\pi}{4}\right]\,.
\end{equation}

\paragraph{Plaquette Weights}
Schr{\"o}dinger functional boundary conditions are not supported for the L{\"u}scher-Weisz Gauge action. As a result, the only weights that are set during initialization are the plaquette weights. In contrast to open boundary conditions, there is no correction in the spatial direction close to the boundary. As a result, the weights are set according to the following principles:
\begin{enumerate}
    \item If one of the links of the plaquette is located on the fixed boundary, that is, the corresponding site is located on timeslice $0$ or $T-1$, or timeslice $1$ with $\mu \neq 0$ and $\nu \neq 0$ then the weight is zero.
    \item If the plaquette is adjacent and perpendicular to the fixed boundary, that is, the corresponding site is located on timeslice $1$ or $T-2$ with $\mu$ or $\nu=0$, the weight is $c_t$
    \item For all other plaquettes, the weight is $1$.
\end{enumerate}

\paragraph{Clover term}
The boundary conditions for the clover term are again imposed by setting
\begin{equation}
C_{\mu\nu}(U,x)\big|_{x_0=0} = C_{\mu\nu}(U,x)\big|_{x_0=1} = 0\, .
\end{equation}

\subsubsection{Spatially Twisted Boundary Conditions}
Additionally, spatially twisted boundary conditions, see \cite{tHooft:1979rtg, Bedaque:2004kc, Sachrajda:2004mi}, defined by a rotation by a symmetry of the action $\Omega_i$
\begin{align}
A_{\mu}(x_0,\bm{x} + L_i\bm{e}_i) = \Omega_iA_{\mu}(x)\Omega_{i}^{\dagger}\\
\phi(x_0,\bm{x} + L_i\bm{e}_i) = \Omega_i\phi(x)\Omega_i^{\dagger}
\label{eq:twistedbc}
\end{align}
for an element of the given gauge group $SU(N_g)$, with the constraint 
\begin{equation}
\Omega_j\Omega_k = e^{\tfrac{2\pi\mathrm{i}}{N}\varepsilon_{jkl}\bm{m}}\Omega_k\Omega_j
\end{equation}
where $\bm{m}=(1,1,1)$ is the magnetic flux, chosen to the maximal symmetric twist, as described in \cite{DelDebbio:2015byq}.

The implementation now relies on the fact that the boundary conditions do not change the fermionic action and can be absorbed into a gauge transformation. The only change is in the gauge action. For the plaquette action, this is as in \cite{DelDebbio:2015byq}:
\begin{equation}
S_{g} = -\dfrac{\beta}{2N}\sum_{x,\mu<\nu}z_{\mu\nu}(x)\mathrm{tr}\,U_{\mu\nu}(x)
\end{equation}
with
\begin{equation}
z_{\mu\nu}(x) = \begin{cases}-1 & x_0 = x_\rho = 1\\ 1 & \mathrm{otherwise}\end{cases}
\end{equation}
for the three distinct spatial directions $\mu,\nu$ and $\rho$.

\section{Observables}\label{sec:observables}

\subsection{Spectroscopy}

\HiRep features methods to estimate two-point correlators relevant to calculating the spectrum of the simulated lattice gauge theories. 

A flag in the input file of HMC simulations allows to switch on online measurements that estimate the flavor non-singlet two-point correlation function at zero momentum for all ultra-local fermionic operators of the form
\begin{align}
    \label{eq:29}
    O_\Gamma(x) = \overline{\Psi}_{f_1}(x) \Gamma  \Psi_{f_2}(x)\,.
\end{align}
where $\Gamma=1,\gamma_5, \gamma_\mu, \gamma_\mu \gamma_5, \dots$. The simulation code uses a stochastic estimation based on $\mathbb{Z}_2 \times \mathbb{Z}_2 $ spin-diluted wall sources randomly located on the lattice. The code averages the stochastic estimator of each channel over the number of hits set in the input file. 

The suite can also estimate, for instance, flavor non-singlet two-point function mesonic correlators using point sources or spin-diluted wall sources. The final output will correspond to the volume-averaged connected correlation function

\begin{equation}
C(t) = \dfrac{1}{XYZ}\sum_{\bm{x}}\langle O_{\Gamma}(t,\bm{x})\overline{O}_{\Gamma'}(0)\rangle\,.
\end{equation}

The executable also calculates baryonic correlators using point sources for $SU(3)$ gauge theories such as QCD or $N_F=2$ with fermions in the symmetric representation \cite{Hansen:2017ejh}. The code also computes disconnected loops of the form $\langle \overline{\psi} \Gamma \psi\rangle$, which are printed to the output without explicit volume and stochastic averaging to allow for more flexibility. 

The suite also offers measurement tools to evaluate Schr{\"o}dinger functional correlators relevant to non-perturbatively tuning the improvement coefficient $c_{sw}$ as defined in \cite{Luscher:1996sc} and some three-point correlators relevant to estimate some form factors of pseudoscalar mesons \cite{Hietanen:2013fya}.

\subsubsection{Connected Contributions}

\paragraph{Point Sources}
 The simplest way to evaluate the connected contributions to the correlation function 
\begin{equation}
 \langle\overline{\Psi}_{f_1}(x)\Gamma \Psi_{f_2}(x)\overline{\Psi}_{f_2}(y)\overline{\Gamma}' \Psi_{f_1}(y)\rangle = -\mathrm{tr}\,[\Gamma S_{f_2}(x,y)\overline{\Gamma}' S_{f_1}(y,x)]
\end{equation} 
 is to calculate the propagator $S_{f}(x,y) = \langle\overline{\Psi}_{f}^{(i)}\Psi_f\rangle$ using point sources. This means that we are placing the source at $y$ and creating the source according to the function
 \begin{equation}
\eta_{ab}^{\alpha\beta}(x,y) = \delta_{x,y}\delta_{ab}\delta_{\alpha\beta} = \begin{cases}\mathds{1}_{ab}\otimes \mathds{1}_{\alpha\beta} & x = y\\ 0 &\mathrm{otherwise}\end{cases}
\end{equation}
with $\mathds{1} = \mathrm{diag}(1,\ldots,1)$ the unit matrix, as the input field to the inversion problem
 \begin{equation}
D_{m,ac}^{\alpha\gamma}(x,y)\widehat{S_{cb}^{\gamma\beta}(y,z)} = \eta^{\alpha\beta}_{ab}(x,z)\,,
 \end{equation}
which provides an estimate for the propagator $\widehat{S_{cb}^{\gamma\beta}(y,z)}$ for each gauge configuration.

The propagator can then be used to calculate the different contractions. In practice we set $y=(0,\bm{0})$. The bottleneck of this is the inversion procedure. For light quark masses, most of the time is spent calculating the propagator.

\paragraph{Dilution of volume sources}
To deal with noise in the measurements, \HiRep supports dilution schemes \cite{Foley:2005ac}. Here the underlying idea is to decompose the inversion domain requiring orthogonality of the corresponding source contributions. These decomposed sources have better overlap with the low-eigenmodes of the system and, therefore, achieve less contaminated ground state estimates. The sources can be simple volume or stochastic sources, which are then decomposed. Take a set of disjoint domains $D_i\in \mathcal{D}$ for $i=1,\ldots,N_{\mathrm{dom}}$ such that
\begin{equation} 
\eta^{(j)}(x) = \sum_{i\in \mathcal{D}}\eta^{(i,j)}(x)
\end{equation}
where each $\eta^{(i,j)}(x)$ has support only in $D_i$ and $j=1,\ldots,N_{\mathrm{stoch}}$ enumerates the stochastic hits. If $N_{\mathrm{stoch}}=1$ we denote $\eta^{(i)} = \eta^{(i,j)}$. Stochastic sources with $N_{\mathrm{stoch}}>1$ are constructed such that these sources are all orthogonal to each other in the stochastic limit $N_{\mathrm{stoch}}\rightarrow \infty$
\begin{align}
\begin{split}
\langle \eta^{(i_1,j)},\eta^{(i_2,j)}\rangle = \sum_{j=0}^{N_{\mathrm{stoch}}}\overline{\eta}^{(i_1,j)}(x)\otimes\eta^{(i_2,j)}(y) = \mathds{1}_{4N_g\times 4N_g}\delta_{x,y}\delta^{i_1,i_2}
\end{split}
\end{align}
The inversion is then performed on the $\eta^{(i,j)}$ instead of $\eta$, yielding a solution $\psi$ for each domain
\begin{equation}
\psi^{(i,j)}(x) = D^{-1}\eta^{(i,j)}(y)\,.
\end{equation}
The reassembling of the full domain is performed at the level of the propagator
\begin{equation}
\widehat{S^{ab}_{\alpha\beta}(x, y)} = \lim_{N_{\mathrm{stoch}}\rightarrow\infty}\dfrac{1}{N_{\mathrm{stoch}}}\sum_{i}^{N_{\mathrm{dom}}}\sum_{j}^{N_{\mathrm{stoch}}}\psi^{(i,j)}_{a\alpha}(x)\eta^{(i,j),\dagger}_{b\beta}(y). 
\label{eq:alltoallprop}
\end{equation}
where $\widehat{S^{ab}_{\alpha\beta}(x, y)}$ denotes the noise-reduced estimate for the propagator obtained from a single gauge configuration.

\paragraph{SEMWall sources}

\HiRep implements Spin-Explicit Method (SEMWall) sources, see, for example, \cite{Boyle:2008rh}. SEMWall sources are stochastic sources with a simple spin dilution scheme. Due to Eq. \ref{eq:alltoallprop}, the sources are simple spinors instead of matrices that have an additional dependence on the dilution domain. Using a $\mathds{Z}_2\otimes \mathds{Z}_2$ noise, a random drawing from the set $\{\dfrac{1}{\sqrt{2}}(\pm 1\pm\mathrm{i})\}$. We create a source for each time slice $t_{\mathrm{slice}}$, 
 chosen at random, meaning each spinor component of spin $\alpha$ on each point $(t_{\mathrm{slice}},x,y,z)$ and fill it with stochastic noise. Formally, a SEMWall source is defined as
\begin{equation}
\eta_{a\alpha}^{(i,j)}(x,t_{\mathrm{slice}}) = \delta_{
\alpha i}\delta_{t,t_{\mathrm{slice}}}K_{a}(x)
\end{equation}
where $K_{a}$ is a color vector, $i=1,\ldots,4$, randomly filled with stochastic noise, and t the 0-th component of $x$. 

These sources are normalized to $1$, identical to the point source. The correlation functions will have the same normalization, and matrix elements can be extracted completely alike.

\paragraph{Extended Propagators}

The code supports the computation of extended propagators, where the correlation function is calculated both with periodic and antiperiodic boundary conditions and superimposed to extend the correlator length to $2T$, as described in \cite{Goode:2011pd}.

\subsubsection{Disconnected Contributions}

\HiRep contains tools dedicated to measuring disconnected loops using various dilution schemes that do not perform the stochastic average. The code calculates disconnected loops of the form $\langle \overline{\Psi}_{f_1} \Gamma_i \Psi_{f_2} \rangle$.  The Hermitian basis of gamma matrices used is given in Tbl.~\ref{tab:Gammabasis}. Each basis element is referred to by an index in $[0,15]$ shown in the table. The code also supports projection at non-zero momenta.

\begin{table}[H]
\centering
\begin{minipage}[c]{\linewidth}
\centering
 \begin{tabular}{p{4mm}p{4mm}p{4mm}p{4mm}p{10mm}p{10mm}p{10mm}p{10mm}}
\toprule
\centering 0 & \centering 1 & \centering 2 & \centering 3 & \centering 4 & \centering 5 & \centering 6 &  \multicolumn{1}{c}{7} \\
$\gamma_5$ & $\gamma_1$  & $\gamma_2$  & $\gamma_3$ & $-i\gamma_0 \gamma_5$  & $-i\gamma_0 \gamma_1$  & $-i\gamma_0 \gamma_2$ &$-i\gamma_0 \gamma_3$\\
\bottomrule
\end{tabular}
\end{minipage}

\vspace{3mm}
\begin{minipage}[c]{\linewidth}
\centering
\begin{tabular}{p{4mm}p{10mm}p{10mm}p{10mm}}
\toprule
 \centering 8 & \centering 9 & \centering 10 & \multicolumn{1}{c}{11}  \\
 $1$  & $-i\gamma_5 \gamma_1$  & $-i\gamma_5 \gamma_2$  & $-i\gamma_5 \gamma_3$ \\
\bottomrule
\end{tabular}
\end{minipage}
 
\vspace{3mm}
\begin{minipage}[c]{\linewidth}
\centering
 \begin{tabular}{p{4mm}p{12mm}p{12mm}p{12mm}}
\toprule
\centering 12 & \centering 13 & \centering 14 & \multicolumn{1}{c}{15} \\
  $\gamma_0$  & $-i\gamma_5\gamma_0\gamma_1 $  & $-i\gamma_5 \gamma_0\gamma_2$  & $-i\gamma_5 \gamma_0 \gamma_3$\\ 
\bottomrule
\end{tabular}
\end{minipage}
\caption{Basis used in compute loops. }
\label{tab:Gammabasis}
\end{table}

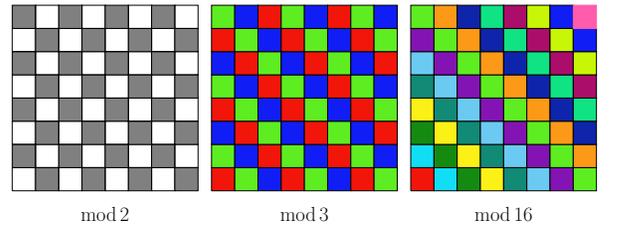
\begin{figure}[H]
\centering
\resizebox{1in}{!}{
\begin{tikzpicture}
    \foreach \i in {0,...,7} {
        \foreach \j in {0,...,7} {
            \pgfmathparse{Mod(\i+\j,2)==0?1:0};
                \ifnum\pgfmathresult>0
                    \draw (\i,\j) -- (\i+1,\j) -- (\i+1,\j+1) -- (\i,\j+1) -- (\i,\j);
                \else
                    \draw[fill=gray] (\i,\j) -- (\i+1,\j) -- (\i+1,\j+1) -- (\i,\j+1) -- (\i,\j);
                \fi
        }
    }
    \node at (4, -1) {\Huge$\mathrm{mod}\,2$};

\end{tikzpicture}
}
\resizebox{1in}{!}{
\begin{tikzpicture}
    \foreach \i in {0,...,7} {
        \foreach \j in {0,...,7} {
            \pgfmathparse{Mod(\i+\j,3)==0?1:0};
                \ifnum\pgfmathresult>0
                    \draw[fill=rgbred] (\i,\j) -- (\i+1,\j) -- (\i+1,\j+1) -- (\i,\j+1) -- (\i,\j);
                \else
                    \pgfmathparse{Mod(\i+\j,3)==1?1:0};
                    \ifnum\pgfmathresult>0
                        \draw[fill=rgbgreen] (\i,\j) -- (\i+1,\j) -- (\i+1,\j+1) -- (\i,\j+1) -- (\i,\j);
                    \else
                        \draw[fill=rgbblue] (\i,\j) -- (\i+1,\j) -- (\i+1,\j+1) -- (\i,\j+1) -- (\i,\j);
                    \fi
                \fi
        }
    }
    \node at (4, -1) {\Huge$\mathrm{mod}\,3$};
\end{tikzpicture}
}
\resizebox{1in}{!}{
\begin{tikzpicture}
    \foreach \i in {0,...,7} {
        \foreach \j in {0,...,7} {
            \pgfmathparse{Mod(\i+\j,16)==0?1:0};
                \ifnum\pgfmathresult>0
                    \draw[fill=palette1] (\i,\j) -- (\i+1,\j) -- (\i+1,\j+1) -- (\i,\j+1) -- (\i,\j);
                \else
                    \pgfmathparse{Mod(\i+\j,16)==1?1:0};
                    \ifnum\pgfmathresult>0
                        \draw[fill=palette2] (\i,\j) -- (\i+1,\j) -- (\i+1,\j+1) -- (\i,\j+1) -- (\i,\j);
                    \else
                        \pgfmathparse{Mod(\i+\j,16)==2?1:0};
                        \ifnum\pgfmathresult>0
                            \draw[fill=palette4] (\i,\j) -- (\i+1,\j) -- (\i+1,\j+1) -- (\i,\j+1) -- (\i,\j);
                        \else
                            \pgfmathparse{Mod(\i+\j,16)==3?1:0};
                            \ifnum\pgfmathresult>0
                                \draw[fill=palette3] (\i,\j) -- (\i+1,\j) -- (\i+1,\j+1) -- (\i,\j+1) -- (\i,\j);
                            \else
                                \pgfmathparse{Mod(\i+\j,16)==4?1:0};
                                \ifnum\pgfmathresult>0
                                    \draw[fill=palette5] (\i,\j) -- (\i+1,\j) -- (\i+1,\j+1) -- (\i,\j+1) -- (\i,\j);
                                \else
                                    \pgfmathparse{Mod(\i+\j,16)==5?1:0};
                                    \ifnum\pgfmathresult>0
                                        \draw[fill=palette7] (\i,\j) -- (\i+1,\j) -- (\i+1,\j+1) -- (\i,\j+1) -- (\i,\j);
                                    \else
                                        \pgfmathparse{Mod(\i+\j,16)==6?1:0};
                                        \ifnum\pgfmathresult>0
                                            \draw[fill=palette6] (\i,\j) -- (\i+1,\j) -- (\i+1,\j+1) -- (\i,\j+1) -- (\i,\j);
                                        \else
                                            \pgfmathparse{Mod(\i+\j,16)==7?1:0};
                                            \ifnum\pgfmathresult>0
                                                \draw[fill=palette8] (\i,\j) -- (\i+1,\j) -- (\i+1,\j+1) -- (\i,\j+1) -- (\i,\j);
                                            \else
                                                \pgfmathparse{Mod(\i+\j,16)==8?1:0};
                                                \ifnum\pgfmathresult>0
                                                    \draw[fill=palette9] (\i,\j) -- (\i+1,\j) -- (\i+1,\j+1) -- (\i,\j+1) -- (\i,\j);
                                                \else
                                                    \pgfmathparse{Mod(\i+\j,16)==9?1:0};
                                                    \ifnum\pgfmathresult>0
                                                        \draw[fill=palette13] (\i,\j) -- (\i+1,\j) -- (\i+1,\j+1) -- (\i,\j+1) -- (\i,\j);
                                                    \else
                                                        \pgfmathparse{Mod(\i+\j,16)==10?1:0};
                                                        \ifnum\pgfmathresult>0
                                                            \draw[fill=palette11] (\i,\j) -- (\i+1,\j) -- (\i+1,\j+1) -- (\i,\j+1) -- (\i,\j);
                                                        \else
                                                            \pgfmathparse{Mod(\i+\j,16)==11?1:0};
                                                            \ifnum\pgfmathresult>0
                                                                \draw[fill=palette14] (\i,\j) -- (\i+1,\j) -- (\i+1,\j+1) -- (\i,\j+1) -- (\i,\j);
                                                            \else
                                                                \pgfmathparse{Mod(\i+\j,16)==12?1:0};
                                                                \ifnum\pgfmathresult>0
                                                                    \draw[fill=palette12] (\i,\j) -- (\i+1,\j) -- (\i+1,\j+1) -- (\i,\j+1) -- (\i,\j);
                                                                \else
                                                                    \pgfmathparse{Mod(\i+\j,16)==13?1:0};
                                                                    \ifnum\pgfmathresult>0
                                                                        \draw[fill=palette15] (\i,\j) -- (\i+1,\j) -- (\i+1,\j+1) -- (\i,\j+1) -- (\i,\j);
                                                                    \else
                                                                        \pgfmathparse{Mod(\i+\j,16)==14?1:0};
                                                                        \ifnum\pgfmathresult>0
                                                                            \draw[fill=blue, palette10] (\i,\j) -- (\i+1,\j) -- (\i+1,\j+1) -- (\i,\j+1) -- (\i,\j);
                                                                        \else
                                                                            \draw[fill=red,palette14] (\i,\j) -- (\i+1,\j) -- (\i+1,\j+1) -- (\i,\j+1) -- (\i,\j);
                                                                        \fi
                                                                    \fi
                                                                \fi
                                                            \fi
                                                        \fi
                                                    \fi
                                                \fi
                                            \fi
                                        \fi
                                    \fi
                                \fi
                            \fi
                        \fi
                    \fi
                \fi
        }
    }
    \node at (4, -1) {\Huge$\mathrm{mod}\,16$};
\end{tikzpicture}
}
\caption{2D Illustration of modular volume dilution schemes implemented. One color corresponds with one of the disjoint domains on which the corresponding source has support.}
\label{fig:moddilute}
\end{figure}

\paragraph{Volume dilution of non-stochastic sources}
\HiRep supports non-stochastic dilution schemes. Here, the volume source is set to one everywhere
\begin{equation}
\eta(x) \equiv 1\,.
\end{equation}
The dilution parameter describes the modulus $m$ of the dilution, that is 
\begin{equation}
\eta^{(i)}(x) = \begin{cases}1 & \sum_{i}x_i\,\mathrm{mod}\,m = 0\\ 0 & \mathrm{otherwise}\end{cases}\,,
\end{equation}
see also illustration in \ref{fig:moddilute}. This can be maximally diluted to $T+X+Y+Z$ independent sources. The code for disconnected contributions supports different dilution schemes for stochastic sources.

\paragraph{Full dilution of stochastic sources}

Disconnected contributions are generally substantially harder to compute because of the large statistical noise in the observables. To reduce this noise, \HiRep supports stochastic $\mathds{Z}_2\otimes \mathds{Z}_2$ volume sources with different optional dilution schemes. In contrast to the connected correlations, the software supplies the user with the intermediate data, specifically the one-point operator measurements given by
\begin{equation}
O_{\Gamma}^{(i,j)}(t) = \sum_{\bm{x}}\mathrm{tr}\left[\eta^{(i,j)\dagger}(x)\Gamma \psi^{(i,j)}(x)\right]
\end{equation}
with $\psi$ and $\eta$ as in Eq. \ref{eq:alltoallprop} and t the 0-th component of $x$. While the user generally has to use this output to subtract the vacuum expectation value, divide by the spatial volume, and perform the stochastic average, combinatoric factors in assembling the correct Wick contractions are up to the user and need to be worked out from the theory.

Various dilution schemes are implemented for the stochastic sources to reduce the noise of the disconnected contributions. For disconnected contributions, volume sources yield better results than point sources. The volume sources can be diluted in time with spin dilution, spin and color dilution, and spin, color and even-odd dilution. It is also possible to dilute the source only in spin, color, and even-odd. This means that the diluted space is given by a disjoint cover of the space $\mathcal{D} = V_4 \otimes S \otimes C$, where $S$ spin space and $C$ color space. We can define sources in terms of random numbers and vectors and the characteristic function of the respective subdomains $D_i$, given by $\theta_{D_i}$.

Time and spin dilution of the stochastic volume sources is here defined as a dilution of the full domain $\mathcal{D} = V_4\otimes S$ into time and spin slices, and the sources are given by 
\begin{equation}
\eta^{(i,j)}_{a\alpha}(x) = \theta_{D_i}(x,\alpha)K^{(j)}_a(x) =\begin{cases}K^{(j)}_{a}(x) & (x,\alpha) \in D_i\\
0 & \mathrm{otherwise}\end{cases}\,,
\end{equation}
with a color vector $K^{(j)}_{a}(x)$ filled with $\mathds{Z}_2\otimes \mathds{Z}_2$ noise entries for each $x$ and $\alpha$, and stochastic source $j$. Adding color dilution to this means we decompose the full domain $V_4\otimes S \otimes C$ into time, spin, and color slices
\begin{align}
\begin{split}
\eta^{\alpha(i,j)}_{a\alpha}(x) &= \theta_{D_i}(x,\alpha, a)k(x,\alpha,a)
\end{split}
\end{align}
for a random number $k(x,\alpha,a)$ drawn from the $\mathds{Z}_2\otimes \mathds{Z}_2$ noise distribution for each $x,\alpha, a$ and $j$. 

The inversion can further feature an even-odd decomposition, where the even and odd parts of the source are inverted separately. 

\subsubsection{Normalization}

Global factors often depend on the theory and the concrete application and should be derived given the specific methods. However, some main points need to be considered. When computing the full correlation function of connected and disconnected contributions for the singlet in a two-flavor theory the interpolating operator is given by
\begin{equation}
O_{\Gamma}(x) = \overline{u}(x)\Gamma u(x) + \overline{d}(x)\Gamma d(x)\,,
\end{equation}
with $x = (t,\bm{x})$. 
When computing the correlation function with point sources, we can only average over the spatial sink points
\begin{equation}
C_{\Gamma}(t) = \dfrac{1}{V_3}\sum_{\bm{x}}\langle O_{\Gamma}(t,\bm{x})\overline{O}_{\Gamma}(0,\bm{0})\rangle
\end{equation}
while when using wall or volume sources, the correlation allows for volume averaging
\begin{align}
\begin{split}
C_{\Gamma}(t) = \dfrac{1}{V_3^2}\sum_{\bm{x},\bm{x}_0}\langle &O_{\Gamma}(x) \overline{O}_{\Gamma}(x_0)\rangle\\ = -\dfrac{1}{V_3^2}\sum_{\bm{x},\bm{x}_0}
\bigg[&N_f\mathrm{tr}[\Gamma S(x,x_0)\overline{\Gamma} S(x_0, x)] - N_f^2\mathrm{tr}[\Gamma S(x,x)]\mathrm{tr}[\overline{\Gamma} S(x_0,x_0)]\bigg]
\label{eq:conndisc}
\end{split}
\end{align}
with $\overline{\Gamma} = \gamma_0\Gamma^{\dagger}\gamma_0$ and any factors of $N_f$ are not included in the measurement output and thus need to be added manually. If the traces were obtained using stochastic sources, the stochastic average needs to be performed as follows
\begin{equation}
\mathrm{tr}[K] = \dfrac{1}{N_{\mathrm{stoch}}}\sum_{i=1}^{N_{\mathrm{stoch}}}\mathrm{tr}[K^{(i)}]\,,
\end{equation}
for a general matrix $K$. An additional factor of two appears if the source has support only on the even sites. This is the case for the computation of the connected piece of the correlation function using SEMWall sources. The stochastic averaging for the connected contributions is done internally, while for the disconnected contributions, the output is given for each stochastic source.

We define the connected part of the correlation function as
\begin{equation}
C_{\Gamma,\mathrm{conn}}(t) = -\dfrac{1}{V_3^2}\sum_{\bm{x},\bm{x}_0}N_f\mathrm{tr}[\Gamma S(x,x_0)\overline{\Gamma} S(x_0, x)]
\end{equation}
when evaluated with SEMWall sources.

For volume sources the output of the disconnected contributions produces
\begin{equation}
L^{(i)}_{\Gamma}(t_0) = \sum_{\bm{x}_0}\mathrm{tr}[\Gamma S^{(i)}(x_0,x_0)]\,, 
\end{equation}
with $x_0 = (t_0,\bm{x}_0)$ for each source $i$, we can reconstruct
\begin{equation}
C_{\Gamma,\mathrm{disc}}^{(t_0,i,j)}(t) = \dfrac{N_f^2}{V_3^2}\sum_{\bm{x},\bm{x_0}}\mathrm{tr}[\Gamma S^{(i)}(x,x)]\mathrm{tr}[\overline{\Gamma}S^{(j)}(x_0,x_0)]\,.
\end{equation}
A noise-reduced estimate can be achieved by subsequently averaging over $t_0$ and combinations of stochastic sources by taking
\begin{equation}
C^{\mathrm{av}}_{\Gamma,\mathrm{disc}}(t) = \lim_{N_{\mathrm{stoch}}\to \infty}\dfrac{N_f^2}{\left(\tfrac{N_{\mathrm{stoch}}}{2}\right)^2 TV_3^2}\sum_{t_0}
\sum_{i=1}^{N_{\mathrm{stoch}}/2}\sum_{j=N_{\mathrm{stoch}}/2+1}^{N_{\mathrm{stoch}}}
\langle L^{(i)}_{\Gamma}(t_0)L^{(j)}_{\overline{\Gamma}}(t+t_0)\rangle\,,
\end{equation}
where the combinations of estimates $i$ and $j$ can also be chosen differently, as long as $i\neq j$ for all combinations.

The full correlation can be assembled as
\begin{equation}
C_{\Gamma,\mathrm{full}}(t) = C_{\Gamma,\mathrm{conn}}(t) + C^{\mathrm{av}}_{\Gamma,\mathrm{disc}}(t)\,.
\end{equation}

\subsection{Scattering methods}

Several tools are implemented to investigate the scattering properties of pseudoscalar mesons in various channels using the L\"uscher method~\cite{Luscher:1986pf,Luscher:1990ux,Luscher:1991cf} in the context of two-flavor gauge theory in which the scattering channels can be worked out in terms of isospin symmetry. As such, the tools discussed here allow us to calculate the scattering amplitude of pseudoscalar mesons in isospin channel $I=0,1$ and $2$. 

\begin{figure*}[ht]
\begin{picture}(0.7\textwidth,135)
\put(0,77){\includegraphics[width=0.92\textwidth,clip]{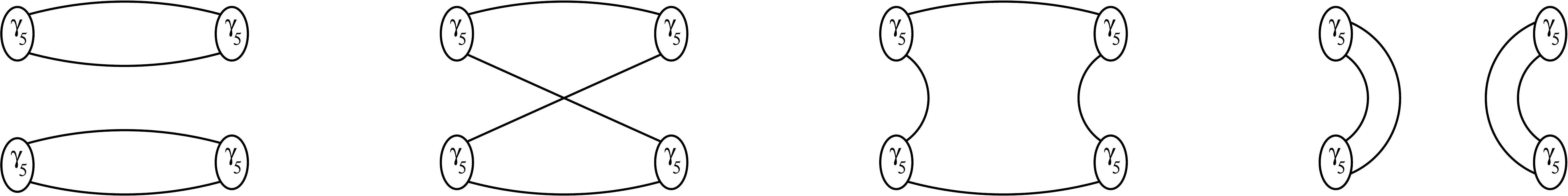} }
\put(0,10){\includegraphics[width=0.92\textwidth,clip]{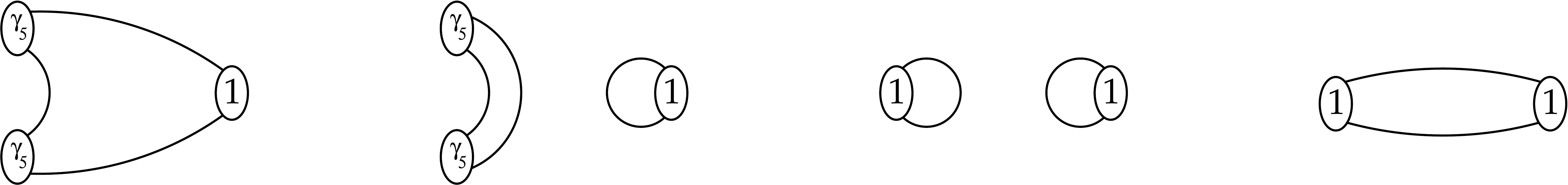} }
\put(17,65){$D(t)$}
\put(117, 65){$X(t)$}
\put(219, 65){$R(t)$}
\put(320, 65){$V(t)$}
\put(17,0){$T(t)$}
\put(117,0){$W(t)$}
\put(219,0){$\Sigma(t)$}
\put(320,0){$B(t)$}
\end{picture} 
\caption{Diagrammatic representation of the Wick contractions to distinguish the $\sigma$ and $\pi\pi$ states, which have the same quantum numbers, see \cite{Drach:2021uhl} for the original publication.}
\label{fig:contractions} 
\end{figure*}

In the $I=2$ channel, the relevant correlators are estimated following the approach presented in \cite{Arthur:2014zda}. In the channels $I=0$ and $1$, the tools aim at estimating a $2\times2$ correlation matrix spanned by one and two ultra-local fermion bilinear operators. The precise definition of the operators can be found in \cite{Drach:2020wux,Drach:2021uhl}. The contractions of the two-particle two-point function feature a number of topologies as defined in Fig.~\ref{fig:contractions}. This figure also explains our naming convention.
 
\HiRep allows to calculate the contributions $D(t)$, $X(t)$, $R(t)$ and a part of $V(t)$. Two methods are implemented to calculate $R(t)$, the default one that uses sequential inversions and an alternative one. \HiRep allows to calculate the contributions $T(t), B(t)$ and $\Sigma(t)$. From the produced output, $W(t)$ can be built offline. Although developed for the case $SU(2)$ with two fundamental flavors, the same topologies will appear in other theories.
\begin{equation}
C_{X \to Y} (\delta t)= \frac{1}{T} \sum_{t} \langle O_X(t+\delta t) O_Y(t)^\dagger\rangle. \label{eq:matrix}
\end{equation}

The quantum numbers of the $\sigma$ and $\pi\pi$ states are identical, so to distinguish them, we need to employ a variational technique using the small scalar operator basis consisting of $\mathds{1}$ and $\gamma_0$ and solve the generalized eigenvalue problem. This allows us to match the two lowest energy states to the corresponding diagrams that appear analytically by finding all Wick contractions from the expression in Eq.~\ref{eq:matrix}:
\begin{align}
\begin{split}
C_{\sigma \to  \sigma}(t) &= -B(t) + 2 \Sigma(t),  \\
C_{\pi\pi \to  \pi\pi}(t) &= 2 D(t) + 3 X(t) - 10 R(t) + 5 V(t), \\
C_{\pi\pi \to  \sigma}(t) &= \sqrt{10} \left( T(t)- W(t) \right). \label{eq:contractions}
\end{split}
\end{align}
Find the corresponding diagrammatic representation in Fig.~\ref{fig:contractions}. The $V$, $W$, and $\Sigma$ are known to dominate the statistical uncertainty. Compare \cite{Drach:2021uhl} for the above discussion.

\subsection{Stochastic estimation of the mode number using spectral projectors}

Following \cite{Giusti:2008vb}, the mode number $\nu(M_{\mathrm{thresh}})$, the number of eigenmodes with eigenvalues lower than a parameter $M_{\mathrm{thresh}}$ can be estimated stochastically using spectral projectors that can be estimated using a rational approximation. The approach has been implemented in \HiRep to estimate the mass anomalous dimension in a mass-deformed conformal field theory~\cite{Patella:2012da}.

 \subsection{Non-perturbative renormalisation}

 A tool allows to calculate non-perturbatively renormalization factors
 in the RI’-MOM scheme (regularisation invariant momentum scheme)~\cite{Martinelli:1994ty}.
 The code allows us to estimate the amputated vertex function:
 \begin{align}xs
\Pi_\Gamma(p)= S(p)^{-1} G_\Gamma(p) S(p)^{-1},
 \end{align}
 where $S^{-1}(p)$ is the inverse propagator in momentum space and
\begin{align}
G_\Gamma(p)= \langle \psi(p) O_{\Gamma}(p) \overline{\psi}(p) \rangle
\end{align}
for each fermion (non-singlet)  bilinear operator $O_{\Gamma}(p)$.
The RI'-MOM scheme is then defined by imposing the conditions that, in
the chiral limit and at a given scale $p^2=\mu^2 $, the inverse
propagator and amputated Green's function $\Pi_\Gamma(p)$ are set to
their tree-level values. The estimation of the correlators is
performed using momentum sources~\cite{Gockeler:1998ye} and the gauge
is fixed using the Landau gauge~\cite{Vink:1992ys}. In addition the
code support twisted boundary
conditions~\cite{Bedaque:2004kc,Sachrajda:2004mi} to interpolate
between the lattice momenta, as described above.

More details can be found in~\cite{Arthur:2016dir}.

\subsection{Glueball interpolators} 

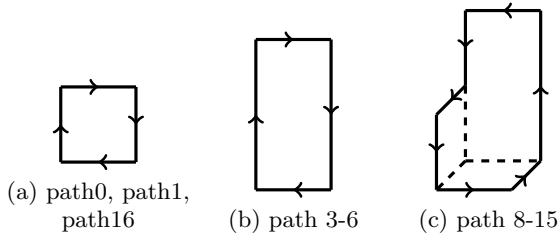
\begin{figure}
\centering
\begin{minipage}[b]{0.3\linewidth}
\centering
\begin{tikzpicture}
\begin{scope}[very thick,decoration={markings, mark=at position 0.5 with {\arrow{>}}}]
    \draw[postaction={decorate}] (0,0,0) -- (0,1,0);
    \draw[postaction={decorate}] (0,1,0) -- (1,1,0);
    \draw[postaction={decorate}] (1,1,0) -- (1,0,0);
    \draw[postaction={decorate}] (1,0,0) -- (0,0,0);
\end{scope}
\end{tikzpicture}
\subcaption{path0, path1, path16}
\end{minipage}%
\begin{minipage}[b]{0.3\linewidth}
\centering
\begin{tikzpicture}
\begin{scope}[very thick,decoration={markings, mark=at position 0.5 with {\arrow{>}}}]
    \draw[postaction={decorate}] (0,0,0) -- (0,2,0);
    \draw[postaction={decorate}] (0,2,0) -- (1,2,0);
    \draw[postaction={decorate}] (1,2,0) -- (1,0,0);
    \draw[postaction={decorate}] (1,0,0) -- (0,0,0); 
\end{scope}
\end{tikzpicture}
\subcaption{path 3-6}
\end{minipage}%
\begin{minipage}[b]{0.3\linewidth}
\centering
\begin{tikzpicture}
\begin{scope}[very thick,decoration={markings, mark=at position 0.5 with {\arrow{>}}}] 
    \draw[dashed] (-1,1,0) -- (-1,0,0);
    \draw[dashed] (-1,0,0) -- (0,0,0);
    \draw[dashed] (-1,0,0) -- (-1,0,1);
    \draw[postaction={decorate}] (0,0,0) -- (0,2,0);
    \draw[postaction={decorate}] (0,2,0) -- (-1,2,0);
    \draw[postaction={decorate}] (-1,2,0) -- (-1,1,0);
    \draw[postaction={decorate}] (-1,1,0) -- (-1,1,1);
    \draw[postaction={decorate}] (-1,1,1) -- (-1,0,1);
    \draw[postaction={decorate}] (-1,0,1) -- (0,0,1);
    \draw[postaction={decorate}] (0,0,1) -- (0,0,0);
\end{scope}
\end{tikzpicture}
\subcaption{path 8-15}
\end{minipage}
\caption{Example default operators to build a variational basis for glueball spectroscopy.}
\label{fig:hireploops}
\end{figure}

Glueball interpolators are constructed by forming appropriate combinations
of rotations of products of closed loops. The states are classified according to
all good finite-volume quantum numbers.
The relevant symmetry group for zero-momentum states in a finite 3D cubic spatial volume is the cover of the octahedral group, denoted \( O_h^D \). This group also includes parity transformations.
For states with a non-zero total momentum \(\mathbf{P}\) in the finite-volume frame, the symmetry group is reduced to the little group of \(\mathbf{P}\). This is the subgroup of \( O_h^D \) consisting of all elements that leave \(\mathbf{P}\) invariant.\\
\HiRep supports the generation of such states for an arbitrary
choice of quantum numbers (irrep charge and momentum). By default, a selected subset of these operators is included in the
code, see Fig.~\ref{fig:hireploops}.
Additionally, the suite supports the automatic addition of generic interpolators defined by user-inputted closed loops and choice of quantum numbers through an offline interface.

\subsection{Gauge Smearing}

\HiRep implements several gauge-smearing algorithms that are described below. Specifically, the use of smeared fields in spectral observables is useful to reduce ultraviolet fluctuations associated with excited state contaminations and improve the overlap with ground state physics. 

\subsubsection{APE smearing}

\begin{figure}[H]
\centering
\begin{minipage}[c]{0.3\linewidth}
\centering
\begin{tikzpicture}
\begin{scope}[very thick]
    \draw[postaction={decorate}, decoration={markings, mark=at position 0.6 with {\arrow{>}}}, opacity=1] (0,0,0) -- (1,0,0);
\end{scope}
\end{tikzpicture}%
\end{minipage}%
\begin{minipage}[c]{0.3\linewidth}
\centering
\begin{tikzpicture}
\begin{scope}[very thick]
    \draw[postaction={decorate}, decoration={markings, mark=at position 0.6 with {\arrow{>}}}, opacity=1] (0,0+0.2,0) -- (0,1+0.2,0);
    \draw[postaction={decorate}, decoration={markings, mark=at position 0.6 with {\arrow{>}}}, opacity=1] (0,1+0.2,0) -- (1,1+0.2,0);
    \draw[postaction={decorate}, decoration={markings, mark=at position 0.6 with {\arrow{>}}}, opacity=1] (1,1+0.2,0) -- (1,0+0.2,0);

    \draw[postaction={decorate}, decoration={markings, mark=at position 0.6 with {\arrow{>}}}, opacity=0.4] (0,0,0+0.2) -- (0,0,1+0.2);
    \draw[postaction={decorate}, decoration={markings, mark=at position 0.75 with {\arrow{>}}}, opacity=0.4] (0,0,1+0.2) -- (1,0,1+0.2);
    \draw[postaction={decorate}, decoration={markings, mark=at position 0.6 with {\arrow{>}}}, opacity=0.4] (1,0,1+0.2) -- (1,0,0+0.2);

    \draw[postaction={decorate}, decoration={markings, mark=at position 0.6 with {\arrow{>}}}, opacity=1] (0,0-0.2,0) -- (0,-1-0.2,0);
    \draw[postaction={decorate}, decoration={markings, mark=at position 0.6 with {\arrow{>}}}, opacity=1] (0,-1-0.2,0) -- (1,-1-0.2,0);
    \draw[postaction={decorate}, decoration={markings, mark=at position 0.6 with {\arrow{>}}}, opacity=1] (1,-1-0.2,0) -- (1,0-0.2,0);

    \draw[postaction={decorate}, decoration={markings, mark=at position 0.6 with {\arrow{>}}}, opacity=0.4] (0,0,0-0.2) -- (0,0,-1-0.2);
    \draw[postaction={decorate}, decoration={markings, mark=at position 0.45 with {\arrow{>}}}, opacity=0.4] (0,0,-1-0.2) -- (1,0,-1-0.2);
    \draw[postaction={decorate}, decoration={markings, mark=at position 0.6 with {\arrow{>}}}, opacity=0.4] (1,0,-1-0.2) -- (1,0,0-0.2);
\end{scope}
\end{tikzpicture}%
\end{minipage}
\caption{3D illustration of APE smearing. The terms contributing are the link and the adjacent staples. }
\label{fig:apesm}
\end{figure}
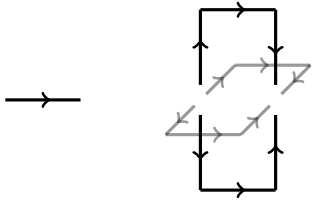

The simplest gauge smearing implementation in \HiRep is APE smearing \cite{APE:1987ehd}, as illustrated in Fig.~\ref{fig:apesm}. Here, the fat links are produced by averaging over the adjacent staples. These are defined by the expression
\begin{align}
\begin{split}
K_{\mu}(x) = \sum_{\nu\neq\mu}\rho_{\mu\nu}\big(&U_{\nu}(x)U_{\mu}(x+\hat{\nu})U_{\nu}^{\dagger}(x+\hat{\mu}) \\&+ U_{\nu}^{\dagger}(x-\hat{\nu})U_{\mu}(x-\hat{\nu})U_{\nu}(x-\hat{\nu}+\hat{\mu})\big)\,\textcolor{purple}{\cancel{.}},
\label{eq:adjacentstaples}
\end{split}
\end{align}
where $\rho_{\mu\nu}$ is, in principle, a freely tunable parameter. However, the implementation in \HiRep does not allow a generic setting of $\rho_{\mu\nu}$ instead it permits an anisotropic selection of the two parameters $\rho_t$ and $\rho_s$, defining $\rho_{\mu\nu}$ as follows
\begin{equation}
\rho_{\mu\nu} = \begin{cases}\rho_t & \rho = 0~\text{or}~\nu = 0\\ \rho_s & \mathrm{otherwise}\end{cases}
\end{equation}

Subsequently, the gauge links are smeared by applying
\begin{equation}
U_{\mu}^{(n+1)}(x) = \mathrm{Proj}_{SU(N_g)}[(1-\alpha)U_{\mu}^{(x)} + \dfrac{\alpha}{6}K_{\mu}(x)]
\end{equation}
with the smearing parameter $\alpha$. 

\subsubsection{Stout smearing}

\begin{figure}[H]
\centering
\begin{tikzpicture}
\begin{scope}[very thick]
    \draw[postaction={decorate}, decoration={markings, mark=at position 0.6 with {\arrow{>}}}, opacity=1] (0,0+0.2,0) -- (0,1+0.2,0);
    \draw[postaction={decorate}, decoration={markings, mark=at position 0.6 with {\arrow{>}}}, opacity=1] (0,1+0.2,0) -- (1,1+0.2,0);
    \draw[postaction={decorate}, decoration={markings, mark=at position 0.6 with {\arrow{>}}}, opacity=1] (1,1+0.2,0) -- (1,0+0.2,0);

    \draw[postaction={decorate}, decoration={markings, mark=at position 0.6 with {\arrow{>}}}, opacity=0.4] (0,0,0+0.2) -- (0,0,1+0.2);
    \draw[postaction={decorate}, decoration={markings, mark=at position 0.75 with {\arrow{>}}}, opacity=0.4] (0,0,1+0.2) -- (1,0,1+0.2);
    \draw[postaction={decorate}, decoration={markings, mark=at position 0.6 with {\arrow{>}}}, opacity=0.4] (1,0,1+0.2) -- (1,0,0+0.2);

    \draw[postaction={decorate}, decoration={markings, mark=at position 0.6 with {\arrow{>}}}, opacity=1] (0,0-0.2,0) -- (0,-1-0.2,0);
    \draw[postaction={decorate}, decoration={markings, mark=at position 0.6 with {\arrow{>}}}, opacity=1] (0,-1-0.2,0) -- (1,-1-0.2,0);
    \draw[postaction={decorate}, decoration={markings, mark=at position 0.6 with {\arrow{>}}}, opacity=1] (1,-1-0.2,0) -- (1,0-0.2,0);

    \draw[postaction={decorate}, decoration={markings, mark=at position 0.6 with {\arrow{>}}}, opacity=0.4] (0,0,0-0.2) -- (0,0,-1-0.2);
    \draw[postaction={decorate}, decoration={markings, mark=at position 0.45 with {\arrow{>}}}, opacity=0.4] (0,0,-1-0.2) -- (1,0,-1-0.2);
    \draw[postaction={decorate}, decoration={markings, mark=at position 0.6 with {\arrow{>}}}, opacity=0.4] (1,0,-1-0.2) -- (1,0,0-0.2);
\end{scope}
\end{tikzpicture}%
\begin{tikzpicture}
\begin{scope}[very thick]
    \draw[postaction={decorate}, decoration={markings, mark=at position 0.9 with {\arrow{>}}}, opacity=1] (1,0+0.2,0) -- (0,0.5+0.2,0) -- (0,1+0.2,0);
    \draw[postaction={decorate}, decoration={markings, mark=at position 0.6 with {\arrow{>}}}, opacity=1] (0,1+0.2,0) -- (1,1+0.2,0);
    \draw[postaction={decorate}, decoration={markings, mark=at position 0.2 with {\arrow{>}}}, opacity=1] (1,1+0.2,0) -- (1,0.5+0.2,0) -- (0,0+0.2,0);

    \draw[postaction={decorate}, decoration={markings, mark=at position 0.9 with {\arrow{>}}}, opacity=0.4] (1,0,0+0.2) -- (0,0,0.5+0.2) -- (0,0,1+0.2);
    \draw[postaction={decorate}, decoration={markings, mark=at position 0.40 with {\arrow{>}}}, opacity=0.4] (0,0,1+0.2) -- (1,0,1+0.2);
    \draw[postaction={decorate}, decoration={markings, mark=at position 0.2 with {\arrow{>}}}, opacity=0.4] (1,0,1+0.2) -- (1,0,0.5+0.2) -- (0,0,0+0.2);

    \draw[postaction={decorate}, decoration={markings, mark=at position 0.9 with {\arrow{>}}}, opacity=1] (1,0-0.2,0) -- (0,-0.5-0.2, 0) -- (0,-1-0.2,0);
    \draw[postaction={decorate}, decoration={markings, mark=at position 0.6 with {\arrow{>}}}, opacity=1] (0,-1-0.2,0) -- (1,-1-0.2,0);
    \draw[postaction={decorate}, decoration={markings, mark=at position 0.2 with {\arrow{>}}}, opacity=1] (1,-1-0.2,0) -- (1,-0.5-0.2,0) -- (0,0-0.2,0);

    \draw[postaction={decorate}, decoration={markings, mark=at position 0.9 with {\arrow{>}}}, opacity=0.4] (1,0,0-0.2) -- (0,0,-0.5-0.2) -- (0,0,-1-0.2);
    \draw[postaction={decorate}, decoration={markings, mark=at position 0.60 with {\arrow{>}}}, opacity=0.4] (0,0,-1-0.2) -- (1,0,-1-0.2);
    \draw[postaction={decorate}, decoration={markings, mark=at position 0.15 with {\arrow{>}}}, opacity=0.4] (1,0,-1-0.2) -- (1,0,-0.5-0.2) -- (0,0,0-0.2);
\end{scope}
\end{tikzpicture}%
\begin{tikzpicture}
\begin{scope}[very thick]
    \draw[postaction={decorate}, decoration={markings, mark=at position 0.6 with {\arrow{>}}}, opacity=1] (0,0+0.3,0) -- (0,1+0.3,0);
    \draw[postaction={decorate}, decoration={markings, mark=at position 0.6 with {\arrow{>}}}, opacity=1] (0,1+0.3,0) -- (1,1+0.3,0);
    \draw[postaction={decorate}, decoration={markings, mark=at position 0.6 with {\arrow{>}}}, opacity=1] (1,1+0.3,0) -- (1,0+0.3,0);
    \draw[postaction={decorate}, decoration={markings, mark=at position 0.7 with {\arrow{>}}}, opacity=1] (1,0+0.3,0) -- (0,0+0.3,0);
    \draw[postaction={decorate}, decoration={markings, mark=at position 0.7 with {\arrow{>}}}, opacity=1]  (0,0+0.2,0) -- (1,0+0.2,0);

    \draw[postaction={decorate}, decoration={markings, mark=at position 0.6 with {\arrow{>}}}, opacity=0.4] (0,0,0+0.3) -- (0,0,1+0.3);
    \draw[postaction={decorate}, decoration={markings, mark=at position 0.75 with {\arrow{>}}}, opacity=0.4] (0,0,1+0.3) -- (1,0,1+0.3);
    \draw[postaction={decorate}, decoration={markings, mark=at position 0.6 with {\arrow{>}}}, opacity=0.4] (1,0,1+0.3) -- (1,0,0+0.3);
    \draw[postaction={decorate}, decoration={markings, mark=at position 0.7 with {\arrow{>}}}, opacity=0.4] (1,0,0+0.3) -- (0,0,0+0.3);
    \draw[postaction={decorate}, decoration={markings, mark=at position 0.7 with {\arrow{>}}}, opacity=0.4]  (0,0,0+0.2) -- (1,0,0+0.2);

    \draw[postaction={decorate}, decoration={markings, mark=at position 0.6 with {\arrow{>}}}, opacity=1] (0,0-0.3,0) -- (0,-1-0.3,0);
    \draw[postaction={decorate}, decoration={markings, mark=at position 0.6 with {\arrow{>}}}, opacity=1] (0,-1-0.3,0) -- (1,-1-0.3,0);
    \draw[postaction={decorate}, decoration={markings, mark=at position 0.6 with {\arrow{>}}}, opacity=1] (1,-1-0.3,0) -- (1,0-0.3,0);
    \draw[postaction={decorate}, decoration={markings, mark=at position 0.7 with {\arrow{>}}}, opacity=1] (1,0-0.3,0) -- (0,0-0.3,0);
    \draw[postaction={decorate}, decoration={markings, mark=at position 0.7 with {\arrow{>}}}, opacity=1]  (0,0-0.2,0) -- (1,0-0.2,0);

    \draw[postaction={decorate}, decoration={markings, mark=at position 0.6 with {\arrow{>}}}, opacity=0.4] (0,0,0-0.3) -- (0,0,-1-0.3);
    \draw[postaction={decorate}, decoration={markings, mark=at position 0.45 with {\arrow{>}}}, opacity=0.4] (0,0,-1-0.3) -- (1,0,-1-0.3);
    \draw[postaction={decorate}, decoration={markings, mark=at position 0.6 with {\arrow{>}}}, opacity=0.4] (1,0,-1-0.3) -- (1,0,0-0.3);
    \draw[postaction={decorate}, decoration={markings, mark=at position 0.7 with {\arrow{>}}}, opacity=0.4] (1,0,0-0.3) -- (0,0,0-0.3);
    \draw[postaction={decorate}, decoration={markings, mark=at position 0.7 with {\arrow{>}}}, opacity=0.4]  (0,0,0-0.2) -- (1,0,0-0.2);
\end{scope}
\end{tikzpicture}%
\begin{tikzpicture}
\begin{scope}[very thick]
    \draw[postaction={decorate}, decoration={markings, mark=at position 0.6 with {\arrow{>}}}, opacity=1] (0,0+0.3,0) -- (0,1+0.3,0);
    \draw[postaction={decorate}, decoration={markings, mark=at position 0.6 with {\arrow{>}}}, opacity=1] (0,1+0.3,0) -- (1,1+0.3,0);
    \draw[postaction={decorate}, decoration={markings, mark=at position 0.6 with {\arrow{>}}}, opacity=1] (1,1+0.3,0) -- (1,0+0.3,0);
    \draw[postaction={decorate}, decoration={markings, mark=at position 0.4 with {\arrow{>}}}, opacity=1] (1,0+0.3,0) -- (0,0+0.3,0);
    \draw[postaction={decorate}, decoration={markings, mark=at position 0.8 with {\arrow{>}}}, opacity=1] (1,0+0.2,0) -- (0,0+0.2,0);

    \draw[postaction={decorate}, decoration={markings, mark=at position 0.6 with {\arrow{>}}}, opacity=0.4] (0,0,0+0.3) -- (0,0,1+0.3);
    \draw[postaction={decorate}, decoration={markings, mark=at position 0.75 with {\arrow{>}}}, opacity=0.4] (0,0,1+0.3) -- (1,0,1+0.3);
    \draw[postaction={decorate}, decoration={markings, mark=at position 0.6 with {\arrow{>}}}, opacity=0.4] (1,0,1+0.3) -- (1,0,0+0.3);
    \draw[postaction={decorate}, decoration={markings, mark=at position 0.4 with {\arrow{>}}}, opacity=0.4] (1,0,0+0.3) -- (0,0,0+0.3);
    \draw[postaction={decorate}, decoration={markings, mark=at position 0.8 with {\arrow{>}}}, opacity=0.4] (1,0,0+0.2) -- (0,0,0+0.2);

    \draw[postaction={decorate}, decoration={markings, mark=at position 0.6 with {\arrow{>}}}, opacity=1] (0,0-0.3,0) -- (0,-1-0.3,0);
    \draw[postaction={decorate}, decoration={markings, mark=at position 0.6 with {\arrow{>}}}, opacity=1] (0,-1-0.3,0) -- (1,-1-0.3,0);
    \draw[postaction={decorate}, decoration={markings, mark=at position 0.6 with {\arrow{>}}}, opacity=1] (1,-1-0.3,0) -- (1,0-0.3,0);
    \draw[postaction={decorate}, decoration={markings, mark=at position 0.4 with {\arrow{>}}}, opacity=1] (1,0-0.3,0) -- (0,0-0.3,0);
    \draw[postaction={decorate}, decoration={markings, mark=at position 0.8 with {\arrow{>}}}, opacity=1] (1,0-0.2,0) -- (0,0-0.2,0);

    \draw[postaction={decorate}, decoration={markings, mark=at position 0.6 with {\arrow{>}}}, opacity=0.4] (0,0,0-0.3) -- (0,0,-1-0.3);
    \draw[postaction={decorate}, decoration={markings, mark=at position 0.45 with {\arrow{>}}}, opacity=0.4] (0,0,-1-0.3) -- (1,0,-1-0.3);
    \draw[postaction={decorate}, decoration={markings, mark=at position 0.6 with {\arrow{>}}}, opacity=0.4] (1,0,-1-0.3) -- (1,0,0-0.3);
    \draw[postaction={decorate}, decoration={markings, mark=at position 0.4 with {\arrow{>}}}, opacity=0.4] (1,0,0-0.3) -- (0,0,0-0.3);
    \draw[postaction={decorate}, decoration={markings, mark=at position 0.8 with {\arrow{>}}}, opacity=0.4] (1,0,0-0.2) -- (0,0,0-0.2);
\end{scope}
\end{tikzpicture}%
\caption{3D Illustration of the different shapes contributing to the fat stout-smeared links. The closed loops are traced, and scalar multiply the adjacent links.}
\label{fig:stoutsm}
\end{figure}
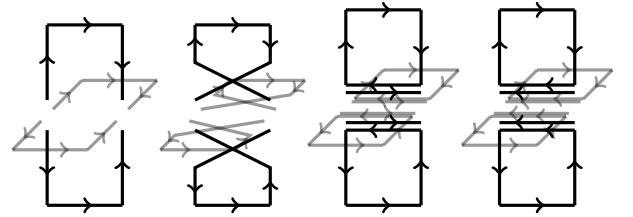

Stout smearing is, as first described in \cite{Morningstar:2003gk} an iterative smearing procedure of the gauge links $U_{\mu}(x)$ by an $SU(N)$ matrix $e^{\mathrm{i}Q_{\mu}(x)}$
\begin{equation}
U_{\mu}^{(n+1)}(x) = e^{\mathrm{i}Q_{\mu}^{(n)}}U_{\mu}^{(n)}(x)
\end{equation}
where the matrix $Q_{\mu}(x)$ is given by a combination of links and sums over adjacent staples
\begin{align}
\begin{split}
Q_{\mu}(x) = \dfrac{\mathrm{i}}{2}\left(\Omega_{\mu}^{\dagger}(x) - \Omega_{\mu}(x)\right) - \dfrac{\mathrm{i}}{2N_g}\mathrm{tr}\,\left(\Omega_{\mu}^{\dagger}(x) - \Omega_{\mu}(x)\right)
\end{split}
\end{align}
with
\begin{equation}
\Omega_{\mu}(x) = K_{\mu}(x)U^{\dagger}_{\mu}(x)\,,
\end{equation}
with $K_{\mu}(x)$ as defined in Eq.~\ref{eq:adjacentstaples}. An illustration of the terms contributing to this type of smearing can be found in Fig.~\ref{fig:stoutsm}. Since the smearing of the links is applied by multiplication with another $SU(N)$ matrix, this smearing is, by definition, gauge covariant. 

\subsubsection{HYP smearing}

Hypercubic smearing \cite{Hasenfratz:2001hp} features three APE-type steps per iteration, where the links belonging to the staples of the target link are first APE smeared themselves. In each of these three steps, a projection to the gauge group $SU(N_g)$ is applied. Using the conventions from \cite{Hasenfratz:2001hp}, this is
\begin{align}
V_{\mu|\rho\nu}(x) =& \mathrm{Proj}_{SU(N_g)}\big[(1-\alpha_3)U_{\mu}(x)\nonumber\\ &+ \dfrac{\alpha_3}{2}\sum_{\pm\eta\neq\rho,\nu,\mu}U_{\eta}(x)U_{\mu}(x+\hat{\eta})U^{\dagger}_{\eta}(x+\hat{\mu})\big]
\end{align}
This step is now repeated twice:
\begin{align}
W_{\mu|\nu}(x) =& \mathrm{Proj}_{SU(N_g)}\big[(1-\alpha_2)U_{\mu}(x)\nonumber\\ &+ \dfrac{\alpha_2}{4}\sum_{\pm\rho\neq\nu,\mu}V_{\rho|\nu\mu}(x)V_{\mu|\rho\nu}(x+\hat{\rho})V^{\dagger}_{\rho|\nu\mu}(x+\hat{\mu})\big]
\end{align}

\begin{align}
X_{\mu}(x) =& \mathrm{Proj}_{SU(N_g)}\big[(1-\alpha_1)U_{\mu}(x)\nonumber\\ &+ \dfrac{\alpha_1}{6}\sum_{\pm\nu\neq\mu}W_{\nu|\mu}(x)W_{\mu|\nu}(x+\hat{\nu})W^{\dagger}_{\nu|\mu}(x+\hat{\mu})\big]
\end{align}

\subsubsection{SU(N) Projection}

The necessary projection for APE and HYP smearing can be chosen arbitrarily. To give some flexibility \HiRep supports multiple projection methods. 

The standard projection method takes the row vectors of the matrix and applies a Gram-Schmidt orthonormalization. This projects the matrix to $U(N)$. The determinant is then fixed by calculating the determinant of the matrix $U$ and dividing each entry by $\mathrm{det}(U)^{-1/N_g}$.
Alternatively, a covariant projection is implemented as described in \cite{Liang:1992cz}.
Another method typically used for APE smearing is the MaxReTr-method \cite{CABIBBO1982387}, which minimizes the action by maximizing a matrix $X\in SU(N)$

\begin{equation}
U^{\mathrm{proj}}_{\mu} = \underset{X\in SU(N)}{\mathrm{arg\,max}}\,\mathrm{Re}\,\mathrm{tr}\,[XU^{\mathrm{smear},\dagger}_{\mu}]\,,
\end{equation}

where $U^{\mathrm{smear}}_{\mu}$ is the smeared link before projection and $U^{\mathrm{proj}}_{\mu}$ is the link after projection to $SU(N)$.

\subsubsection{Wilson Flow}\label{sec:wilsonflow}

\paragraph{Implementation}
Measurements on configurations using the Wilson flow can also be performed. The Wilson flow, as proposed in \cite{Luscher:2010iy}, to obtain observables from the smeared, renormalized version of a given gauge configuration. For this, the original gauge field $A_{\mu}$ is the initial zero-flow time $t=0$ version of a smeared field $B_{\mu}$
\begin{equation}
B_{\mu}\big|_{t=0} = A_{\mu}
\end{equation}
which is smoothed by a numerical integration of the differential equation
\begin{equation}
\dot{B}_{\mu} = D_{\nu}G_{\nu\mu}
\end{equation}
where the dot denotes differentiation with respect to the flow time $t$ and
\begin{equation}
D_{\mu} = \partial_{\mu} + [B_{\mu}, \cdot]\,,
\end{equation}
\begin{equation}
G_{\mu\nu} = \partial_{\mu}B_{\nu} - \partial_{\nu}B_{\mu} + [B_{\mu},B_{\nu}]\,.
\end{equation}

This numerical integration can be performed using a simple Euler integration or a 3rd-order Runge-Kutta integration scheme. We also support an adaptive version of the 3rd-order Runge-Kutta, which increases the evaluation speed of the numerical integration by allowing for variable step size. This does not affect the measurements, which can be evaluated at a fixed step size specified in the input file.

\paragraph{Observables}

Specific to this code is the evaluation of the Wilson flow for higher representations, relying on a correctly normalized definition of the corresponding observables. We denote in the following the unsmeared fundamental gauge field $U(\mu, x)$ and its flowed version as $V(\mu, x, t)$.

We define the plaquette in the $\mu\nu$-plane at site $x$ as in Eq.~\ref{eq:plaquette} and evaluate the average plaquette on a single configuration as
\begin{equation}
\overline{P_{\mu\nu}}(U) = \dfrac{\sum_{x,\mu < \nu}\mathrm{Re\,tr\,}P_{\mu\nu}(U, x)}{6N_g T^{\mathrm{eff}}XYZ}
\end{equation}
where $T, X, Y$ and $Z$ are the dimensions of the global lattice. For periodic, anti-periodic and twisted boundary conditions $T^{\mathrm{eff}}=T$, while for open boundary conditions we have $T^{\mathrm{eff}} = T - 1$ and $T^{\mathrm{eff}} = T - 2$ for Schr{\"o}dinger functional boundary conditions.

In each measurement step, we evaluate the energy density
\begin{align}
E &= \dfrac{2\sum_{x,\mu<\nu}\mathrm{Re}\,\mathrm{tr}[1-P_{\mu\nu}(V, x)]}{T^{\mathrm{eff}}XYZ} \\ &= 2N_g(1 - 6\overline{P_{\mu\nu}}(V))
\end{align}
and the symmetric version of the energy density, which is defined by a discretization of the field strength tensor as given in Eq.~\ref{eq:field-strength-tensor}. Given the trace normalization $T_{R}$, which is defined for $SU(N_g)$ by
\begin{equation}
T_{FUND} = \dfrac{1}{2}\,,\qquad T_{ADJ} = N_g\,,
\end{equation}
\begin{equation}
T_{SYM} =\dfrac{N+2}{2}\,,\qquad T_{AS} = \dfrac{N_g-2}{2}\,,
\end{equation}
the algebra vector obtained by the projection of $C_{\mu\nu}(U,x)$ is given by
\begin{equation}
C_{\mu\nu}^{a}(U,x) = -\dfrac{\mathrm{i}}{T_{R}}\mathrm{Re}\,\mathrm{tr}[T^{a}_{R}C_{\mu\nu}(U,x)]
\end{equation}
which defined a symmetrized energy density
\begin{equation}
E_{\mathrm{sym}} = \dfrac{\sum_{x}C_{\mu\nu}^{a}(V,x)C_{\mu\nu}^{a}(V,x)}{4T^{\mathrm{eff}}XYZ}\,.
\end{equation}
The software displays the corresponding quantities $t^2E$ and $t^2E_{\mathrm{sym}}$ for simplicity. Finally, the estimate of the topological charge is
\begin{equation}
Q = \sum_{x}\dfrac{T_{FUND}}{4\pi^2}\varepsilon_{\mu\nu\rho\sigma}\mathrm{tr}[C_{\mu\nu}(V,x)C_{\rho\sigma}(V,x)]\,.
\end{equation}
As a result, independent of the choice of the fermion representation, the topological charge evaluated  in this way on periodic boundary conditions will approximate integers, even for higher representations \cite{Luscher:2010iy, Fodor:2009nh, DellaMorte:2025tks}, and will typically be rounded to the nearest integer during analysis.
Note that the above formulae are all working with a smeared fundamental field and the only used trace normalization is the fundamental one. 
\section{Algorithmic tests}\label{sec:tests}

\begin{figure*}
\begin{minipage}[t]{0.45\linewidth}
\centering
\includegraphics[width=2.5in]{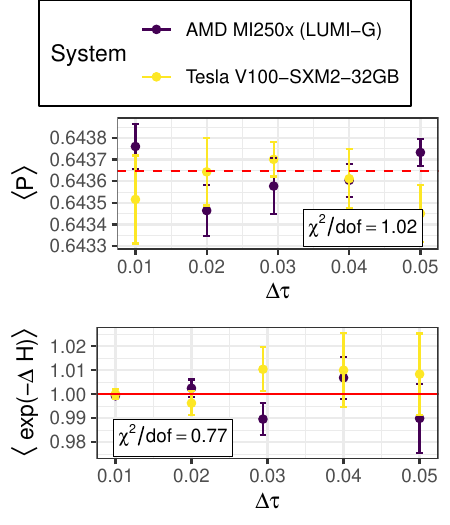}
\subcaption{(Top) Testing the independence of the average plaquette depending on the step size in the 2nd-order Omelyan integrator (Bottom) Test that the Creutz equality is fulfilled. Some of this data was already published in \cite{Martins:2024dew}.}
\end{minipage}%
\hfill
\begin{minipage}[t]{0.45\linewidth}
\centering
\includegraphics[width=2.5in]{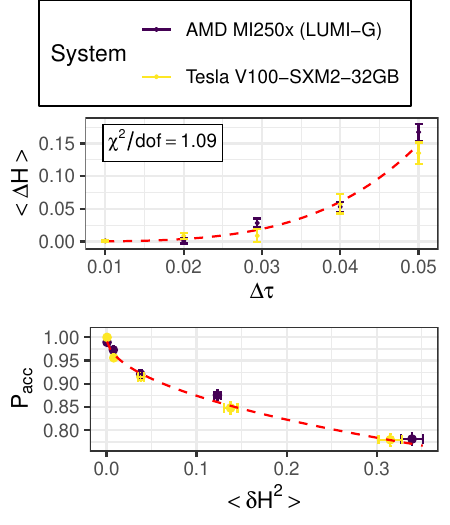}
\subcaption{(Top) Testing the scaling of the integrator by examining the dependence of the size of the Hamiltonian violations depending on the integrator step size (Bottom) Testing the behavior of the acceptance rate depending on the Hamiltonian violations against the asymptotically analytically known behavior. Some of this data was already published in \cite{Martins:2024dew}.}
\end{minipage}
\caption{Algorithmic tests of analytically known quantities for an $SU(2)$ gauge theory using fermions in the adjoint representation simulated using different GPUs.}
\label{fig:algo}
\end{figure*}

\HiRep contains a library of unit tests continuously integrated with the code so that unit tests run on each commit. The algorithmic tests performed in \cite{DelDebbio:2008zf} are reproduced in this section, see Fig.~\ref{fig:algo}. For this, we test several properties of the generation algorithm using ensembles of the SU($N_g$) theory with two fermions in the adjoint representation generated using the GPU version of the code. We tested the Creutz equality \cite{Creutz:1988wv}, stating that
\begin{equation}
\langle\mathrm{exp}(-\Delta H)\rangle = 1\,,
\end{equation}
a property of the (R)HMC algorithm. Further, since the algorithm is exact, the expectation value of the average plaquette has to be independent of the step size in the integrator used in the molecular dynamics evolution.
We should also verify that the integration errors, in this case, the 2nd-order Omelyan integrator, scale in agreement with the analytic prediction \cite{OMELYAN2003272, Takaishi:1999bi}:
\begin{equation}
\langle\Delta H\rangle = \delta \tau ^4\,.
\end{equation}
Further, we see the agreement of the acceptance rates depending on the expectations of Hamiltonian violations matching the asymptotic analytical relation
\begin{equation}
P_{\mathrm{acc}} = \mathrm{erfc}\left(\sqrt{\dfrac{\langle\delta H^2\rangle}{8}}\right)\,,
\end{equation}
see \cite{Gupta:1990ka}, that applies to high acceptance rates.

\begin{figure}
\centering
\includegraphics[width=2.3in]{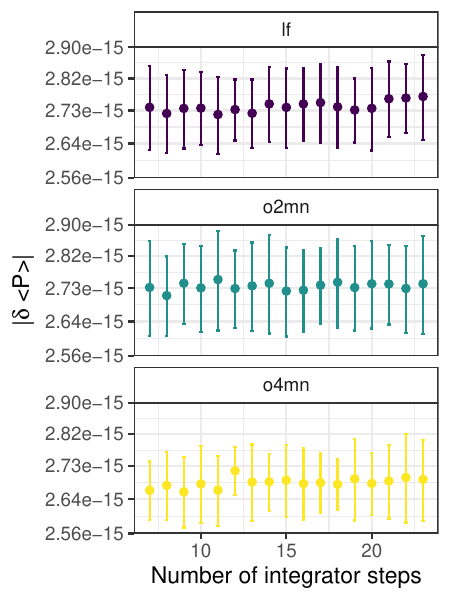}
\caption{Reversibility of the algorithm. Here we show that integrating over a trajectory and then flipping the momenta to reverse the trajectory yields the same initial plaquette value up to the expected error in double precision.
}
\label{fig:reversibility}
\end{figure}

\begin{figure}
\centering
\includegraphics[width=2.6in]{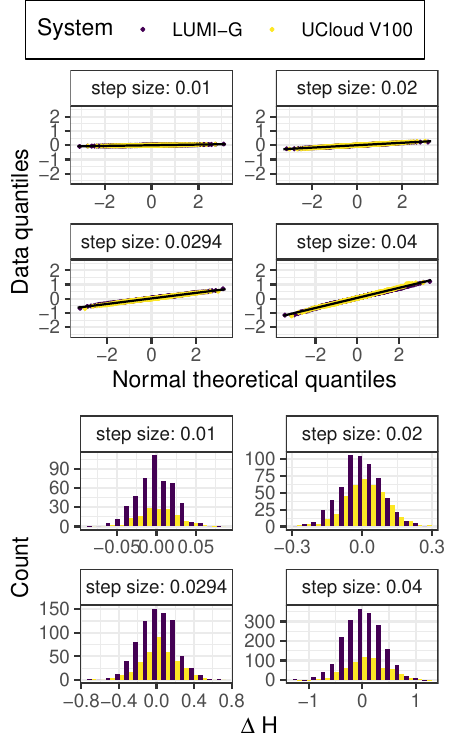}
\caption{Quantile-quantile plots and histograms to show the Gaussianity of the distribution of the Hamiltonian violations. Some of this data was already published in \cite{Martins:2024dew}.}
\label{fig:QQ}
\end{figure}

The tested ensembles do not display any non-gaussianities in the regions explored. The below histograms and quantile-quantile plots demonstrate that the recorded values are normally distributed, see Fig.~\ref{fig:QQ}.
The algorithm is also reversible up to inverter and computer precision. \cite{DelDebbio:2008zf} tests this by checking that integrating a trajectory using the 2nd-order Omelyan integrator, then flipping the momenta and reversing the trajectory yields absolute Hamiltonian violations around $10^{-7}$, given by the relative inverter precision. We can examine the same property by checking the absolute violation of reversibility in the average plaquette $|\delta \langle P \rangle|$ and the independence of the integrator types and number of steps. We find, as shown in Fig.~\ref{fig:reversibility} that the algorithm is reversible using the GPU-accelerated version of the code for all integrators and shows no dependence on the number of inversion steps. This test was performed at a relative squared inverter precision of $10^{-28}$.
\section{Benchmarks}\label{sec:benchmarks}
In this section, we present results for representative benchmarks of the \HiRep v2 simulation code. In particular we test the performance of the linear algebra routines and the Wilson-Dirac operator and their scaling properties to a large number of processing units. Benchmark tests for GPUs are performed on both AMD and NVIDIA GPUs.

\subsection{Linear Algebra}

\begin{figure}
\centering
\includegraphics[width=4in]{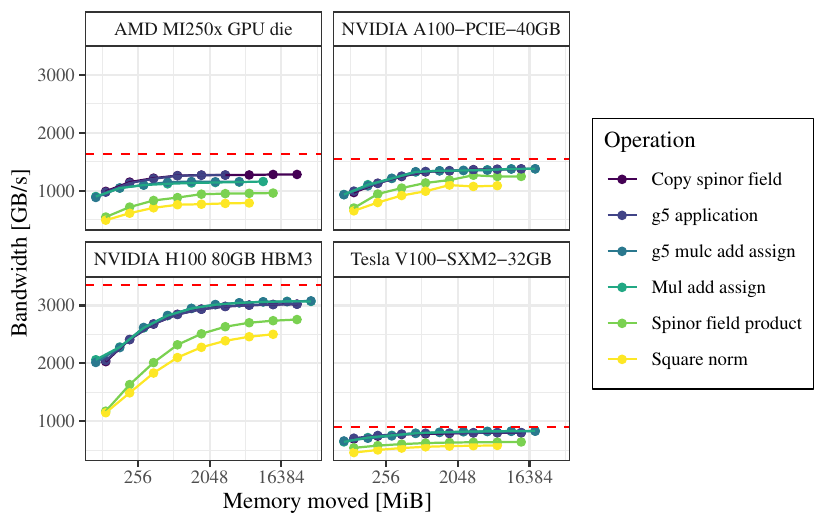}
\caption{Bandwidth of $SU(3)$ linear algebra operations on different GPU types, tested using each a single GPU. The dashed red line shows the peak performance on the respective card. Apart from the reductions, the performance relative to the peak bandwidth will be identical because these linear algebra operations are trivially parallelizable.}
\label{fig:linabenchmarks}
\end{figure}

We study the performance of linear algebra operations by looking at the effective bandwidth achieved by the linear algebra kernels. Figure~\ref{fig:linabenchmarks} shows the achieved bandwidth of the linear algebra operations as a function of the total amount of memory moved. For sufficiently large local lattices, we achieve up to 90\% of the peak theoretical bandwidth listed by the manufacturer.

\subsection{Single-GPU Efficiency}
\label{sec:singlegputests}

\begin{figure}
\centering
\includegraphics[width=3in]{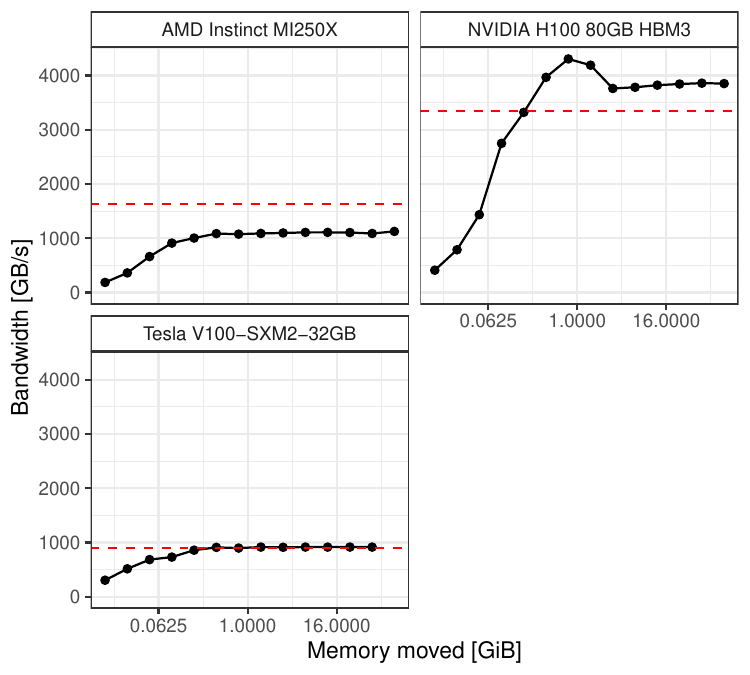}
\caption{Single GPU bandwidths reached depending on the memory moved}
\label{fig:singlegpudirac}
\end{figure}

The performance the kernel reaches is given in terms of the effective bandwidth, which can reach values higher than the maximum peak bandwidth in kernels with data reusage due to caching. However, using the \texttt{Nsight Compute} profiler, we can determine the actual bandwidth reached in global memory access. For the typical settings of $N_c=2$ and $N_c=3$ and the best-performing kernel, we always reach bandwidths of $80-90\%$ of the peak performance on the NVIDIA V100, A100, and H100, which shows that any performance improvement could at most achieve another $10-20\%$ speedup, which in most practical scenarios is only a minimal improvement.
 
\begin{figure}
\centering
\includegraphics[width=3.5in]{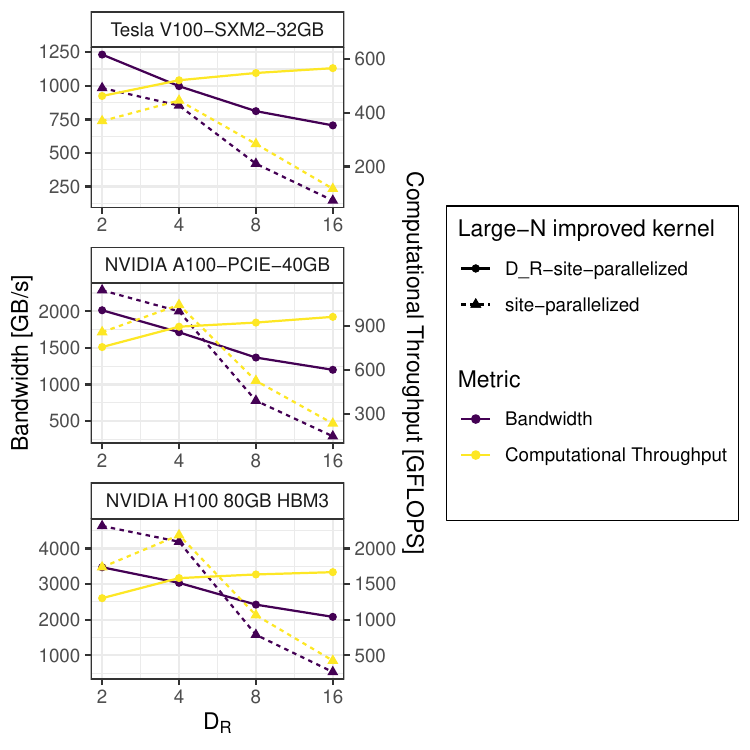}
\caption{Single GPU bandwidths reached depending on the fermion dimension $D_{R}$ using different GPU types. For large $N_g$, we generally see that the site-parallelized kernel experiences a sudden decrease in bandwidth due to register spilling. Further, the oldest card, NVIDIA V100, suffers the most from long scoreboard warp stalls, which the $D_R$-site-parallelized kernel can avoid better. Note that the lines between the measurements are only drawn to guide the eye.}
\label{fig:bwlargeN}
\end{figure}

When we increase the number of colors for a fixed lattice size, the amount of memory moved by the kernel will increase with $N_c^2$. The reached bandwidths also depend on the amount of memory moved, meaning, if the amount of memory is too low, it may not reach the peak performance because the kernel call overhead dominates the computation. This memory dependence can be observed for different GPUs in Fig.~\ref{fig:singlegpudirac}. For the tests in Fig.~\ref{fig:bwlargeN}, we ensured that we were always located in the saturated region.

In addition, we aim to simulate theories with arbitrary numbers of colors without loss of efficiency. When applying the Dirac operator moving a constant amount of total memory, we aim at no dependence on the dimension of the fermion representation. However, due to the matrix multiplication with the represented gauge field matrices in the Dirac operator, the operation on the spinor components cannot be trivially parallelized and requires rereading part of the memory. In addition, the missing L3 cache limits performance if the memory requirements of a single thread become too large due to register spilling. As a result, a kernel parallelized in the fermion dimension $D_R$ and lattice sites behaves better for large-$N_g$, while a parallelized only in sites might still outperform the improved version for smaller $N_g$. The exact critical $N_g$ depends on the specific hardware.

The register spilling in the GPU Wilson-Dirac operator is reduced because of the following: In the Dirac operator, applications of $SU(D_R)$ matrices to vectors take for a complex representation $D_R^2 + D_R$ complex numbers and $4D_R^2$ double precision floating point operations. Depending on the available registers, we can achieve efficiency only up to a specific $N_c$. In contrast, the $D_R$-site-parallelized implementation splits the computation so that a GPU thread only computes one vector component of the result for one site. As a result, the local operations are not matrix-vector multiplications but scalar products of two $SU(N_g)$ vectors, the original vector and a row vector of the matrix, corresponding to $2D_R$ complex numbers and $4D_R$ double precision floating point operations. For $SU(3)$, this already needs only half the memory per thread. The downside is that the vector components in the matrix multiplication need to be read from each thread corresponding to a row of the $SU(D_R)$ matrix. However, the cache locality of the components in this parallelization is good, so the overhead compared to the standard implementation is expected to be marginal. While without spilling, the naive kernel is often faster, when spilling is present, using the parallelized kernel can speed up the execution of the Dirac operator substantially.

The only exception are cards with lower compute capabilities that have slower global memory access, which can result in long scoreboard stalling of the warp execution, which is worse for the site-parallelized kernel than it is for the $D_R$-site-parallelized kernel because less memory is read per thread which reduces memory dependencies and associated stalls. The NVIDIA V100 GPU shows such a case: Here the stalling in the site-parallelized kernel is sufficiently severe for the $D_R$-site-parallelized kernel to achieve higher bandwidth for any $N_g$.

For the default block size 256, we tested the NVIDIA H100, where the Large-$N_c$ $D_R$-site-parallelized outperforms the site-parallelized kernel for $N_c>5$. On an NVIDIA A100, the $D_R$-site-parallelized only outperforms the site-parallelized for $N_c>6$ and a V100 profits from the $D_R$-site-parallelized for all $N_c$. The AMD Instinct MI250X and NVIDIA V100 are only efficient when the $D_R$-site-parallelized kernel is used. 

Fig.~\ref{fig:bwlargeN} shows the behavior of the bandwidth and computational throughput reached for the application of the Wilson-Dirac operator at different $N_g$. Given only a site-parallelized kernel, one can see that the bandwidth and computational throughput are decreasing for $N_g > 4$ due to the register spilling. However, when using the Large-$N_g$ $D_R$-site-parallelized kernel, one observes the expected behavior that the kernel stops being memory bound and becomes compute bound. 

\subsection{Scaling on large systems}
\begin{figure}
\centering
\includegraphics[width=2in]{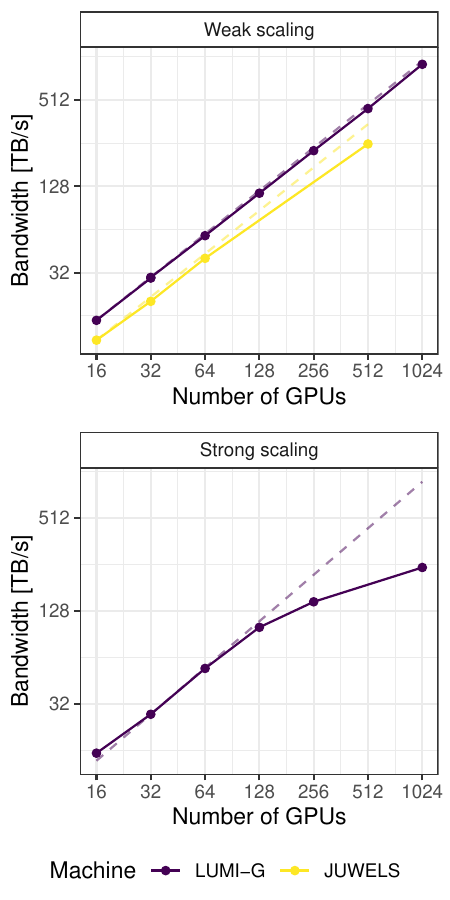}
\caption{Strong and weak scaling on LUMI-G with all dimensions parallelized up to 1024 GPUs for an SU(3) gauge group with fundamental fermions.}
\label{fig:thousand}
\end{figure}
Fig.~\ref{fig:thousand} shows the ability of the code to scale to a considerable number of GPUs. Here, we scale up to 1024 GPUs, reaching a peak bandwidth of $908~\mathrm{TB/s}$ for weak scaling $SU(3)$ with fermions in the fundamental representation. While the weak scaling is close to perfect, the strong scaling shows some loss of efficiency for the smallest local lattices. This is not a problem of the code because the computation is becoming too fast to mask the communications.

\begin{figure}
\begin{minipage}[b][16cm][b]{0.48\linewidth}
\centering
\includegraphics[width=0.8\linewidth,trim={0 0 0 3.5cm},clip]{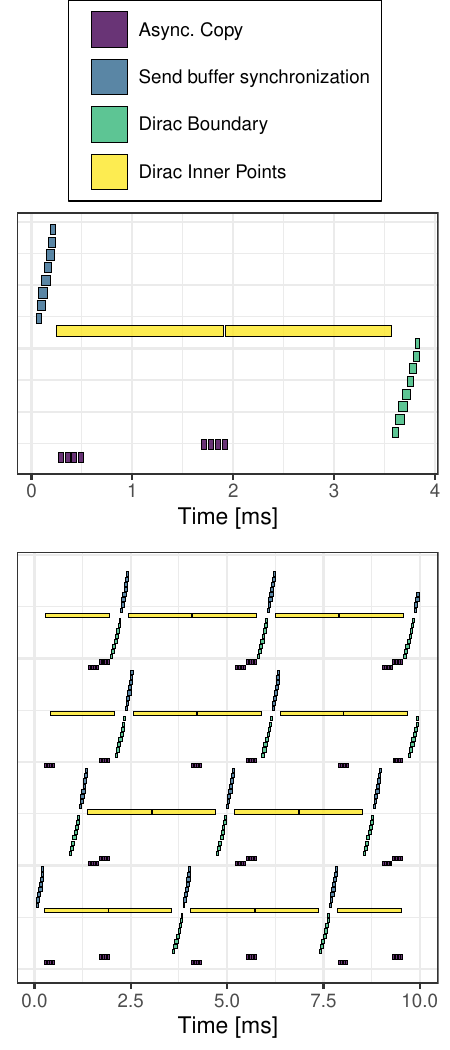}
\subcaption{NVIDIA Nsight profile of the Dirac operator application for an SU(3) gauge group with fermions in the fundamental representation on a GPU node with four NVIDIA H100 GPUs using a $48^4$ local lattice, parallelized in two dimensions}
\end{minipage}
\hfill
\begin{minipage}[b][16cm][b]{0.48\linewidth}
\centering
\includegraphics[width=0.8\linewidth]{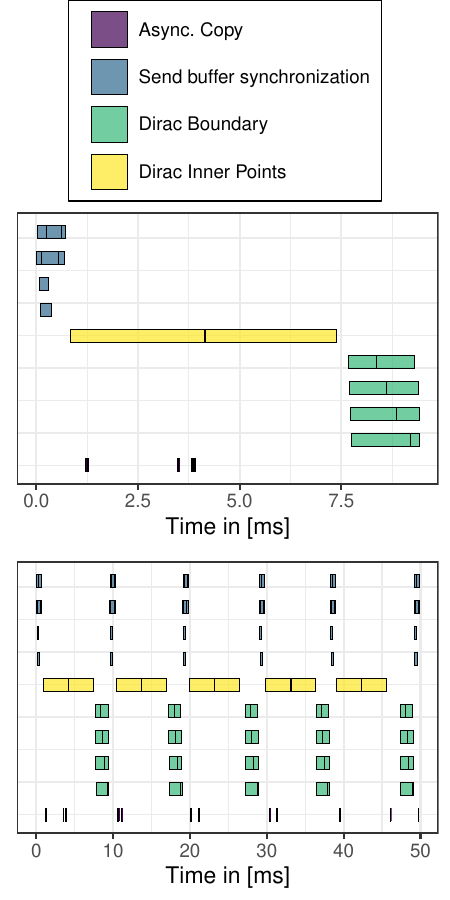}
\subcaption{rocprof ROCm profile of the Dirac operator application for an SU(3) gauge group with fermions in the fundamental representation on a GPU node with four AMD Instinct MI250x GCDs using a $48^4$ local lattice, parallelized in two dimensions}
\end{minipage}
\caption{Tracing profiles on different GPUs.}
\label{fig:tracingprofiles}
\end{figure}

Fig.~\ref{fig:tracingprofiles} shows the masking of communications using a tracing profiler. Here, the execution of the points of the inner Dirac operator takes the most time. The communications are executed in parallel to the inner Dirac kernels. The synchronization to the send buffer is executed for all parallel dimensions for the even and odd send buffer and placed as much as possible in parallel using GPU streams. The boundary operations are also executed in parallel using an atomic write to avoid the race condition in the corners of the local lattice. Because this race condition concerns only a small number of sites, the overhead through the atomic operation is much more negligible and entirely offset by the increase in performance due to the additional parallelism.

\begin{table}
\centering
\caption{Bandwidths for strong and weak scaling on LUMI-G with all dimensions parallelized up to 1024 GPUs for an SU(3) gauge group with fundamental fermions. }
\begin{tabular}{rrrrrrrrrl}
\toprule
$T/a$ & $X/a$ & $Y/a$ & $Z/a$ & GPUs & Machine & \multicolumn{2}{l}{Bandw. [$\mathrm{TB/s}$] /[\%]} \\
\midrule
48 & 48 & 48 & 48 & 16 & JUWELS & 10.83935  & 44\%\\
96 & 48 & 48 & 48 & 32 & JUWELS & 20.21290 & 41\%\\
192 & 48 & 48 & 48 & 64 & JUWELS & 40.31304 & 41\%\\
1536 & 48 & 48 & 48 & 512 & JUWELS & 252.62902  & 32\%\\
\addlinespace
96 & 96 & 96 & 96 & 16 & LUMI-G & 14.90487 & 58\%\\
192 & 96 & 96 & 96 & 32 & LUMI-G & 29.36911 & 57\%\\
192 & 96 & 96 & 96 & 32 & LUMI-G & 29.68592 & 58\%\\
384 & 96 & 96 & 96 & 64 & LUMI-G & 57.98584 & 57\%\\
768 & 96 & 96 & 96 & 128 & LUMI-G & 114.62270 & 56\%\\
1536 & 96 & 96 & 96 & 256 & LUMI-G & 226.98568 & 55\%\\
3072 & 96 & 96 & 96 & 512 & LUMI-G & 445.84729 & 54\%\\
6144 & 96 & 96 & 96 & 1024 & LUMI-G & 908.00002 & 55\%\\
\addlinespace
96 & 96 & 96 & 96 & 16 & LUMI-G & 15.32197 & 60\%\\
96 & 96 & 96 & 96 & 32 & LUMI-G & 27.37527 & 53\%\\
96 & 96 & 96 & 96 & 64 & LUMI-G & 54.18906 & 53\%\\
96 & 96 & 96 & 96 & 128 & LUMI-G & 100.06425 & 49\%\\
96 & 96 & 96 & 96 & 256 & LUMI-G & 146.36395 & 36\%\\
96 & 96 & 96 & 96 & 1024 & LUMI-G & 244.15889 & 15\%\\
\bottomrule
\end{tabular}
\end{table}
\section{The Large-$N_g$ limit using \HiRep}\label{sec:largeN}
\begin{figure}
\centering
\includegraphics[width=2.5in]{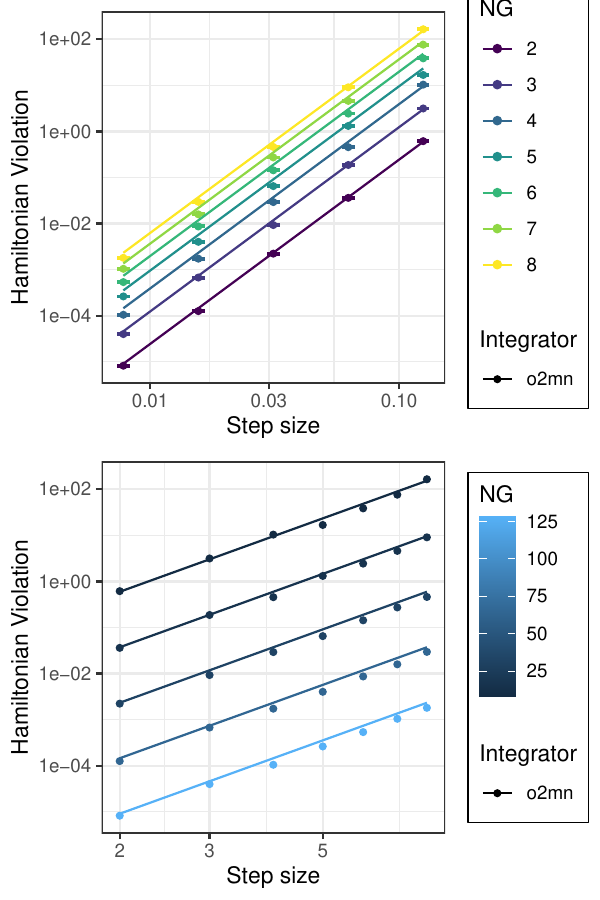}
\caption{Scaling of the Hamiltonian violations with both $N_g$ and $\delta\tau$}
\label{fig:hamvio}
\end{figure}
\begin{figure}
\centering
\includegraphics[width=2.5in]{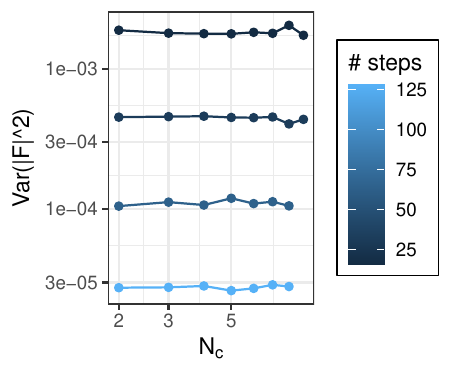}
\caption{Scaling of the variance of the gauge force in the molecular dynamics evolution with $N_g$.}
\label{fig:gaugeforce}
\end{figure}
The large-color limit of $SU(N_g)$ theories has many interesting phenomenological properties and can be explored using \texttt{HiRep}. Here, we demonstrate that \HiRep scaling with $N_g$ is consistent with analytical formulas for the integrators used.

The scaling of the integrator can be characterized by its Hamiltonian violations, more precisely, their relation to the difference of the energy given by the Hamiltonian after integration to an exactly conserved shadow Hamiltonian for the specific integrator \cite{Clark:2008gh, Clark:2010qw, Clark:2011ir, Kennedy:2012gk}. These violations are the difference in the action of the gauge configuration between the beginning and the end of the molecular dynamics evolution. They can be characterized by analytical differences to an exactly conserved shadow Hamiltonian and most easily analyzed in a pure gauge setup. Here we use the Hamiltonian
\begin{equation}
H = \dfrac{1}{2}T_{\mathrm{FUND}}\sum_{x,a,\mu}\pi_{\mu}^{a}(x)^2 + \beta \sum_{x,\mu<\nu}\left( 1 - \dfrac{1}{N_g}\mathrm{Re\,tr}P_{\mu\nu}(x)\right)
\end{equation} 
where $\pi$ are the momenta and $P_{\mu\nu}$ the plaquette as defined in Eq.~\ref{eq:plaquette}.

The integration with any symplectic integrator is only exact up to terms of order $\mathcal{O}(\delta\tau^{k})$, where $\delta\tau$ is the discrete step size and $k$ depends on the order of the integrator. Therefore, the integrator does not conserve the true Hamiltonian exactly, but instead conserves an associated shadow Hamiltonian. In the particular case of a 2nd-order Omelyan integrator with the free parameter chosen as $\alpha = 1/6$, the exactly conserved shadow Hamiltonian is \cite{Bussone:2018mzi}
\begin{equation}
\widetilde{H} = H + \dfrac{\delta\tau^2}{72}F^2 + \mathcal{O}(\delta\tau^4)\,,
\end{equation}
where $F$ denotes the driving force of the gauge dynamics.

Correspondingly, the exact expression for these violations as in \cite{Bussone:2018mzi} for a 2nd-order Omelyan integrator in a pure gauge theory is
\begin{equation}
\Delta H := \widetilde{H} - H = \dfrac{\delta \tau^2}{72}\sum_{x,a}T_{\mathrm{FUND}}|F^{a}_{\mathrm{G}}(x)|^2 + \mathcal{O}(\delta \tau^4)\,.
\end{equation}

Consequently, we find
\begin{equation}
\Delta H = \dfrac{\delta \tau^2\beta^2}{72N_g^2T_{\mathrm{FUND}}}\sum_{x,\mu,a}\mathrm{Re}\,\mathrm{tr}\, [\mathrm{i}T_{\mathrm{FUND}}^aUC^{\dagger}]^2\,.
\end{equation}

The difference between the Hamiltonian and shadow Hamiltonian, $\Delta H = \widetilde{H} - H$, and the difference between the Hamiltonian $H$ at the start of the trajectory and the Hamiltonian $H'$ at the end of the trajectory coined Hamiltonian violation $\delta H = H' - H$ are not directly related. However, their variances are sampled approximately by the same distribution \cite{Clark:2008gh} so that we can conclude
\begin{equation}
\mathrm{Var}(\delta H) = \mathrm{Var}(\Delta H) = \dfrac{\delta \tau ^4}{(72)^4}\mathrm{Var}(\mathcal{F}_{G})
\end{equation}
with 
\begin{equation}
|\mathcal{F}_{G}|^2 = \sum_{x}T_F|F_G(x)|^2\,.
\end{equation}
Further, the Creutz equality $\langle e^{-\delta H}\rangle = 1$ constrains the expectation of the Hamiltonian violations to approximately coincide with their variance
\begin{equation}
\langle \delta H\rangle = \mathrm{Var}(\delta H) \,+\,\mathcal{O}(\langle\delta H^3\rangle)+\mathcal{O}(\langle \delta H \rangle^2)\,.
\end{equation}

The string constant of the quark potential determines the variance of the gauge forces. Given an approximate 't Hooft rescaling of the bare coupling $\beta \propto N^2$, this string tension is kept constant to first order in the strong coupling expansion of the potential. As a result, we can find the scaling of the Hamiltonian violations
\begin{equation}
\Delta H \propto \delta \tau^2 N_g^4 +\mathcal{O}(\delta \tau^4)\,.
\end{equation}
Both the behavior of the variances of the forces and the Hamiltonian violations with respect to the number of colors and the step size can be numerically confirmed with \texttt{HiRep}, as shown in Fig.~\ref{fig:hamvio} and \ref{fig:gaugeforce}.

Note that this was partly developed during the project \cite{Ziegler:2021nbl, Jaeger:2022ypq, DellaMorte:2023ylq, Martins:2023kcj, DellaMorte:2025tks}

\section{Spectroscopy of adjoint fermions with an SU(2) gauge group}\label{sec:su2adjoint}

We want to check specifically, that the spectroscopy code is giving consistent results between CPU and GPU. For this, we are using a theory where fermions are in a higher representation, so the representation of the gauge field is not trivial. Further, the adjoint representation of $SU(2)$ is a real representation, so this demonstrates, in addition, the correctness of using real numbers instead of complex numbers for the represented gauge field. We are using unimproved Wilson fermions for the results shown below. These results should be regarded as integration tests of the software.

\begin{table}
\centering
\begin{tabular}{lllllp{1.3cm}p{1.3cm}}
Name & $\beta$ & m & $m_{\mathrm{PCAC}}$ & Lattice size & Statistics [Configs] & Separation [MDU]\\
\midrule
T2B11 & 2.3 & -1.34 & 0.0168(12) & $32\times 24^3$ & 16 & 50\\
\end{tabular}
\caption{Some features of the ensemble T2B11 with parameters from \cite{DelDebbio:2008zf} but repeated on a larger lattice.}
\end{table}

\begin{figure}
\centering
\includegraphics[width=2in]{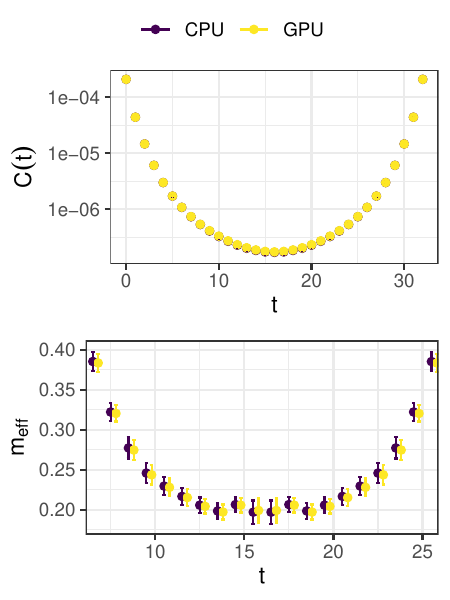}
\caption{Correlation function and effective mass of the pseudoscalar meson channel evaluated with the GPU measurement code for the ensemble above at a $32\times 24^3$ lattice}
\label{fig:effmps}
\end{figure}

\begin{figure}
\centering
\includegraphics[width=2.5in]{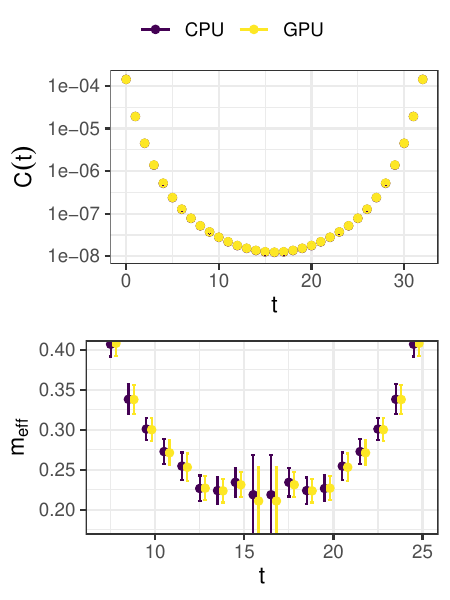}
\caption{Correlation function and effective mass of the vector meson channel evaluated with the GPU measurement code for the ensemble above at a $32\times 24^3$ lattice}
\label{fig:effmv}
\end{figure}

A theory with an $SU(2)$ gauge group and two mass-degenerate fermions in the adjoint representation is given by the Lagrangian
\begin{equation}
\mathcal{L} = -\dfrac{1}{4}G_{\mu\nu}^{a}G_{\mu\nu}^{a} + u(i\slashed{D} - m)u + d(i\slashed{D} - m)d
\end{equation}
with $u$ and $d$, the adjoint quarks in the theory. Using the configurations obtained for the ensemble studied in section \ref{sec:tests}, we can evaluate the effective masses for the pseudoscalar and vector mesons, effective masses shown in Figs.~\ref{fig:effmps} and \ref{fig:effmv}. 
%

This theory is approaching a conformal theory in the chiral limit. As a result, despite a relatively small pseudoscalar and partially conserved axial current (PCAC) mass, the ratio between the pseudoscalar and vector mass is close to one. This is consistent with the physics close to the conformal fixed point: The ratio stays constant instead of diverging. 

\begin{figure}[H]
\centering
\includegraphics[width=2.5in]{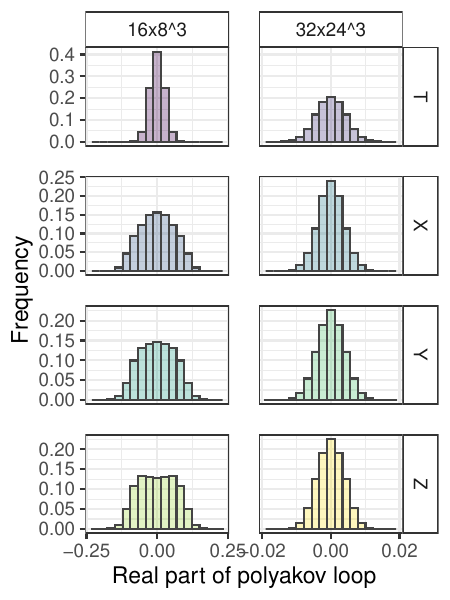}
\caption{Polyakov loop distribution}
\label{fig:polyakov}
\end{figure}

We also examined the distribution of the Polyakov loop, see Fig.~\ref{fig:polyakov}, which for the smaller lattices in \cite{DelDebbio:2008zf} of size $16\times 8^3$ is spatially broken and therefore admits two metastable equilibria. We excluded this effect in the algorithmic tests performed using the data from this ensemble by using a sufficiently large lattice to show a symmetric distribution around zero.

\section{Conclusion}\label{conclusion}

\HiRep~v2 is a software suite for lattice field theory simulations with a rich set of features. Its primary focus is the efficient numerical simulations of Beyond-the-Standard-Model physics by allowing the explicit computation of observables in non-perturbative regimes of novel strongly coupled theories. In \HiRep~v2, we added many features to improve the precision of spectroscopic studies, such as different implementations of actions, including Symanzik improvement and Hasenbusch acceleration, sources for the calculation of propagators, smearing, boundary conditions, and special algorithms such as multi-level updates, featuring a large amount of automatically evaluated observables. The software suite is suitable for high-performance computing on modern large supercomputers, supporting efficient parallelization with OpenMPI, GPU acceleration, and OpenMP multithreading. 

Future work on this extensive library could include the support for mixed representations and domain-decomposition methods and multi-grid inverters. Modern noise reduction techniques, such as Wuppertal smearing or distillation, would be great additions to the software.

\section*{CRediT authorship contribution statement}

\textbf{Vincent Drach:} Conceptualization, Formal analysis, Software, Writing -- review \& editing. \textbf{Sofie Martins:} Software, Investigation, Validation, Visualization, Writing -- original draft, Writing -- review \& editing.
\textbf{Claudio Pica:} Conceptualization, Methodology, Funding acquisition, Project administration, Resources, Software, Supervision, Validation, Writing -- review \& editing.
\textbf{Antonio Rago:} Conceptualization, Software, Supervision, Writing -- review \& editing.

\section*{Acknowledgments}
This project has received funding from the European Union’s Horizon 2020 research and innovation program under the Marie Skłodowska-Curie grant agreement №813942. Testing, development, and benchmarking of this software was possible using resources on LUMI-G provided by the Danish eInfrastructure Consortium under grant application number DeiC-SDU-N5-2024055 and NVIDIA V100, A100, and H100 nodes provided by the UCloud DeiC Interactive HPC system managed by the eScience Center at the University of Southern Denmark. This work used the DiRAC Extreme Scaling service (Tursa) at the University of Edinburgh, managed by the Edinburgh Parallel Computing Centre on behalf of the STFC DiRAC HPC Facility (www.dirac.ac.uk). The DiRAC service at Edinburgh was funded by BEIS, UKRI and STFC capital funding and STFC operations grants. DiRAC is part of the UKRI Digital Research Infrastructure. Some of the work was carried out using the computational facilities of the High Performance Computing Centre at the University of Plymouth.

\appendix

\section{C/C++ Generic Linear Algebra}\label{app:lina}
\begin{figure}
\centering
\includegraphics[width=3in]{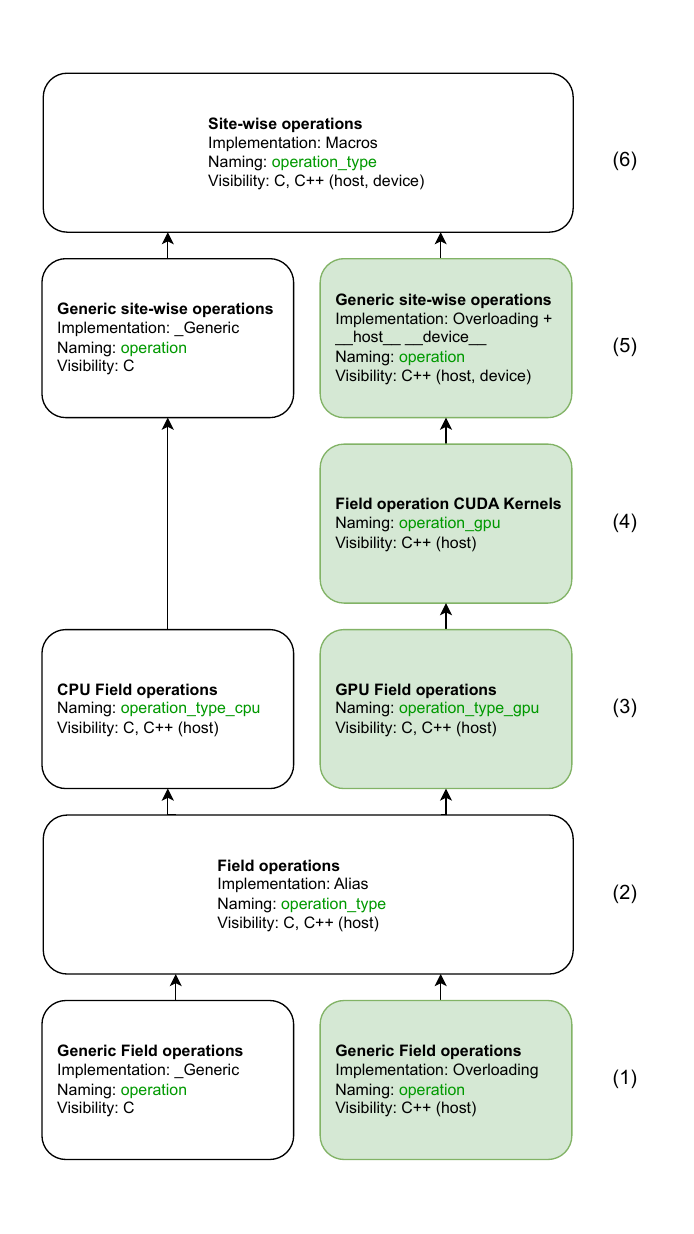}
\caption{Implementation of generic functions for availability in both C and CUDA C when compiling with GPU support}
\label{fig:genericgpu}
\end{figure}
In this appendix, we describe how generic functions for the basic linear algebra operations in \HiRep are implemented.  

While the CUDA C++ files of the library can use overloaded function definitions to call the core library functions, the C code needs an alternative implementation using the keyword \texttt{\_Generic} for generic macros. This is implemented as illustrated in Fig.~\ref{fig:genericgpu}. Here, operations applied on the elementary site types are implemented using manually unrolled operations, which are automatically generated during compilation using a \texttt{perl} script. These macros, step (6) in Fig.~\ref{fig:genericgpu}, implement operations such as multiplication of a spinor with a scalar addition of two $SU(N_g)$ vectors. Generic C functions are then defined using the \texttt{\_Generic} keyword, see step (5). This, for example, allows us to call the operation \texttt{mul} with an input and output spinor and a real number, which will then call the macro \texttt{\_spinor\_mul\_f}. Since \texttt{\_Generic} and the elementary type operations are all macros, all functions will be fully inlined, and function call overhead is avoided.

While the elementary-type macros are still available, the generic version allows us to define type-independent templates. At the field operations level, we can now define C-templates referencing the generic version, step (3) in Fig.~\ref{fig:genericgpu}. All linear algebra functions will be implemented using the scheme \texttt{operation\_type\_architecture}. For each operation that specifies an architecture and alias exists with the name \texttt{operation\_type} that reroutes to the function with the correct architecture, which is the function ending in \texttt{\_cpu} when compiling without GPU acceleration and the function ending in \texttt{\_gpu} if not. Note that the GPU-accelerated functions are templates that internally call kernels, which are implemented separately in step (4) in Fig.~\ref{fig:genericgpu}.

To give an example for the generics at the field level, we implement functions that calculate the real part of the scalar product between two spinor fields. This function will be called \texttt{prod\_re\_spinor\_field\_cpu} for the version that executed the operation on CPUs. Internally, this is a reduction operation over the sites of the two fields, where for each site, we call a generic operation \texttt{prod\_re}, and \texttt{\_Generic} will resolve this call to the correct elementary site type. 

To call generic functions at the field operations level, we now add the resolution of \texttt{prod\_re} when called with spinor fields to \texttt{prod\_re\_spinor\_field\_cpu}. This simplifies the code dramatically in implementing high level algorithms, for example, inverters, because calling the generic operation will always automatically resolve the type and architecture.

When adding GPU acceleration, achieving this is more complicated. The code now features C and CUDA C (\texttt{C++}) source files. In contrast, \texttt{C++} supports generics by function overloading and is incompatible with the keyword \texttt{\_Generic}. As a result, we need to duplicate each generic function with an overloaded function definition in the header. This function has to be located in the header to be correctly inlined by the compiler and avoid function call overhead in the kernels. 

Further, in contrast to the \texttt{\_Generic} definitions, the \texttt{C++} function definitions need to be available from the host and the device. This logic is implemented by employing the \HiRep native keyword \texttt{visible}, which is equivalent to the keyword \texttt{\_\_host\_\_ \_\_device\_\_} if compiling with GPU acceleration and has no meaning when compiling without it.





\bibliographystyle{elsarticle-num}
\bibliography{literature}







\end{document}